\documentclass[a4paper,11pt]{article}
\pdfoutput=1
\usepackage{jheppub}
\usepackage{amsmath}
\usepackage{marvosym}
\usepackage{hyperref}
\usepackage[normalem]{ulem}
\usepackage{amsthm}
\usepackage{slashed}
\usepackage{amsfonts}
\usepackage{amssymb}
\usepackage{textcomp}
\usepackage{graphicx}
\usepackage{bm}
\usepackage{colortbl}
\usepackage{verbatim}
\usepackage{graphicx,color}
\usepackage{epsfig}

\usepackage{graphicx}
\usepackage{dcolumn}
\usepackage{bm}

\def\coeff#1#2{{\textstyle {\frac {#1}{#2}}}}
\def\MO{{M_5}}
\def\aBB{{\chi_{B}}}

\DeclareMathAlphabet{\mathpzc}{OT1}{pzc}{m}{it}

\definecolor{Gray}{gray}{0.9}

\renewcommand{\vec}[1]{\boldsymbol{#1}}

\usepackage{color}
\PassOptionsToPackage{usenames, dvipsnames}{color}

\usepackage{xcolor}

\begin{document}

\title{
Chiral hydrodynamics in strong 
external magnetic fields
}%

\author[a]{Martin Ammon,}%
\author[a,b,c]{Sebastian Grieninger,}%
\author[d,e]{Juan Hernandez,}
\author[f]{Matthias Kaminski,}
\author[f]{Roshan Koirala,}
\author[a]{Julian Leiber,}%
\author[f]{Jackson Wu}
\preprint{IFT-UAM/CSIC-20-179}

\affiliation[a]{Theoretisch-Physikalisches Institut, Friedrich-Schiller-Universit\"at Jena,
Max-Wien-Platz 1, D-07743 Jena, Germany.} 
\affiliation[b]{Institute for Theoretical Physics, Vienna University of Technology (TU Wien), Wiedner Hauptstra\ss e 8-10, A-1040 Vienna, Austria.}
\affiliation[c]{Instituto de F\'isica T\'eorica UAM/CSIC and Departamento de F\'isica T\'eorica, Universidad Aut\'onoma de Madrid, Campus de Cantoblanco, C/Nicolás Cabrera 13–15, ES-28049 Madrid, Spain}
\affiliation[d]{Perimeter Institute for Theoretical Physics, 31 Caroline St N, Waterloo, ON N2L 2Y5, Canada}
\affiliation [e]{Department of Physics and Astronomy, University of Waterloo, 200 University Avenue West, Waterloo, Ontario, Canada N2L 3G1}
 \affiliation[f]{Department of Physics and Astronomy, University of Alabama, 514 University Boulevard,Tuscaloosa, AL 35487, USA }

\emailAdd{martin.ammon@uni-jena.de}
 \emailAdd{sebastian.grieninger@gmail.com}
\emailAdd{jhernandez@perimeterinstitute.ca}
\emailAdd{mski@ua.edu}
\emailAdd{rkoirala1@ua.edu}
\emailAdd{jmwu@ua.edu}

\date{\today}

\abstract{
We construct the general hydrodynamic description of (3+1)-dimensional chiral charged (quantum) fluids subject to a strong external magnetic field with effective field theory methods. 
We determine the constitutive equations for the energy-momentum tensor and the axial charge current, in part from a generating functional. 
Furthermore, we derive the Kubo formulas which relate two-point functions of the energy-momentum tensor and charge current to 27 transport coefficients:  
8 independent thermodynamic, 4  independent non-dissipative hydrodynamic, and 10 independent dissipative hydrodynamic transport coefficients. 
Five  Onsager relations render 5 more transport coefficients dependent.  
We uncover four novel transport effects, which are encoded in what we call the shear-induced conductivity, the two expansion-induced longitudinal conductivities and the shear-induced Hall conductivity. Remarkably, the shear-induced Hall conductivity constitutes a novel non-dissipative transport effect. 
As a demonstration, we compute all transport coefficients explicitly in a strongly coupled quantum fluid via holography. 
}
\maketitle
\section{Introduction}
\label{sec:introduction}
Hydrodynamics is a universal effective field theory description of collective phenomena in (quantum) systems with many degrees of freedom.
The hydrodynamic description of relativistic fluids has facilitated the physical interpretation of data from heavy ion collisions. 
It has also allowed to describe effects in condensed matter systems from 
charge conduction in metals to more recent descriptions of Weyl semimetals and graphene. 

In this work, we consider the hydrodynamic description of a (3+1)-dimensional chiral charged thermal fluid  subject to a strong external magnetic field. 
This is not to be confused with magnetohydrodynamics, in which the magnetic field satisfies Maxwell's equations and hence is {\it dynamical}.
Our {\it external} magnetic field is {\it not dynamical} in that sense. 
This magnetic field and the associated gauge potential $A_\mu$ can either be related to a vector $U(1)_V$ symmetry assumed to be preserved, or to an axial $U(1)_A$ symmetry which may be broken by a chiral anomaly. We point out the differences in the hydrodynamic description depending on which type of the two $U(1)$ symmetries is considered. 
Our focus, however, is the anomalous axial $U(1)_A$ case. 
We limit our considerations to zeroth and first order in the hydrodynamic derivative expansion. The magnetic field is referred to as {\it strong} if it is defined to be of zeroth order in derivatives. 
On a fundamental level, no axial magnetic fields exist in Nature. However, in low energy electronic descriptions for condensed matter systems such as Weyl-semimetals effectively axial magnetic fields and axial potentials can be created~\cite{Cortijo_2015,Pikulin_2016,Grushin_2016,Cortijo:2016wnf}. 
Thus we focus here on exploring the effects of an axial $U(1)_A$ symmetry. 
Based on previous results~\cite{Hernandez:2017mch,Son:2009tf,Neiman:2010zi}, we expect that adding a conserved vector current associated with a dynamical gauge field will not qualitatively change the physical effects, merely distribute them over different currents, introducing copies of analogous effects.  
Therefore, the hydrodynamic description derived here can be extended to be applied to the quark-gluon-plasma generated in heavy-ion-collisions. 

Our main result are the Kubo formulas for the transport coefficients of a charged chiral thermal fluid subject to a strong magnetic field. 
These are derived in part from a generating functional and in part from constitutive relations which we construct in all generality in section~\ref{sec:hydrodynamics}. 
A nonzero strong magnetic field necessarily leads to anisotropic equilibrium states. In addition, an axial chemical potential breaks parity on the level of the state. 
The combination of these two broken symmetries explains the large number of transport coefficients. 
Specifically, we find 8 
independent thermodynamic transport coefficients, and initially 19 hydrodynamic transport coefficients. 
Among the hydrodynamic transport coefficients we find 5 Onsager relations, which leaves 14 independent hydrodynamic transport coefficients. Of those hydrodynamic transport coefficients 3 are non-dissipative and 11 are dissipative. All independent transport coefficients are listed in tables~\ref{tab:transportCoeffsThermo}, \ref{tab:transportCoeffsHydroNonDiss} and \ref{tab:transportCoeffsHydroDiss}. 

The first subset of the 8 
independent thermodynamic transport coefficients are 
the well-known chiral conductivities: vortical $\xi=\xi_{TB}$, chiral magnetic $\xi_B$,  and chiral thermal $\xi_T$. 
Referring to $\xi_B$ as ``chiral magnetic effect'' is a slight abuse of language.  
In most of this paper $\xi_B$ will measure the response of an axial current to an axial $U(1)_A$ magnetic field. The term ``chiral magnetic effect'' was coined for the response of such a current to a $U(1)_V$ vector magnetic field~\cite{Kharzeev:2004ey}, see also~\cite{Jensen:2013vta}. 
The remaining 5 thermodynamic transport coefficients are 
4 newly non-vanishing susceptibilities:  
magneto-thermal $M_1$, perpendicular magneto-vortical $M_2$, magneto-acceleration $M_3$, magneto-electric $M_4$\footnote{$M_1,\, M_2,\, M_3,\, M_4$ have to vanish in a parity-preserving microscopic theory when coupled to an external vector $U(1)_V$ gauge field.  However, they are allowed to be nonzero when the coupling to an axial $U(1)_A$ is considered.}, 
and the previously discussed~\cite{Kovtun:2016lfw,Hernandez:2017mch} magneto-vortical susceptibility~$M_5$.\footnote{Note that this $M_5$ was previously referred to as $M_4=M_\Omega$~\cite{Hernandez:2017mch} but considering only a coupling to a external vector $U(1)_V$ gauge field, while we also consider coupling to an axial $U(1)_A$ here. Our $M_5$ was also referred to as $M_{20}$ in~\cite{Kovtun:2016lfw}.} 
The magneto-acceleration susceptibility $M_3$ vanishes in conformal field theories. 
Note that in addition there are three susceptibilities $\chi_{B}\, , \chi_{33},\, \chi_{13}$ which we do not include in our counting since they are thermodynamic partial derivatives of the pressure, $p$. 
In stark contrast to this, the $M_n$ ($n=1,\,2,\,3,\,4,\,5$) can not be defined as thermodynamic derivatives of the pressure, thus they are independent coefficients. 

There are initially 19 hydrodynamic transport coefficients: the shear viscosities perpendicular and longitudinal to the magnetic field $\eta_\perp,\, \eta_{||}$, the perpendicular and longitudinal Hall viscosities $\tilde\eta_\perp,\, \tilde\eta_{||}$, the bulk viscosities $\zeta_1$, $\zeta_2$, $\eta_1$, $\eta_2$, the perpendicular and longitudinal charge conductivities $\sigma_{\perp}$, $\sigma_{||}$, the Hall conductivity $\tilde\sigma_\perp$ (which can alternately be expressed in terms of the charge resistivities $\rho_\perp, \, \rho_{||},\, \tilde\rho_\perp$), and the novel $c_3$, $c_4$, $c_5$, $c_8$, $c_{10}$, $c_{14}$, $c_{15}$, $c_{17}$. 
Due to their effect on the fluid to be interpreted in section~\ref{sec:transportInterpretation}, we name 
$c_8$ the {\it shear-induced conductivity}, 
$c_{10}$ the {\it shear-induced Hall conductivity}, 
as well as $c_4$ and $c_5$ the {\it expansion-induced longitudinal conductivities}. 
Only 4 of the $c_n$'s (with $n=3, 4, 5, 8, 10, 14, 15, 17$) are independent. Only 3 of the bulk viscosities are independent. This follows from 5 Onsager relations derived in this work. 

In section~\ref{sec:holography}, we prove the existence of these transport coefficients by direct computation of their nonzero values within the specific example of a strongly coupled $\mathcal{N}=4$ Super-Yang-Mills (SYM) theory at a large number of colors, $N_c\to\infty$, coupled to an external axial $U(1)_A$ gauge field. This computation is facilitated by holography~\cite{Maldacena:1997re}. 
In order to allow for a charged thermal state subject  to a strong magnetic field, the charged magnetic black brane solutions are considered~\cite{D'Hoker:2009bc}. 
Within the classical gravity dual to SYM theory, we compute the frequency or momentum dependent fluctuations around the branes, which are holographically dual to field theory correlation functions of the energy momentum tensor and the axial current. 
Applying the Kubo formulas derived in  section~\ref{sec:hydrodynamics}, 
we obtain nonzero values for most of the transport coefficients. 
An exception are the transport coefficients $c_3$, $M_4,\, \tilde\eta_{\perp}$, $\zeta_1$ and $\zeta_2$ which vanish in the holographic model. 
The status of $M_1$ and $M_3$ is unclear within the holographic model as we only have Kubo relations for their derivatives. However, $M_3$ is expected to vanish due to conformal invariance. 

The effect of chiral anomalies in hydrodynamics\footnote{In a non-hydrodynamic context, the effect of chiral anomalies on currents was pioneered by~\cite{Vilenkin:1978is,Vilenkin:1979ui,Vilenkin:1980zv}.} 
was first found through holographic calculations which yielded nonzero anomalous transport in~\cite{Erdmenger:2008rm,Banerjee:2008th,Torabian:2009qk}. More generally, the existence of anomalous transport as a consequence of chiral anomalies was elucidated in terms of a local version of the second law of thermodynamics in~\cite{Son:2009tf}. Subsequent studies of anomalous hydrodynamics include \cite{Neiman:2010zi,Kharzeev:2009pj,Kharzeev:2010gd,Kharzeev:2013ffa,Jensen:2012kj,Jensen:2013kka,Jensen:2013vta}. The equilibrium partition function formulation of relativistic hydrodynamics was first introduced in~\cite{Jensen:2012jh,Banerjee:2012iz} and subsequently used in a variety of settings~\cite{Bhattacharyya:2012xi,Bhattacharya:2012zx,Bhattacharyya:2013ida,Kovtun:2016lfw,Hernandez:2017mch,Kovtun:2018dvd,Armas:2018zbe,Kovtun:2019wjz}. In particular, the equilibrium generating functional approach was used to formulate the hydrodynamic framework of (parity preserving) fluids subject to strong external as well as the framework of magnetohydrodynamics (when the electromagnetic field is dynamical) in~\cite{Hernandez:2017mch}. The relation between the frameworks of hydrodynamics with strong magnetic fields and magnetohydrodynamics, as well as the dual formulation of magnetohydrodynamics in terms of two-form fields introduced in~\cite{Grozdanov:2016tdf} was elucidated in~\cite{Hernandez:2017mch}. The anomaly inflow generating functionals have been used for anomalous hydrodynamics in equilibrium in~\cite{Jensen:2013kka,Jensen:2012kj} and for out of equilibrium hydrodynamics in~\cite{Haehl:2013hoa}. 
Dispersion relations of hydrodynamic modes within the system under consideration in this work have been computed previously at weak magnetic fields of first order in the hydrodynamic derivative expansion~\cite{Ammon:2017ded,Kalaydzhyan:2016dyr,Abbasi:2016rds}.  
Anisotropic hydrodynamics has been discussed in the context of heavy ion collisions, see for example~\cite{Martinez:2010sc,Martinez:2010sd,Ryblewski:2010bs,Ryblewski:2011aq,Ryblewski:2012rr,Florkowski:2012lba,Strickland:2014pga,Huang:2011dc,Ammon:2017ded}. 

Holographic duals of quantum field theories with a chiral anomaly and subject to weak electromagnetic fields (of first order in the hydrodynamic derivative expansion) have received much attention due to 
a host of applications that ranges from condensed matter physics 
to heavy ion collisions. 
Specific interest was focused on the analytically known~\cite{Son:2009tf} chiral conductivities: chiral magnetic effect~\cite{Newman:2005hd,Kharzeev:2004ey,Kharzeev:2007jp,Fukushima:2008xe,Kharzeev:2009pj}, 
the chiral vortical effect~\cite{Vilenkin:1978is,Vilenkin:1979ui,Erdmenger:2008rm,Banerjee:2008th}, and later the chiral thermal conductivity, see e.g.~\cite{Neiman:2010zi,Ammon:2017ded}.  
These (DC) conductivities have been shown to be exact in a multitude of holographic models~\cite{Gursoy:2014boa,Grozdanov:2016ala}, and based on field theory arguments~\cite{Fukushima:2008xe}. (Non-)renormalization of these chiral conductivities was addressed  holographically~\cite{Gursoy:2014ela,Jimenez-Alba:2014iia,Gallegos:2018ozs} and field theoretically~\cite{Golkar:2012kb}. 
The frequency dependent (AC) chiral conductivities have been discussed in~\cite{Amado:2011zx,Landsteiner:2013aba,Li:2018srq}, and from the field theory side in~\cite{Kharzeev:2009pj}. 
At nonzero value of the anomaly and without a strong magnetic field, analytic results for helicity-1 correlators in the hydrodynamic approximation have been obtained in~\cite{Matsuo:2009xn}. 
Without the anomaly, in strong magnetic field in an uncharged state Kubo formulas for seven transport coefficients have been derived and values were calculated numerically~\cite{Grozdanov:2017kyl}. 
The shear viscosities have been calculated in~\cite{Critelli:2014kra,PhysRevD.96.019903} under the assumption of the validity of the membrane paradigm. 
Dispersion relations of hydrodynamic and non-hydrodynamic modes within the system under consideration in this work have been computed from quasinormal modes previously at weak magnetic fields of first order in the hydrodynamic derivative expansion~\cite{Ammon:2017ded}. 
Quasinormal modes of magnetic black branes were calculated in~\cite{Ammon:2016fru,Janiszewski:2015ura,Ammon:2017ded,Baggioli:2020edn}. 
In~\cite{Grozdanov:2017kyl} dynamical gauge fields in the dual field theory are considered within a two-form field formalism which is distinct from ours. 
See also~\cite{Grozdanov:2016tdf,Hernandez:2017mch} for the relation between the two formalisms. 
Anisotropic effects not related to magnetic fields have also been included in holography in the hydrodynamic approximation~\cite{Rebhan:2009vc,Erdmenger:2010xm,Erdmenger:2014jba,Garbiso:2020puw}.


\section{Hydrodynamics}
\label{sec:hydrodynamics}
In this section, the constitutive equations, Kubo formulas, equilibrium generating functionals, as well as symmetry constraints, Onsager relations, and the entropy constraints are derived for a charged fluid subjected to a strong external magnetic field. Chemical potentials and magnetic fields associated with either an axial $U(1)_A$-symmetry or a vector $U(1)_V$-symmetry are considered. Quantities can be classified according to their charge under a parity transformation of the three spatial coordinates in a field theory fluid state. It is helpful to notice that there are three potential sources for parity breaking in the fluids we consider: the chiral anomaly in the microscopic field theory, the external magnetic field associated with a vector $U(1)_V$-symmetry, or the axial chemical potential if a global axial $U(1)_A$-symmetry is considered.\footnote{In most of this work we consider such an axial $U(1)_A$; exceptions are clearly marked. The vector chemical potential associated with a vector $U(1)_V$ does not break parity, neither does the magnetic field associated with an axial $U(1)_A$.} 
In order to derive constitutive relations and Kubo formulas, we will use generating functionals among other methods. Note that the generating functionals~\eqref{eq:Ws}, \eqref{eq:WA} and \eqref{eq:Wcov} in presence of a global axial $U(1)_A$-symmetry transforms even under charge-inversion, parity, and time-reversal, i.e.~it has $(C,  P, T)$-eigenvalues (+/+/+). 

\subsection{Thermodynamics}
\label{sec:thermo}
%
\subsubsection{Generating functional and equilibrium constitutive relations}
Following the procedure proposed in~\cite{Banerjee:2012iz,Jensen:2012jh}, we begin by considering the equilibrium constraints on the hydrodynamic framework arising from the existence of a static generating functional. These constraints arise from considering a system with a time-like Killing vector field $V$ (i.e. ${\cal L}_V = 0$) coupled to an external metric $g$ and gauge field $A$. For systems that 
\begin{itemize}
    \item have finite correlation lengths,
    \item are in equilibrium (${\cal L}_V=0)$,
    \item have sources $g,A$ that vary on scales much longer than the correlation lengths,
\end{itemize}
the generating functional $W_s[g,A] = -i\, {\rm ln}Z[g,A]$ is a local functional of the Killing vector field and the sources. The equilibrium generating functional can be systematically expanded in a derivative expansion. The temperature $T$, the chemical potential $\mu$ and the fluid velocity $u^\mu$, which are traditionally considered the only zero derivative terms and are defined in terms of the Killing vector field and the sources
\begin{equation}
\label{eq:Thermo-def}
    T= \frac{T_0}{\sqrt{-V^2}}\,,\quad u^\mu = \frac{V^\mu}{\sqrt{-V^2}}\,,\quad \mu = \frac{V^\mu A_\mu + \Lambda_V}{\sqrt{-V^2}}\,,
\end{equation}
where $T_0$ is a constant setting the normalization of the temperature, and $\Lambda_V$ is a gauge parameter which ensures that $\mu$ is gauge invariant~\cite{Jensen:2013kka}. In addition, for a system subject to a strong magnetic field $B^\mu = \frac12 \epsilon^{\mu\nu\rho\sigma}u_\nu F_{\rho \sigma}\sim {\cal O}(1)$, the scalar $B^2 = B^\mu B^\nu g_{\mu\nu}$ is order zero in derivatives as well. In this paper, we assume the counting $T,\mu,u^\mu, B^\mu \sim {\cal O}(1)$ and all other terms with a derivative as ${\cal O}(\partial)$. For example, $E_\mu = F_{\mu\nu}u^\nu \sim {\cal O}(\partial)$ and $R^\mu_{\ \nu\rho\sigma} \sim \partial \partial g \sim {\cal O}(\partial^2)$.
In addition, the derivatives of the fluid velocity such as the vorticity $\Omega^\mu =  \epsilon^{\mu\nu\rho\sigma}u_\nu\partial_\rho u_\sigma$ and the fluid acceleration $a^\mu = u^\nu\nabla_\nu u^\mu$ are $\mathcal{O}(\partial)$. In table~\ref{tab:CPT}, the behavior of hydrodynamic fields, sources and other quantities under charge conjugation $C$, parity $P$, and time reversal $T$ are collected. Note that in this subsection on equilibrium states, the only parity violation stems from the axial gauge field and the associated axial chemical potential.\footnote{ Naively, vector magnetic fields do also break parity. However, the magnetic field is the only vector valued quantity characterizing the equilibrium state. Hence, only the scalar $B^2$ enters into the pressure $p$ and the transport coefficients (such as $M_n$). Note that $B^2$ transforms even under $P$ (as well as under $C$ and $T$).}

With this derivative counting, the equilibrium generating functional for a hydrodynamic system coupled to external gauge field and metric subject to strong magnetic fields can be expanded as~\cite{Hernandez:2017mch}
\begin{equation}
\label{eq:Ws}
W_s = \int d^4x \sqrt{-g}\left(p(T,\mu,B^2) + \sum_{n=1}^{5} M_n(T,\mu,B^2) s_n + O(\partial^2)\right)\,,
\end{equation} 
where $p$ is the homogeneous equilibrium pressure and $s_n$ are the first order 
equilibrium scalars, that is, $s_n \sim {\cal O}(\partial)$. Their definitions are listed in table~\ref{tab:T1-2}, 
together with their transformation properties under charge conjugation, parity, time reversal and Weyl transformations. The magnetovortical susceptibility $M_5 = M_\Omega$ is the only nonzero first order thermodynamic function for a parity-preserving theory coupled to vector gauge fields. On the other hand, for a system coupled to an axial gauge field, the axial chemical potential $\mu$ breaks parity and the other $M_n$ can in principle be nonzero. These will appear along with the pressure in the thermodynamic/hydrostatic constitutive relations. We stress that, in absence of the chiral anomaly, the microscopic theory does not break the parity symmetry, but rather the state in question breaks parity, provided there is a nonzero axial chemical potential.\footnote{A nonzero vector magnetic field can also break parity. However, the only zeroth order scalar with odd parity is an axial chemical potential. A parity odd zeroth order scalar is what in turn allows $M_n$ to be nonzero while remaining parity odd. This is essential for a parity preserving microscopic system to have nonzero $M_n$ accompanying the parity odd first order scalars $s_n$.
} 
Table~\ref{tab:T1-2} highlights the symmetry properties of the equilibrium scalars when defined in terms of axial gauge fields. For this paper, we will focus in the case where our system is coupled to axial gauge fields. This requires us to include the terms that are usually considered to be parity-violating when considering vector gauge fields. That is, the $M_1$, $M_2$, $M_3$ and $M_4$ can be nonzero for a parity preserving system coupled to an external axial gauge field. In this section, we elaborate on the modifications of the hydrodynamic framework when these terms are included.

We will write the energy-momentum tensor using the decomposition with respect to the timelike velocity vector $u^\mu$,
\begin{align}
\label{eq:TT1}
   T^{\mu\nu} = {\cal E} u^\mu u^\nu + {\cal P}\Delta^{\mu\nu} + 
    {\cal Q}^\mu u^\nu + {\cal Q}^\nu u^\mu + {\cal T}^{\mu\nu}\,,
\end{align}
where $\Delta^{\mu\nu} \equiv g^{\mu\nu} + u^\mu u^\nu$ is the transverse projector, the energy current ${\cal Q}^\mu$ is transverse to $u_\mu$, and  ${\cal T}^{\mu\nu}$ is transverse to $u_\mu$, symmetric, and traceless. Explicitly, the coefficients are ${\cal E}\equiv u_\mu u_\nu T^{\mu\nu}$, ${\cal P}\equiv \coeff{1}{3} \Delta_{\mu\nu}T^{\mu\nu}$, ${\cal Q}_\mu \equiv -\Delta_{\mu\alpha} u_\beta T^{\alpha\beta}$ and ${\cal T}_{\mu\nu}\equiv \coeff12(\Delta_{\mu\alpha}\Delta_{\nu\beta} + \Delta_{\nu\alpha} \Delta_{\mu\beta} - \coeff{2}{3} \Delta_{\mu\nu} \Delta_{\alpha\beta}) T^{\alpha\beta}$. Similarly, we will write the current as
\begin{align}
\label{eq:JJ1}
  J^\mu = {\cal N} u^\mu + {\cal J}^\mu\,,
\end{align}
where the charge density is ${\cal N}\equiv -u_\mu J^\mu$, and the spatial current is ${\cal J}_\mu \equiv \Delta_{\mu\lambda} J^\lambda$. We also decompose the field strength tensor with respect to $u^\mu$,
\begin{table}
\begin{center}
\def\arraystretch{1.2}
\setlength\tabcolsep{4pt}
\begin{tabular}{|c|c|c|c|c|c|}
 \hline
 \hline
 $n$ & 1 & 2 & 3 & 4 & 5\\ 
 \hline
 \hline
 $s_n$
 & $B^\mu \partial_\mu (\frac{B^2}{T^4})$   %
 & $\epsilon^{\mu\nu\rho\sigma} u_\mu B_\nu \nabla_{\!\rho} B_\sigma$  %
 & $B{\cdot}a$
 & $B{\cdot}E$ 
 & $B{\cdot}\Omega$
 \\
 \hline
  $(C,P,T)_{\textrm{axial}}$  & $+/+/-$ & $+/-/+$ &  $+/+/-$ & $+/-/-$ & $+/-/+$ \\
  \hline
  $(C,P,T)_{\textrm{vector}}$  & $-/-/-$ & $+/-/+$ &  $-/-/-$ & $+/-/-$ & $-/+/+$ \\
  \hline
   W  & 3 & 5 & n/a & 4 & 3\\
  \hline
\end{tabular}
\end{center}
\caption{Independent nonzero $O(\partial)$ invariants in equilibrium in 3+1 dimensions for an axial gauge field. We have used the fluid acceleration $a^\mu=u^\lambda\nabla_\lambda u^\mu$ and the vorticity $\Omega^\mu = \epsilon^{\mu\nu\rho\sigma} u_\nu \partial_\rho u_\sigma$. An axial chemical potential $\mu$ is $C$-even and $P$-odd. In case of an axial chemical potential we expect $M_1$ and $M_3$ to be even functions of $\mu$ while $M_2$, $M_4$ and $M_5$ should be odd functions of $\mu$. 
Here the last row labelled ``W'' indicates the charge under Weyl-transformations and ``n/a'' indicates that the tensor structure $s_3$ does not have a definite behavior under Weyl-transformations. 
}
\label{tab:T1-2}
\end{table}
 \begin{equation}
F_{\mu\nu} = u_\mu E_\nu - u_\nu E_\mu - \epsilon_{\mu\nu\rho\sigma} u^\rho B^\sigma \,,
 \end{equation}
where $E_\mu = F_{\mu\nu} u^\nu$  is the electric field and $B^\mu = \coeff12 \epsilon^{\mu\nu\rho\sigma} u_\nu F_{\rho\sigma}$ is the magnetic field. We use the convention $\epsilon^{\mu\nu\rho\sigma} = \varepsilon^{\mu\nu\rho\sigma}/\sqrt{-g}$, where $\varepsilon^{0123} = 1$. We also use the vorticity $\Omega^\mu = \epsilon^{\mu\nu\rho\sigma} u_\nu \partial_\rho u_\sigma$.

The equilibrium constitutive relations are found by varying the generating functional with respect to the metric and the gauge field
\begin{equation}\label{eq:Wsvar}
\delta W_s[A,g] = \int d^4x \sqrt{-g} \left( \coeff12 T^{\mu\nu}_{\rm eq.} \delta g_{\mu\nu} + J^\mu_{\rm eq.} \delta A_\mu \right)\,.
\end{equation}
This was done in~\cite{Hernandez:2017mch} for a parity-preserving theory coupled to a vector gauge field. The new terms allowed when considering an axial gauge field come from the variation of $M_i(T,\mu,B^2) S_i$ for $i\neq 5$ and are given by
\begin{equation}
\label{eq:PVdecomp}
  {\cal E}_{{\rm eq.\, new}}  = \sum_{n=1}^4 \epsilon_n s_n\,, \quad {\cal P}_{\rm eq.\, new} =  \sum_{n=1}^4 \pi_n s_n\,, \quad {\cal N}_{\rm eq.\, new} = \sum_{n=1}^4 s_n \, \quad {\cal T}^{\mu\nu}_{{\rm eq.\, new}} = \sum_{n=1}^{10} \theta_n \tau^{\mu\nu}_n\,,
\end{equation}
where 
 \begin{equation}
 \label{eq:PVconst}
 \begin{aligned}
  & \epsilon_3 = - \frac{4 B^2}{T^4} \epsilon_1 = 3 \pi_3 = -4 B^2 \theta_3 = \frac{4B^2}{T^4} \left( M_1 - T M_{1,T} - \mu M_{1,\mu} - 4B^2 M_{1,B^2}  - T^4 M_{3,B^2} \right)\,, \\[5pt]
  & \epsilon_2 = \theta_6 = -M_2 + T M_{2,T} + \mu M_{2,\mu}\,, \quad \epsilon_4 = T M_{4,T} + \mu M_{4,\mu} + \frac{4B^2}{T^4} M_{1,\mu} + M_{3,\mu} \,, \\[5pt]
    & \pi_2 = -\coeff23 M_2 - \coeff43 B^2 M_{2,B^2} \,, n\quad \pi_4 = -\coeff43 B^3 \theta_4 = \frac{4B^2}{3 T^4}\phi_1= \frac{4B^2}{3 T^4}\left( M_{1,\mu} -T^4 M_{4,B^2}\right) \,,\\[5pt]
  & \phi_2 = M_{2,\mu} \,, \quad \phi_3 = \epsilon_4 - 3 \pi_4 \,,\quad \theta_2 = M_{2,B^2} \,, \quad \theta_5 = 2M_2\,,\quad \theta_7 = M_{2,B^2}\,, \quad \theta_8 = -M_{2,\mu}\,,\\[5pt]
  & \pi_1 = \phi_4 = \theta_1 = 0\,.
 \end{aligned}
 \end{equation}
The comma subscript denotes the derivative with respect to the argument that follows, and we are using $(T,\mu,B^2)$ as our three independent variables. Hence, for example, $M_{1,T} = \left(\frac{\partial M_1}{\partial T} \right)_{\mu,B^2}$. The equilibrium vectors and tensors are defined in table~\ref{tab:T2}. The equilibrium spatial current $\mathcal{J}^\mu$ and energy current $\mathcal{Q}^\mu$ do not receive contributions to $O(\partial)$ from the novel thermodynamic transport coefficients $M_1$, $M_2$, $M_3$ and $M_4$.
 \begin{table}
\begin{center}
\def\arraystretch{1.2}
\setlength\tabcolsep{4pt}
\begin{tabular}{|c|c|c|c|c|}
 \hline
 \hline
 $n$ & 1 & 2 & 3 & 4 \\ 
 \hline
 \hline
 $v_n^\mu$
 & $\epsilon^{\mu\nu\rho\sigma} u_\nu \partial_\sigma B_\rho$   %
 & $\epsilon^{\mu\nu\rho\sigma} u_\nu B_\rho \partial_\sigma T /T$  %
 & $\epsilon^{\mu\nu\rho\sigma} u_\nu B_\rho \partial_\sigma B^2$
 & $\epsilon^{\mu\nu\rho\sigma} u_\nu E_\rho B_\sigma$ \\
  \hline
\end{tabular}

\bigskip
\begin{tabular}{|c|c|c|c|c|c|}
 \hline
 \hline
 $n$ & $1-4$ & 5 & 6 & 7 & 8  \\ 
 \hline
 \hline
 $t_n^{\mu\nu}$
 & $s_n B^{\langle \mu} B^{\nu\rangle}$   %
 & $v_1^{\langle \mu} B^{\nu\rangle}$
 & $v_2^{\langle \mu} B^{\nu\rangle}$
 & $v_3^{\langle \mu} B^{\nu\rangle}$ 
 & $v_4^{\langle \mu} B^{\nu\rangle}$ 
\\
  \hline
\end{tabular}

\end{center}
\caption{Top: nonzero transverse $O(\partial)$ vectors that appear in the parity-violating equilibrium energy flux~${\cal Q}^\mu$ and in the equilibrium spatial current ${\cal J}^\mu$. The vector $v_4^{\mu}$ is the Poynting vector. Bottom: nonzero symmetric transverse traceless $O(\partial)$ tensors that appear in the equilibrium stress ${\cal T}^{\mu\nu}$. For any two transverse vectors $X^\mu$ and $Y^\mu$, the angular brackets stand for $X^{\langle \mu} Y^{\nu\rangle} \equiv X^\mu Y^\nu + X^\nu Y^\mu - \coeff23 \Delta^{\mu\nu} X{\cdot}Y$. 
}
\label{tab:T2}
\end{table}

For a diffeomorphism and gauge invariant theory, invariance of the generating functional gives the following hydrodynamic equations
\begin{subequations}
\label{eq:Cons}
\begin{align}
& \nabla_\nu T^{\mu\nu} = F^{\mu\nu}J_\nu\,, \\[5pt]
& \nabla_\mu J^\mu = 0\,.
\end{align}
\end{subequations}
The definition of the equilibrium energy-momentum tensor and conserved currents ensure that the equations of motion are satisfied in equilibrium. 

For completeness, let us summarize the equilibrium constitutive relations for the energy-momentum tensor and the current. The equilibrium energy-momentum tensor is given by
\begin{subequations}
\label{eq:TTF-eq}
\begin{align}
  {\cal E}_{\rm eq.} & =  -p + T\, p_{,T} + \mu\, p_{,\mu}
  +\left(T {\MO}_{,T} + \mu {\MO}_{,\mu} -2 \MO \right) B{\cdot}\Omega \nonumber\\[5pt]
   &\ \ \ \, +\left( T M_{1,T} + \mu M_{1,\mu} + 4B^2 M_{1,B^2} + T^4 M_{3,B^2}-M_1\right) s_1\nonumber \\[5pt]
  &\ \ \ \, + \left(T M_{2,T} + \mu M_{2,\mu}-M_2\right) s_2 \nonumber \\[5pt]
  &\ \ \ \, + \frac{4B^2}{T^4} \left( M_1 - T M_{1,T} - \mu M_{1,\mu} - 4B^2 M_{1,B^2}-T^4 M_{3,B^2} \right)s_3 \nonumber\\[5pt]
  &\ \ \ \, + \left(T M_{4,T} + \mu M_{4,\mu} + \frac{4B^2}{T^4} M_{1,\mu} + M_{3,\mu}\right) s_4\,,
\\[5pt]
  {\cal P}_{\rm eq.} & = p - \coeff43\, p_{,B^2}B^2
  -\coeff13 (\MO + 4{\MO}_{,B^2}B^2)B{\cdot}\Omega  - \coeff23 \left(M_2 + 2 B^2 M_{2,B^2} \right) s_2
  \nonumber\\[5pt]  
  &\ \ \ \, + \frac{4B^2}{3T^4} \left(M_1 - T M_{1,T} - \mu M_{1,\mu} - 4B^2 M_{1,B^2} -T^4 M_{3,B^2} \right)s_3 \nonumber \\[5pt]
  &\ \ \ \, + \frac{4B^2}{3T^4}\left(M_{1,\mu}-T^4 M_{4,B^2}\right) s_4 \,,
\end{align}
\begin{align}
  {\cal Q}^\mu_{\rm eq.} & = - \MO \epsilon^{\mu\nu\rho\sigma} u_\nu \partial_\sigma B_\rho
  +(2 \MO - T {\MO}_{,T}- \mu {\MO}_{,\mu}) \epsilon^{\mu\nu\rho\sigma} u_\nu B_\rho \partial_\sigma T /T
  \nonumber \\[5pt]
  & \ \ \ \, - {\MO}_{,B^2} \epsilon^{\mu\nu\rho\sigma} u_\nu B_\rho \partial_\sigma B^2
  +({\MO}_{,\mu} - 2p_{,B^2}) \epsilon^{\mu\nu\rho\sigma} u_\nu E_\rho B_\sigma\\[5pt]
  {\cal T}^{\mu\nu}_{\rm eq.} & = 2p_{,B^2} \left( B^\mu B^\nu -\coeff13 \Delta^{\mu\nu}B^2\right)
  +  B^{\langle\mu} B^{\nu\rangle}\left( {\MO}_{,B^2} B{\cdot}\Omega +M_{2,B^2}s_2 + (M_{4,B^2} - \coeff{1}{T^4} M_{1,\mu})s_4\right) \nonumber \\[5pt]
  & \ \ \ \, + B^{\langle\mu} B^{\nu\rangle} \frac{1}{T^4} \left(T M_{1,T} +\mu M_{1,\mu} +4B^2 M_{1,B^2} - M_1 + T^4 M_{3,B^2}\right)s_3\nonumber + \MO B^{\langle\mu}\Omega^{\nu\rangle}\\[5pt]
  &\ \ \ \, + 2M_2 B^{\langle \mu} \epsilon^{\nu\rangle \rho \sigma \alpha} u_\rho \partial_\sigma B_\alpha + \left(T M_{2,T} + \mu M_{2,\mu}-M_2 \right) B^{\langle \mu}\epsilon^{\nu\rangle \alpha \rho \sigma}u_\alpha B_\rho \partial_\sigma T/T\nonumber \\[5pt]
  & \ \ \ \, + M_{2,B^2}B^{\langle \mu}\epsilon^{\nu\rangle \alpha \rho \sigma}u_\alpha B_\rho \partial_\sigma B^2 - M_{2,\mu} B^{\langle \mu}\epsilon^{\nu\rangle \rho \sigma \alpha} u_\rho E_\sigma B_\alpha \,,
\end{align}
\end{subequations}
where we used the vorticity $\Omega^\mu = \epsilon^{\mu\nu\rho\sigma} u_\nu \partial_\rho u_\sigma$. The current is given by 
\begin{subequations}
\label{eq:JTF-eq}
\begin{align}
  {\cal N}_{\rm eq.} & = p_{,\mu} - \nabla{\cdot}\mathfrak{p} + \mathfrak{p}{\cdot}a - \mathfrak{m} {\cdot} \Omega  + \left(M_{1,\mu} -T^4 M_{4,B^2}\right) s_1 + M_{2,\mu} s_2  \nonumber \\[5pt]
  &\ \ \ \, +\left( M_{3,\mu} + T M_{4,T} + \mu M_{4,\mu} + 4B^2 M_{4,B^2}\right) s_3 + {\MO}_{,\mu}s_5 \,, \\[5pt]
  {\cal J}^\mu_{\rm eq.} & = \epsilon^{\mu\nu\rho\sigma} u_\nu \nabla_{\!\rho} \mathfrak{m}_\sigma
  +\epsilon^{\mu\nu\rho\sigma} u_\nu a_\rho \mathfrak{m}_\sigma \,,
\end{align}
\end{subequations}
where $a^\mu=u^\lambda\nabla_\lambda u^\mu$ defines the acceleration and the (electric) polarization vector is $\mathfrak{p}_\mu= \frac{1}{\sqrt{-g}} \frac{\delta W_s}{\delta E^\mu}=M_4 B_\mu$.
The current is written in terms of the magnetic polarization vector $\mathfrak{m}_\mu = \frac{1}{\sqrt{-g}}\frac{\delta W_s}{\delta B^\mu}$\footnote{Careful comparison with~\cite{Kovtun:2018dvd} shows an agreement with their (2.19a) and (2.19b) in the $B ={\cal O}(\partial) $ limit. Note that the $M_1$ and $M_2$ would be pushed to higher derivative order and none of the $M_n$ would be functions of $B^2$. Similarly, $p_{,B^2}$ would be a second order term which corresponds to their $f_6$.}
\begin{equation}
\begin{aligned}
\label{eq:m-vector}
  \mathfrak{m}^\mu &= \left( 2\,p_{,B^2} + 2 \sum_{n=2}^5 M_{n,B^2} s_n + \frac{2}{T^4} \left(M_1 - T M_{1,T} - \mu M_{1,\mu} - 4 B^2 M_{1,B^2}\right) B{\cdot}\partial T/T \right)B^\mu \\
 &\ \ \ \,+ \MO \Omega^\mu + M_3 a^\mu + M_4 E^\mu +M_1 \Delta^{\mu\nu} \partial_\nu \frac{B^2}{T^4} - M_{2,\mu} \epsilon^{\mu\nu\rho\sigma} u_\nu E_\rho B_\sigma +  M_{2,B^2} \epsilon^{\mu\nu\rho\sigma} u_\nu B_\rho \partial_\sigma B^2\\[5pt]
&\ \ \ \,  +  \left(T M_{2,T} + \mu M_{2,\mu} - M_2 \right) \epsilon^{\mu\nu\rho\sigma} u_\nu B_\rho \partial_\sigma T/T  + 2 M_2 \epsilon^{\mu\nu\rho\sigma} u_\nu \partial_\rho B_\sigma \,.
\end{aligned}
\end{equation}
Note that we are keeping O($\partial^2$) thermodynamic terms in the current (coming from
the variation of $\sum_{n=1}^5 M_n s_n$ in the generating functional) that are needed to ensure that the conservation laws~\eqref{eq:Cons} are satisfied to $O(\partial^2)$ for time-independent background fields. Including the $O(\partial^2)$ thermodynamic terms in the energy-momentum tensor will ensure these are satisfied identically,  but we omit them here for simplicity.

\subsubsection{Incorporating the chiral anomaly}\label{sec:chiralanomalyhydro}
For a theory with a chiral anomaly subject to external axial gauge fields, the generating functional is no longer gauge invariant,\footnote{In curved spacetime, the gauge non-invariance of the generating functional~\eqref{eq:WA-gauge} includes some curvature terms proportional to the square of the Riemann tensor. However, in this paper we restrict our attention to the derivative counting $\partial g \sim {\cal O}(\partial)$ so that these terms are of order four in derivatives. We therefore neglect these terms for the rest of the paper. Strictly speaking, the corresponding gravitational Chern-Simons contribution to eq.~\eqref{eq:WCS} includes curvature terms which cannot be taken as ${\cal O}(\partial^2)$ in the bulk spacetime $\cal M$. These terms give rise to the effects we will find by including the term multiplying $c_1$ in the consistent generating functional \eqref{eq:WA}. See, for example, \cite{Chen:2012ca,Jensen:2012kj,Landsteiner:2011cp,Stone:2018zel,Landsteiner:2011iq} for more careful treatments of the mixed anomaly term. \label{footy-anom}
} 
\begin{equation}
\label{eq:WA-gauge}
\delta_\alpha W_{cons} =  \frac{C}{24} \int d^4x \sqrt{-g}\, \alpha \, \epsilon^{\mu\nu\rho\sigma}F_{\mu\nu}F_{\rho\sigma}  \equiv {\cal A}  \,,
\end{equation}
leading to the following 
\begin{subequations}
\label{eq:ConsViol}
\begin{align}
& \nabla_\mu T^{\mu\nu}_A = F^{\mu\nu}J^{cons}_\nu -A^\nu \nabla_\mu J^\mu_{cons} \,, \\[5pt]
& \nabla_\mu J_{cons}^\mu = - \frac{C}{24}\epsilon^{\mu\nu\rho\sigma}F_{\mu\nu} F_{\rho\sigma} = \frac{C}{3} E{\cdot}B\,,
\end{align}
\end{subequations}
where $J_{cons}^\mu = \frac{1}{\sqrt{-g}} \frac{\delta W_{cons}}{\delta A_\mu}$ is the gauge dependent consistent current. The fact that it is gauge dependent follows from the commuting of $\frac{\delta}{\delta A_\mu}$ with the BRST operator $s = \int d^4x \partial_\mu c \frac{\delta}{\delta A_\mu}$ generating gauge transformations, from which we get $\delta_\alpha J^\mu_{cons} =  \frac{1}{\sqrt{-g}}\frac{\delta}{\delta A_\mu} {\cal A} =  \coeff16 C \epsilon^{\mu\nu\rho\sigma} \partial_\nu \alpha F_{\rho\sigma}$. Noting that ${\cal A}$ is independent of the metric, a similar argument shows that the consistent energy-momentum tensor $T^{\mu\nu}_A = \frac{2}{\sqrt{-g}} \frac{\delta W_{cons}}{\delta g_{\mu\nu}}$ is gauge invariant. It is possible to add a Chern-Simons current $J^\mu_{BZ}=-\coeff16 C \epsilon^{\mu\nu\rho\sigma} A_\nu F_{\rho\sigma}$, also known as a Bardeen-Zumino polynomial, to the consistent current to get a gauge invariant current $J^\mu_{cov}$, usually named covariant current. The equations of motion \eqref{eq:ConsViol} then take the manifestly gauge covariant from
\begin{subequations}
\label{eq:CovViol}
\begin{align}
& \nabla_\nu T^{\mu\nu}_A = F^{\mu\nu}J^{cov}_\nu \,, \\[5pt]
& \nabla_\mu J^\mu_{cov} = - \frac{C}{8} \epsilon^{\mu\nu\rho\sigma}F_{\mu\nu} F_{\rho\sigma} = C E{\cdot}B\,.
\end{align}
\end{subequations}
Note that the covariant energy-momentum tensor is the same as the consistent energy-momentum tensor. See~\cite{Landsteiner:2016led} for a recent review on anomalous currents. 

To understand how this gauge anomaly affects the hydrodynamic description, we construct the equilibrium generating functionals for the consistent and for the covariant currents using the anomaly inflow mechanism~\cite{Callan:1984sa}. The anomaly inflow generating functionals have been used for anomalous hydrodynamics in equilibrium in~\cite{Jensen:2013kka,Jensen:2012kj} and for out of equilibrium hydrodynamics in~\cite{Haehl:2013hoa}.

The gauge dependent generating functional for the consistent current of a 3+1 dimensional theory is given by\footnote{Note that $c_1$ and $c_2$ defined here do not depend on the thermodynamic quantities. They are properties of the microscopic theory. Hence, they are entirely different from the transport coefficients which we will name $c_n$ later in the text.}
\begin{equation}
\label{eq:WA}
W_{cons} = W_s + \int d^4 x\sqrt{-g} \left( c_1 T^2 \Omega{\cdot}A + c_2 T\left(B{\cdot}A + \mu \Omega{\cdot}A\right) + \frac{C}{3}\mu \left( B{\cdot}A+\coeff{1}{2} \mu \Omega{\cdot}A \right)\right)\,,
\end{equation}
where $W_s$ is the generating functional for a theory without anomalies~\eqref{eq:Ws}. We refer to $W_{cons}$ as the consistent generating functional. The vectors $T^2 \Omega^\mu$ and $TB^\mu + T\mu \Omega^\mu$ have vanishing divergence and do not contribute to the gauge anomaly, unless the 3+1-dimensional theory has a boundary. The gauge dependence of the consistent generating functional~\eqref{eq:WA-gauge} comes from the non-conservation of the vector $\nabla_\mu \left(\mu B^\mu + \coeff12 \mu^2 \Omega^\mu\right) = B{\cdot}E$. Note that the term multiplying $c_2$ breaks CPT symmetry and is therefore not allowed for Lorentz invariant theories~\cite{Jensen:2013vta}. The coefficient $c_1$ is related to the mixed gauge-gravitational anomaly $c_m$ by~\cite{Chen:2012ca,Jensen:2012kj,Landsteiner:2011cp,Stone:2018zel,Landsteiner:2011iq}
\begin{equation}\label{eq:gaugeGravitationalAnomalyCoeffs}
c_1=-8 \pi^2 c_m\,.
\end{equation}

The variation of the consistent generating functional yields the energy-momentum tensor and the consistent current. We now focus on the new terms coming from $W_{cons}-W_s$ and write $\Delta J_{cons}^\mu = J^\mu_{cons} - J^\mu$ where $J^\mu$ is the current found  in eq.~\eqref{eq:Wsvar} by varying $W_s$. Similarly, we write $\Delta T^{\mu\nu} = T^{\mu\nu}_A-T^{\mu\nu}$ where $T^{\mu\nu}$ comes from varying $W_s$. Taking the source variations we find
\begin{subequations}
\begin{align}
\label{eq:DeltaT}
\Delta T^{\mu\nu} &= \xi_T u^{(\mu}  \Omega^{\nu)} + \xi_{TB}\, u^{(\mu} B^{\nu)} \,,\\
\Delta J^\mu_{cons} & = \coeff13 C B{\cdot}A u^\mu +\xi\, \Omega^\mu + \left(\xi_B - \coeff13 C\mu \right) B^\mu  + \coeff13 C \epsilon^{\mu\nu\rho\sigma}A_\nu u_\rho E_\sigma\,,
\end{align}
\end{subequations}
where\footnote{The following conventions for anomalous transport coefficients are in the thermodynamic frame used, for example, in~\cite{Jensen:2013vta}. This corresponds in~\cite{Amado:2011zx} to the ``no drag frame'' coefficients $\sigma_B$ and $\sigma_V$.}
\begin{align}
\xi &= \coeff12 C \mu^2 + c_1 T^2 + 2 c_2 T\mu  \,, \quad &\xi_B = C\mu + 2 c_2T\,, \nonumber \\
\label{eq:xis}
\xi_T &= \coeff13 C \mu^3 + 2c_1 T^2 \mu + 2c_2 T \mu^2 \,, \quad &\xi_{TB} =  \coeff12 C \mu^2 + c_1 T^2 + 2 c_2 T\mu  \,.
 \end{align}
The consistent current $J^\mu_{cons}$ and energy-momentum tensor $T^{\mu\nu}_A$ satisfy the consistent equations of motion~\eqref{eq:ConsViol} derived from the diffeomorphism invariance and gauge-non-invariance of the consistent generating functional $W_{cons}$. From the consistent current, we can construct the covariant current $J^\mu_{cov}$ by adding to it the Bardeen-Zumino/Chern-Simons current $J^\mu_{BZ} = -\coeff16 C \epsilon^{\mu\nu\rho\sigma}A_\nu F_{\rho\sigma}$,
\begin{equation}
\label{eq:cons+BZ}
J^\mu_{cov} = J^\mu_{cons} + J^\mu_{BZ}\,.
\end{equation}

Alternatively, we can construct a covariant generating functional $W_{cov}$ by adding a Chern-Simons functional to the consistent generating functional

\begin{equation}
\label{eq:Wcov}
W_{cov} = W_{cons} + W_{CS}\,,
\end{equation}
where\footnote{This Chern-Simons functional contains only gauge terms since we are omitting the gravitational anomalies which appear at higher order in hydrodynamic derivatives. See footnote~\ref{footy-anom}.}
\begin{equation}\label{eq:WCS}
W_{CS} = -\frac{C}{6} \int A \wedge F \wedge F = -\frac{C}{24}\int d^5x \sqrt{-G} \epsilon^{mnopq} A_m F_{no} F_{pq}\,.
\end{equation}
We take our Chern-Simons theory to live in a 4+1 dimensional space-time ${\cal M}$ with a boundary $\partial {\cal M}$ which corresponds to the space-time where $W_{cons}$ is defined. The five dimensional field strength $F_{mn} = \partial_m A_n - \partial_n A_m$ is defined in terms of the five dimensional gauge field $A_m$. We take the gauge field $A_\mu$ appearing in $W_{cons}$ as the induced gauge field on $\partial {\cal M}$ from $A_m$. The Chern-Simons functional is independent of the five dimensional metric $G_{m n}$, and we use the convention $\epsilon^{mnopq} = \varepsilon^{mnopq} / \sqrt{-G}$, where $\varepsilon^{0123z} = 1$ and z is the coordinate normal to $\partial {\cal M}$. We also take $G_{mn}$ so that the induced metric in $\partial{\cal M}$ is $g_{\mu\nu}$, the metric used in the consistent generating functional. The Chern-Simons theory is gauge invariant up to a boundary term
\begin{equation}
\delta_\alpha W_{CS} = -\frac{C}{24} \int_{\partial{\cal M}} d^4x \sqrt{-g}\alpha \epsilon^{\mu\nu\rho\sigma} F_{\mu\nu} F_{\rho\sigma}\,,
\end{equation}
which cancels the gauge dependence of $W_{cons}$. Taking source variations of the covariant generating functional gives the covariant energy-momentum tensor and current as well as the bulk current $J^m_H$,
\begin{equation}
\delta W_{cov} = \int_{\partial {\cal M}} d^dx \sqrt{-g} \left( \coeff12 T^{\mu\nu}_A \delta g_{\mu\nu} + J^\mu_{cov}\delta A_\mu \right) + \int_{\cal M} d^5x \sqrt{-G}J^m_H \delta A_m\,.
\end{equation}

Note that the bulk energy-momentum tensor $T^{mn}_H = \frac{2}{\sqrt{-G}} \frac{\delta W_{cov}}{\delta G_{mn}}$ vanishes since $W_{CS}$ is independent of $G_{mn}$. Variations of the Chern-Simons functional give the Bardeen-Zumino current and the bulk current
\begin{equation}
\delta W_{CS} = \int_{\partial {\cal M}} d^4x \sqrt{-g}J^\mu_{BZ}\delta A_\mu + \int_{\cal M} d^5x \sqrt{-G} J^m_H \delta A_m\,.
\end{equation}
Explicitly, these currents are
\begin{subequations}
\begin{align}
J^\mu_{BZ} &= -\frac{C}{6} \epsilon^{\mu\nu\rho\sigma} A_\nu F_{\rho\sigma}\,,\\
J^m_H & = - \frac{C}{8} \epsilon^{mnopq} F_{no} F_{pq}\,.
\end{align}
\end{subequations}

Notice that $J_H^z = -\frac{C}{8} \epsilon^{\mu\nu\rho\sigma}F_{\mu\nu}F_{\rho\sigma} = C B{\cdot}E$. Diffeomorphism and gauge invariance of $W_{cov}$ then lead to the covariant equations of motion~\eqref{eq:CovViol} together with 
\begin{equation}
\nabla_m J^m_H = 0\,,
\end{equation}
which follows directly from the Bianchi identity. The covariant current can be found from~\eqref{eq:cons+BZ}. Using $\coeff12 \epsilon^{\mu\nu\rho\sigma} A_\nu F_{\rho\sigma} = B{\cdot}Au^\mu + \epsilon^{\mu\nu\rho\sigma} A_\nu u_\rho E_\sigma - \mu B^\mu$, we get

\begin{equation}
\label{eq:Jcov}
J^\mu_{cov} = J^\mu + \xi\,\Omega^\mu + \xi_B\, B^\mu  
\end{equation}

Equations~\eqref{eq:DeltaT} and \eqref{eq:Jcov} show how the covariant current and the energy-momentum tensor have to be modified in the presence of a chiral anomaly. The transport coefficients determined by the anomaly coefficient $C$ first appeared in holographic calculations~\cite{Erdmenger:2008rm,Banerjee:2008th}. Their first derivation in the hydrodynamic framework was done in~\cite{Son:2009tf} using entropy current arguments. In~\cite{Neiman:2010zi}, the result was generalized for theories with general triangle anomalies and the coefficients $c_1$ and $c_2$ appear as integration constants from solving the entropy constraints. These results were then derived using equilibrium generating functionals in~\cite{Banerjee:2012iz,Jensen:2013kka}. The anomaly induced transport terms found in the thermodynamic frame are exact~\cite{Loganayagam:2011mu} and can be brought to the Landau-Lifshitz frame by a redefinition of the hydrodynamic variables, using $u^\mu \to u^\mu + \delta u^\mu$ so that ${\cal Q}_A^\mu = 0$.

\subsubsection{Thermodynamic correlation functions and Kubo formulas}
\label{subsec:Kubo}
The Kubo formulas relate the transport coefficients to two-point functions of conserved currents and stress tensors of the underlying microscopic theory. For a system in equilibrium $({\cal L}_V=0)$, the static correlation functions can be found by taking second order variations of the generating functional $W_s[g,A]$ with respect to the external sources $g_{\mu\nu}$ and $A_\mu$. Concretely, for $A_\mu \to A_\mu + \delta A_\mu$ and $g_{\mu\nu} \to g_{\mu\nu} + \delta g_{\mu\nu}$ such that ${\cal L}_V\delta A_\mu= {\cal L}_V \delta g_{\mu\nu}=0$\,, we have
\begin{equation}
    \delta W_s^{(2)} = \int \sqrt{-g} \left( \coeff14 G_{T^{\mu\nu} T^{\rho\sigma}} \delta g_{\mu\nu} \delta g_{\rho\sigma} +\coeff12 G_{T^{\mu\nu} J^\rho} \delta g_{\mu\nu} \delta A_\rho + G_{J^\mu J^\nu} \delta A_\mu \delta A_\nu \right)\,,
\end{equation}
where $\delta W_s^{(2)}$ is the second order variation\footnote{The first order variation is simply eq.~\eqref{eq:Wsvar}.} of $W_s$ in eq.~\eqref{eq:Ws} and $G_{J^\rho T^{\mu\nu}} = G_{T^{\mu\nu} J^\rho}$. Note that this is equivalent to taking the first order variations of the equilibrium current $J^\mu[g,A]$ and stress tensor $T^{\mu\nu}[g,A]$ in eq.~\eqref{eq:Wsvar} with respect to the sources.
From here on, we work within an equilibrium state defined in flat space with metric $\eta_{\mu\nu}=\text{diag}(-1,1,1,1)$, with background magnetic field $B=(0,0,0,B_0)$, and in the fluid rest frame with velocity $u^\mu=(1,0,0,0)$. 
We then consider plane wave fluctuations ($\delta A,\delta g \sim {\rm exp}\left(i k{\cdot}x\right)$) parallel ($k=(0,0,0,k_z)$) and perpendicular ($k=(0,0,k_y,0)$) about such a background.\footnote{The fact that these correlation functions are evaluated at zero frequency ensures that the fluctuations satisfy the equilibrium constraint (${\cal L}_V=0$).} 

Let us begin with the Kubo formulas for a thermodynamic transport coefficient which was previously considered, $M_5$, and a novel one $M_2$. Both are expressed in terms of static correlation functions 
as follows 
\begin{equation}
\label{eq:M2M5Kubo}
\begin{aligned}
\frac{1}{k_z} {\rm Im}\, G_{T^{xz} T^{yz}}(\omega = 0, k_z \mathbf{\hat{z}}) = -2\, B_0^2\, M_2\,, \\
\frac{1}{k_z} {\rm Im}\, G_{T^{tx} T^{yz}}(\omega = 0, k_z \mathbf{\hat{z}}) = - B_0 \, M_5 \,,
\end{aligned}
\end{equation}
in the limit of first setting $\omega = 0$, and then taking $k_z\to 0$, and $\mathbf{\hat z}$ is the unit vector in $z$-direction. In what follows, we take this limit in all the Kubo relations for thermodynamic transport coefficients.

For zero background magnetic field, it is still possible to find Kubo formulas for the magneto-vortical susceptibility\footnote{This Kubo formula agrees with (2.26) of \cite{Kovtun:2018dvd}}
\begin{equation}
\label{eq:M5-Kubo}
    \frac{1}{k_z^2} G_{J^xT^{tx}}(\omega=0,k_z \hat{k}) = M_5\,.
\end{equation}
While in principle the second order expression~\eqref{eq:M5-Kubo} could require corrections from ${\cal O}(\partial^2)$ thermodynamic transport coefficients which we have omitted here (such as a coefficient multiplying $E{\cdot}a$ in the generating functional), this was shown not to be the case in~\cite{Kovtun:2018dvd}.\footnote{One might worry that the anomaly could cause~\eqref{eq:M5-Kubo} to receive other ${\cal O}(\partial^2)$ corrections. However, thermodynamic Kubo formulas are ``protected'' from the anomaly in the sense that one can write a static generating functional $W_s$ which includes the pressure and the $M_n$ and simply add gauge dependent term in~\eqref{eq:WA} to account for the anomaly. The resulting equilibrium correlation functions, which are simply variations of $W_{cons}$ with respect to the sources, keep the anomalous sector separate from the other thermodynamic transport coefficients.}

The remaining thermodynamic transport coefficients $M_1,\, M_3,$ and $M_4$ are also expressed in terms of static correlation functions. However, in terms of two-point functions, we only find Kubo relations involving thermodynamic derivatives of the transport coefficients
\begin{equation}
\label{eq:M1M3M4Kubo}
\begin{aligned}
&\frac{1}{k_z} {\rm Im}\, G_{J^t T^{xx}}(\omega = 0, k_z \mathbf{\hat{z}}) = - \frac{2\,B_0^3}{T_0^4}\, \frac{\partial M_1}{\partial \mu}\,,\\
&\frac{1}{k_z} {\rm Im}\, G_{J^t J^t}(\omega = 0, k_z \mathbf{\hat{z}}) = B_0 \frac{\partial M_4}{\partial \mu}\,,\\
&\frac{1}{k_z} {\rm Im}\, G_{J^t T^{tt}}(\omega = 0, k_z \mathbf{\hat{z}}) = -B_0 \left(\frac{\partial M_3}{\partial \mu} + \frac{4 B_0^2}{T_0^4} \frac{\partial M_1}{\partial \mu} \right)\, . 
\end{aligned}
\end{equation}

The transport coefficient $M_1$ can be found without derivatives in the following combination
\begin{equation}
     \frac{1}{k_z} {\rm Im}\,G_{T^{tt}T^{xx}}(\omega=0,k_z \hat{k}) = 2\frac{B_0^3}{T_0^4} \left( M_1-T_0 \frac{\partial M_1}{\partial T}-\mu \frac{\partial M_1}{\partial \mu}-4 B_0^2 \frac{\partial M_1}{\partial B^2} - T_0^4\frac{\partial M_3}{\partial B^2} \right)\,.
\end{equation}

The susceptibility matrix may be defined as
\begin{equation}
    \chi_{ab} = \frac{\delta \langle \varphi_a \rangle }{\delta \lambda^b}\,,
\end{equation}
where $\varphi_{a} = (T^{tt}, T^{ti},J^t)$, and $\lambda^a = ( \delta T/T, u^i, T \delta \frac{\mu}{T})$.\footnote{Note that since the magnetic field breaks rotation invariance, we should have separated $T^{ti}$ into $T^{t\perp}$ and $T^{tz}$ where $\perp$ labels the orthogonal part of the momentum to the magnetic field. However, we find that at ${\cal O}(1)$ they coincide, and therefore simply use $T^{ti}$ here.} Explicitly, we have
\begin{equation}
    \chi_{ab} = \begin{pmatrix} 
    T \left(\frac{\partial \epsilon}{\partial T}\right)_{\mu/T} & 0 & \left(\frac{\partial \epsilon}{\partial \mu}\right)_T\\
    0 & w_0 & 0 \\
    T \left(\frac{\partial n}{\partial T} \right)_{\mu/T} & 0 & \left( \frac{\partial n}{\partial \mu}\right)_T
    \end{pmatrix}\,.
\end{equation}
The susceptibility matrix is symmetric since $T \left(\frac{\partial n}{\partial T} \right)_{\mu/T} = \left(\frac{\partial \epsilon}{\partial \mu}\right)_T$. The Kubo formulas for these terms are 
\begin{subequations}
    \begin{align}
      G_{J^tJ^t}(\omega=0, \mathbf{k}\to 0) = \chi_{33}\,,\\
      G_{T^{tt} J^t}(\omega=0, \mathbf{k}\to 0) = \chi_{13}\,,\\
      G_{T^{tt}T^{tt}}(\omega=0, \mathbf{k}\to 0) = \chi_{11}\,.
    \end{align}
\end{subequations}
These susceptibilities simplify some of the expressions for the transport coefficients. 
The enthalpy $w_0$ can be read off from the one point functions $\langle T^{tt} \rangle + \langle T^{zz} \rangle = w_0 $. In addition, the magnetic susceptibility $\aBB=2p_{,B^2}$ can be found by
\begin{equation}
    \frac{1}{k_z} {\rm Im}\,G_{J^x T^{yz}}(\omega=0,k_z\hat{k}) = - B_0\, \aBB\,. \label{eq:magsuc}
\end{equation}
As discussed in appendix~\ref{sec:noteOnChiB}, we can interpret $\aBB$ as a susceptibility. 

The anomalous transport coefficients can be found from static correlation functions~\cite{Amado:2011zx, Jensen:2013vta}. For example, in flat space with constant temperature, constant chemical potential and constant magnetic field in the $z$-direction, we find the following static correlation functions at small momentum\footnote{We write our Kubo formulas in terms of the covariant-consistent correlation functions. These can also be written in terms of the consistent-consistent correlation functions, which we summarize in appendix~\ref{app:jconsjcons}.}
\begin{equation}
\label{eq:chiral1}
\begin{aligned}
    &\langle J_{cov}^x(\mathbf{k}) T^{tz}(-\mathbf{k}) \rangle = - i \xi\, k_y\,, \quad \langle J_{cov}^x(\mathbf{k}) J_{cons}^z(-\mathbf{k})) \rangle = -i \xi_B\, k_y\,, \\
    &\langle T^{tx}(\mathbf{k}) T^{tz}(-\mathbf{k}) \rangle = -i \xi_T\, k_y \,, \quad \langle T^{tx}(\mathbf{k}) J_{cons}^z(-\mathbf{k}) \rangle = -i \xi_{TB}\, k_y \,,
\end{aligned}
\end{equation}
where we take the momentum in $y$-direction which is perpendicular to the magnetic field. We can instead take the momentum to point in the direction of the magnetic field, in which case we find
\begin{equation}
\label{eq:chiral2}
\begin{aligned}
    &\langle J_{cov}^x(\mathbf{k}) T^{ty}(-\mathbf{k}) \rangle = - i \xi\, k_z\,, \quad \langle J_{cov}^x(\mathbf{k}) J_{cons}^y(-\mathbf{k})) \rangle = -i \xi_B\, k_z\,, \\
    &\langle T^{tx}(\mathbf{k}) T^{ty}(-\mathbf{k}) \rangle = -i \xi_T\, k_z \,, \quad \langle T^{tx}(\mathbf{k}) J_{cons}^y(-\mathbf{k}) \rangle = -i \xi_{TB}\, k_z \,.
\end{aligned}
\end{equation}

\subsubsection{A comment about thermodynamic Kubo formulas}\label{sec:commentKubo}
In our equilibrium setup with homogeneous magnetic fields, the thermodynamic functions, $M_{1,3,4}$, unfortunately cannot be isolated using only first order, static two-point functions. Nevertheless, it is still possible to isolate their derivatives with respect to the chemical potential, $M_{n,\mu}$.

Now if a given $M_n$ is parity odd such that $M_{n,\mu} = 0$, then $M_n = 0$. The reason for this is quite simple. In a microscopic system where parity is broken only by the presence of some axial chemical potential, $\mu$, the generating functional $W_s$, see eq.~\eqref{eq:Ws}, is parity invariant. The coefficient in front of a parity odd $M_n$ must then also be parity odd, and with $\mu$ the only parity breaking term in the hydrodynamic system, we must have $M_n(\mu = 0) = 0$. Thus for finite $\mu$, we can write any parity odd $M_n$ as
\begin{equation}\label{eq:parityOddMnIntegral}
    M_n(\mu) = \int_0^\mu d\mu' M_{n,\mu'}(\mu')\, ,
\end{equation}
and the statement follows. 
As shall be seen, $M_{1,\tilde\mu} =  M_{3,\tilde\mu} =  M_{4,\tilde\mu} = 0$ in our holographic model, where $\tilde\mu = \mu/T$ for fixed $T$, see section~\ref{sec:thermoCoeffsHolo}. 
Since we are dealing with external axial gauge fields, only $M_4 $ is odd in $\mu$, and we conclude that $M_4 = 0$ by the argument above. However, if the system were coupled to vector gauge fields, the same argument would hold for $M_{1,3,5} = 0$.

\subsection{Hydrodynamics}
%
\subsubsection{Non-equilibrium constitutive relations}
With the equilibrium terms out of the way, the next step is to add the non-equilibrium terms to our constitutive relations. 
The non-equilibrium terms are the scalar, vector and tensor structures which are required to vanish in equilibrium by the constraint ${\cal L}_V=0$\footnote{See table 3 in~\cite{Hernandez:2017mch} for an exhaustive list.}. These terms can be derived from a non-local effective Schwinger-Keldysh action. Recent reviews on the non-equilibrium formalism for hydrodynamics can be found in~\cite{Haehl:2018lcu,Glorioso:2018wxw,Jensen:2018hse}. For the purposes of our analysis, we use the effective field theory 
approach of adding all the non-equilibrium terms allowed by our symmetries to the constitutive relations, and constraining their transport coefficients via the Onsager relations and the entropy constraints.

The definition of the thermodynamic quantities~\eqref{eq:Thermo-def} is ambiguous when out of equilibrium. The redefinition of $T$, $\mu$ and $u^\mu$ are referred to as hydrodynamic frame transformations. An introductory review of this ambiguity in the hydrodynamic framework can be found in~\cite{Kovtun:2011np}, implications of frame-choice on the stability of hydrodynamics were discussed recently~\cite{Kovtun:2019hdm,Hoult:2020eho,Bemfica:2017wps,Bemfica:2019knx,Bemfica:2020zjp}, and the modifications required for fluids in strong magnetic fields are explained in~\cite{Hernandez:2017mch}. For our purposes, we use the approach in~\cite{Hernandez:2017mch} 
to add the non-equilibrium terms in a systematic way to the hydrodynamic frame invariants.

We begin by isolating $O(1)$ and $O(\partial)$ contributions to the energy-momentum tensor~\eqref{eq:TT1} and the current~\eqref{eq:JJ1}. 
The spatial part of the current ${\cal J}^\mu$ has no ${\cal O}(1)$ term, neither does the energy current ${\cal Q}^\mu$. So we are left with the ${\cal O}(1)$ quantities 
\begin{align*}
  & {\cal E} = \epsilon(T,\mu,B^2) + f_{\cal E}\,,\\[5pt]
  & {\cal P} = \Pi(T,\mu,B^2) + f_{\cal P}\,,\\[5pt]
  & {\cal N} = n(T,\mu,B^2) + f_{\cal N}\,,\\[5pt]
  & {\cal T}^{\mu\nu} = \aBB(T,\mu,B^2) \left( B^\mu B^\nu - \coeff13 \Delta^{\mu\nu} B^2 \right)
    +f^{\mu\nu}_{\cal T}\,,
\end{align*}
where $\epsilon = -p+ T(\partial p/\partial T) + \mu (\partial p/\partial\mu)$, $\Pi=p-\coeff23 \aBB B^2$, $n=\partial p/\partial\mu$, and the magnetic susceptibility is $\aBB = 2\partial p/\partial B^2$. The terms $f_{\cal E}$, $f_{\cal P}$, $f_{\cal N}$, $f^{\mu\nu}_{\cal T}$, ${\cal Q}^\mu$, and ${\cal J}^\mu$ are all $O(\partial)$, and contain both equilibrium and non-equilibrium contributions, $f_{\cal E} = \bar f_{\cal E} + f_{\cal E}^\textrm{non-eq.}$ etc, where the bar denotes $O(\partial)$ contributions coming from the variation of $W_s$.

We can then write down the following quantities which are invariant under hydrodynamic frame transformations
\begin{subequations}
\label{eq:fi}
\begin{align}
  & f \equiv f_{\cal P} - \left(\frac{\partial\Pi}{\partial\epsilon}\right)_{\!n} f_{\cal E}
    - \left(\frac{\partial\Pi}{\partial n}\right)_{\!\epsilon} f_{\cal N}\,,\\[5pt]
  & \ell \equiv \frac{B^\alpha}{B} \left( {\cal J}_\alpha - \frac{n}{\epsilon+p}{\cal Q}_\alpha \right)\,,\\[5pt]
  & \ell^\mu_\perp \equiv \mathbb{B}^{\mu\alpha} \left( {\cal J}_\alpha - \frac{n}{\epsilon+p-\aBB B^2}{\cal Q}_\alpha \right) \,,\\[5pt]
  & t^{\mu\nu} \equiv f^{\mu\nu}_{\cal T} - \left( B^\mu B^\nu - \coeff13 \Delta^{\mu\nu}B^2 \right) \left[ \left(\frac{\partial \aBB}{\partial \epsilon}\right)_{\!n} f_{\cal E} + \left(\frac{\partial \aBB}{\partial n}\right)_{\!\epsilon} f_{\cal N}  \right]\,.
\end{align}
\end{subequations}
Here $\mathbb{B}^{\mu\nu} \equiv \Delta^{\mu\nu} - B^\mu B^\nu/B^2$ is the projector onto a plane orthogonal to both $u^\mu$ and $B^\mu$, all thermodynamic derivatives are evaluated at fixed $B^2$, and $B\equiv \sqrt{B^2}$. When the magnetic susceptibility $\aBB$ is $T$- and $\mu$-independent, the stress $f^{\mu\nu}_{\cal T}$ is frame-invariant.

Following the notation of~\cite{Hernandez:2017mch} with the slight modification $c_{14} \to c_{14}-c_{15}$, the terms in the non-equilibrium frame invariants are
\begin{subequations}
\label{eq:PV-non-eq}
\begin{align}
f_{\text{non-eq.}} &= -\zeta_1\, \nabla{\cdot}u - \zeta_2 \,{b^\mu b^\nu}\nabla_{\!\mu} u_\nu + c_3\, b{\cdot} V \,,\\[5pt]
\ell_{\text{non-eq.}} &= \sigma_\parallel\, b{\cdot}V + c_4\, \nabla {\cdot} u + c_5\, b^\mu b^\nu \nabla_\mu u_\nu\,,\\[5pt]
\ell_{\perp,\text{non-eq.}}^\mu &= \sigma_\perp V^\mu_\perp + \tilde{\sigma}_\perp \tilde{V}^\mu + c_8 \Sigma^\mu + c_{10} \tilde{\Sigma}^\mu\,,\\[5pt]
\tau^{\mu\nu}_{\text{non-eq.}} &= - \eta_\perp \sigma^{\mu\nu}_\perp
  -\eta_\parallel (b^\mu \Sigma^\nu + b^\nu \Sigma^\mu) 
  - b^{\langle \mu}b^{\nu\rangle} \left(\eta_1 \nabla{\cdot}u + \eta_2 b^\alpha b^\beta \nabla_\alpha u_\beta - c_{14} b{\cdot V} \right)\nonumber\\[5pt]
  &- \tilde\eta_\perp \tilde\sigma^{\mu\nu}_\perp 
  - \tilde\eta_\parallel (b^\mu \tilde\Sigma^\nu + b^\nu \tilde\Sigma^\mu) + c_{15} (b^{\mu} V_\perp^{\nu} + b^\nu V^\mu_\perp) + c_{17}( b^{\mu}\tilde{V}^{\nu} +b^\nu \tilde{V}^\mu) \,,
\end{align}
\end{subequations}
where $b^\mu = B^\mu/\sqrt{B^2}$, $V^\mu = E^\mu - T \Delta^{\mu\nu} \partial_\nu \frac{\mu}{T}$, $V_\perp^\mu  = (\Delta^{\mu\nu}  - b^\mu b^\nu) V_\nu$, $\Sigma^\mu = (\Delta^{\mu\nu}  - b^\mu b^\nu)\sigma_{\nu\rho}b^\rho$ and for any vectors $\tilde{v}^\mu = \epsilon^{\mu\nu\rho\sigma} u_\nu b_\rho v_\sigma$,  $v^{(\mu}_1 v^{\nu)}_2 = v^\mu_1 v^\nu_2 + v^\nu_1 v^\mu_2$ and $v_1^{\langle\mu}v_2^{\nu\rangle} = v_1^\mu v_2^\nu + v_2^\mu v_1^\nu - \coeff23 \Delta^{\mu\nu} v_1{\cdot} v_2$. The shear tensor and the projector orthogonal to $u^\mu$ have the usual definitions $\sigma^{\mu\nu} = \Delta^{\mu\alpha}\Delta^{\nu\beta}(\nabla_\alpha u_\beta + \nabla_\beta u_\alpha - \coeff23 g_{\alpha\beta} \nabla{\cdot} u)$ and $\Delta^{\mu\nu} = g^{\mu\nu} + u^\mu u^\nu$. The transverse component of the shear tensor is $\sigma_\perp^{\mu\nu} = \frac12 \left( \mathbb{B}^{\mu\alpha}\mathbb{B}^{\nu\beta} + \mathbb{B}^{\mu\beta}\mathbb{B}^{\nu\alpha} - \mathbb{B}^{\mu\nu}\mathbb{B}^{\alpha\beta}\right) \sigma_{\alpha_\beta}$ and the tilded version is $\tilde{\sigma}_\perp^{\mu\nu} = \frac12 \left(\epsilon^{\mu\alpha\beta\gamma} u_\alpha b_\beta \sigma_{\perp\gamma} ^\nu + \epsilon^{\nu\alpha\beta\gamma} u_\alpha b_\beta \sigma_{\perp\gamma} ^\mu  \right)$. The coefficients in front of the first order hydrodynamic terms are hydrodynamic transport coefficients. These are functions of the ${\cal O}(1)$ hydrodynamic quantities (For example, $\zeta_1 = \zeta_1(T,\mu,B^2)$). These transport coefficients will be subject to four equality constraints coming from the Onsager relations, as well as some inequality constraints coming from the entropy/correlation function argument. 

Furthermore, considering the parity eigenvalues of the quantities in front of the transport coefficients, we can predict the parity eigenvalue of the transport coefficient themselves, since the combination of the two must match the parity of the stress tensor or axial current. 
Since $\mu$ is the only parity pseudo-scalar, this allows us to constraint these transport coefficients as even or odd functions of $\mu$. From the previously explored transport coefficients, the tilded ones ($\tilde{\eta}_\perp$, $\tilde{\eta}_\parallel$, and $\tilde{\sigma}_\perp$) are odd functions of the chemical potential, while the rest ($\sigma_\parallel$, $\rho_\perp$, $\eta_\perp$, $\eta_\parallel$, $\zeta_1$, $\zeta_2$, $\eta_1$ and $\eta_2$) are even functions of the chemical potential. From the previously unexplored transport coefficients, $c_{10}$ and $c_{17}$ are even functions of the chemical potential, while $c_3$, $c_4$, $c_5$, $c_8$, $c_{14}$ and $c_{15}$ are odd functions of the chemical potential.

For completeness, let us summarize the constitutive relations for a parity-violating theory in the thermodynamic frame. The energy-momentum tensor is given by
\begin{subequations}
\label{eq:TTF}
\begin{align}
  {\cal E} & = {\cal E}_{\rm eq.}\,, \quad {\cal Q}^\mu  ={\cal Q}^\mu_{\rm eq.}\, \quad  {\cal P} ={\cal P}_{\rm eq.} -\zeta_1 \nabla{\cdot}u - \zeta_2 {b^\mu b^\nu}\nabla_{\!\mu} u_\nu + c_3 b{\cdot} V \,,\\[5pt]
  {\cal T}^{\mu\nu} & ={\cal T}^{\mu\nu}_{\rm eq.} - \eta_\perp \sigma^{\mu\nu}_\perp
  -\eta_\parallel (b^\mu \Sigma^\nu + b^\nu \Sigma^\mu) 
  - b^{\langle \mu}b^{\nu\rangle} \left(\eta_1 \nabla{\cdot}u + \eta_2 b^\alpha b^\beta \nabla_\alpha u_\beta - c_{14} b{\cdot} V \right)\nonumber\\[5pt]
  & \ \ \ \,  
  - \tilde\eta_\perp \tilde\sigma^{\mu\nu}_\perp 
  - \tilde\eta_\parallel (b^\mu \tilde\Sigma^\nu + b^\nu \tilde\Sigma^\mu) + c_{15} (b^{\mu} V_\perp^{\nu} + b^\nu V^\mu) + c_{17}( b^{\mu}\tilde{V}^{\nu} +b^\nu \tilde{V}^\mu) \,,
\end{align}
\end{subequations}
where $\sigma^{\mu\nu}_\perp \equiv \coeff12 \left(\mathbb{B}^{\mu\alpha} \mathbb{B}^{\nu\beta} + \mathbb{B}^{\nu\alpha} \mathbb{B}^{\mu\beta} - \mathbb{B}^{\mu\nu} \mathbb{B}^{\alpha\beta}\right) \sigma_{\alpha\beta}$ is the part of the shear tensor transverse to the magnetic field, and $\tilde{\sigma}_\perp ^{\mu\nu} = \frac{1}{2} \left( \epsilon^{\mu\lambda\alpha\beta} u_\lambda b_\alpha \sigma_{\perp\beta}^{\ \ \ \nu} + \epsilon^{\nu\lambda\alpha\beta} u_\lambda b_\alpha \sigma_{\perp\beta}^{\ \ \ \mu}\right)$. We used the projection orthogonal to the magnetic field and the fluid velocity $\mathbb{B}^{\mu\nu} = \Delta^{\mu\nu} - b^\mu b^\nu$. The current is given by 
\begin{subequations}
\label{eq:JTF}
\begin{align}
  {\cal N} & = {\cal N}_{\rm eq.}\,, \\[5pt]
  {\cal J}^\mu & = {\cal J}^\mu_{\rm eq.} + \sigma_\perp V^\mu_\perp +  \tilde\sigma_\perp\, \tilde V^\mu + b^\mu (\sigma_{||}b{\cdot} V + c_4 \nabla {\cdot} u + c_5 b^\alpha b^\beta \nabla_\alpha u_\beta) + c_8 \Sigma^\mu + c_{10} \tilde{\Sigma}^\mu\,.
\end{align}
\end{subequations}
The magnetic polarization vector is given in~\eqref{eq:m-vector} and the polarization vector is $p_\mu = M_4 B_\mu$. 

\subsubsection{Hydrodynamic correlation functions}
\label{sec:correlators}
We can find the two point correlation functions of energy-momentum and conserved currents by varying the one-point functions given by the constitutive relations in the presence of external sources with respect to the external sources. To do this, we solve the hydrodynamic equations in the presence of plane wave external source perturbations $\delta A, \delta g$ (proportional to $\exp(-i \omega t + i \mathbf{k\cdot x})$) to find $\delta T[A,g], \delta \mu[A,g], \delta u^\mu[A,g]$, then vary the resulting on-shell expressions $T^{\mu\nu}_{\rm on-shell}[A,g]$ and $J^\mu_{\rm on-shell}[A,g]$ with respect to $g_{\mu\nu}$ and $A_\mu$ to find the retarded hydrodynamic correlation functions
\begin{subequations}
\label{eq:corr-funcs}
\begin{align}
  & G^R_{T^{\mu\nu} T^{\alpha\beta}} = \frac{2}{\sqrt{-g}} \frac{\delta}{\delta g_{\alpha\beta}} \left( \sqrt{-g}\, T^{\mu\nu}_\textrm{on-shell}[A,g] \right)\,,
  &&  G^R_{J^\mu T^{\alpha\beta}} = \frac{2}{\sqrt{-g}} \frac{\delta}{\delta g_{\alpha\beta}} \left( \sqrt{-g}\, J^{\mu}_\textrm{on-shell}[A,g] \right)\,,\\[5pt]
  & G^R_{T^{\mu\nu} J^\alpha} = \frac{\delta}{\delta A_{\alpha}}   T^{\mu\nu}_\textrm{on-shell}[A,g]  \,,
  &&  G^R_{J^\mu J^\alpha} = \frac{\delta}{\delta A_{\alpha}}  J^{\mu}_\textrm{on-shell}[A,g] \,,
\end{align}
\end{subequations}
where the source perturbations $\delta g$ and $\delta A$ are set to zero after the variation. The above expressions are to be understood as
\begin{equation}
\label{eq:corr-funcspr}
  \delta(\sqrt{-g}\, T^{\mu\nu}_\textrm{on-shell}) 
  = \coeff12 \sqrt{-g} \,  G^R_{T^{\mu\nu} T^{\alpha\beta}} \, \delta g_{\alpha\beta} (\omega, {\bf k})\,,
\end{equation}
etc. This provides a direct method to evaluate the retarded functions, and allows both to find constrains due to the Onsager relations and to derive Kubo formulas for transport coefficients. 

\subsubsection{Symmetry constraints and Onsager relations}
\label{sec:onsager}
Time reversal covariance adds additional constraints to the transport coefficients, called the Onsager relations~\cite{Onsager1,Onsager2}. We consider a state characterized by a density matrix $\rho(\chi)$ and an anti-unitary operator $\Theta$ such that
\begin{equation}
    \Theta^{-1} \rho(\chi) \Theta = \rho (-\chi)\,,
\end{equation}
where $\chi$ are some $\Theta$ symmetry breaking parameters of the state associated with the density matrix $\rho(\chi)$. 
Recall that the expectation values in this state are given by
\begin{equation}
    \langle {\cal O} \rangle_{\rho(\chi)} = {\rm Tr}\left(\rho(\chi){\cal O}\right)\,,
\end{equation}
and the retarded two point correlation functions are given by 
\begin{equation}
    G^R_{\varphi_a \varphi_b}(t,\mathbf{x};\chi) = i \theta(t) \langle [\varphi_a(t,\mathbf{x}),\varphi_b(0,\mathbf{0})]\rangle_{\rho(\chi)}\,.
\end{equation}
Now, for states that are homogeneous in space-time (i.e. with space-time translation invariance), the transformation properties of $\rho(\chi)$ under $\Theta$ leads to 
\begin{equation}
    G^R_{\varphi_a \varphi_b}(t,\mathbf{x};\chi) = \eta_{\varphi_a} \eta_{\varphi_b} G^R_{\varphi_b^\dagger \varphi_a^\dagger}(t,-\mathbf{x};-\chi) \,,
\end{equation}
where $\eta_{\varphi_a}$ is the $\Theta$ eigenvalue of $\varphi_a$ and similarly for $\varphi_b$.

This relation can be translated to the Fourier basis correlators
\begin{equation}
    G^R_{\varphi_a\varphi_b}(\omega,\mathbf{k};\chi) = \int d^4x\, e^{-i\omega t + i \mathbf{k}{\cdot}\mathbf{x}} G^R_{\varphi_a \varphi_b}(t,\mathbf{x};\chi)\,,
\end{equation}
where we find
\begin{equation}
\label{eq:Ons}
    G^R_{\varphi_a \varphi_b}(\omega,\mathbf{k};\chi) = \eta_{\varphi_a} \eta_{\varphi_b} G^R_{\varphi_b^\dagger \varphi_a^\dagger}(\omega,-\mathbf{k};-\chi)\,.
\end{equation}

To derive the Onsager relations in our system we have the option of using $\Theta = {\cal T}$, $\chi = B_0$ or $\Theta = {\cal PT}$, $\chi = \mu$. The Onsager relations are derived by using~\eqref{eq:Ons} on two point functions of energy-momentum and currents. 

A similar argument using the unitary parity operator ${\cal P}$ also gives the constraint

\begin{equation}
\label{constraint:parity}
    G^R_{\varphi_a\varphi_b}(\omega,\mathbf{k};\chi) = \epsilon_{\varphi_a}\epsilon_{\varphi_b} G^R_{\varphi_a\varphi_b}(\omega,-\mathbf{k};-\chi)\,,
\end{equation}
where $\epsilon_{\varphi_a}$ is the ${\cal P}$ eigenvalue of $\varphi_a$ and in this case $\chi = (B_0,\mu)$. We will refer to the constraints derived from eq.~\eqref{constraint:parity} as the parity constraints. These constraints are the same that can be derived from considering the parity eigenvalue of the terms in the constitutive relations: $\tilde{\eta}_\perp$, $\tilde{\eta}_\parallel$, $\tilde{\sigma}_\perp$, $c_3$, $c_4$, $c_5$, $c_8$, $c_{14}$ and $c_{15}$ are odd functions of the chemical potential, while $\sigma_\parallel$, $\rho_\perp$, $\eta_\perp$, $\eta_\parallel$, $\zeta_1$, $\zeta_2$, $\eta_1$, $\eta_2$, $c_{10}$ and $c_{17}$ are even functions of the chemical potential.

\subsubsection{Hydrodynamic Kubo formulas for systems in strong magnetic fields}
\label{sec:kubo}
The Kubo fomulas for the non-equilibrium transport coefficients can be found by evaluating the zero spatial momentum, low frequency limit of the retarded functions in flat space-time. For parity preserving systems coupled to strong vector magnetic fields, only the viscosities ($\zeta_1$, $\zeta_2$, $\eta_1$, $\eta_2$, $\eta_\parallel$, $\tilde{\eta}_\parallel$, $\eta_\perp$ and $\tilde{\eta}_\perp$) and the conductivities ($\sigma_\parallel$, $\sigma_\perp$ and $\tilde{\sigma}_\perp$) appear in the constitutive relations. The two-point function of the longitudinal current $J^z$ gives the longitudinal conductivity,\footnote{Note that we drop the superscript ``$R$'' for all retarded Green's functions from here on in order to declutter the notation.}
\begin{subequations}
\label{eq:Kubo-r1}
\begin{align}
\label{eq:Kubo-r1a}
   \coeff{1}{\omega} {\rm Im}\, G_{J^z J^z}(\omega,{\bf k}{=}0) =  \sigma_\parallel + \cdots\,,
\end{align}
in the limit of first setting ${\bf k} = 0$, and then taking $\omega\to 0$. In what follows, we take this limit in all the Kubo relations for hydrodynamic transport coefficients. 
The ellipsis denote terms that vanish for $B_0 \ll T_0^2$ or when $M_1=M_3=M_4=0$. The Kubo formulas for the transverse conductivities simplify when written in terms of the transverse resistivities. We define the $2\times2$ conductivity matrix in the plane transverse to ${\bf B}_0$ as $\sigma_{ab} \equiv \sigma_\perp \delta_{ab} + \left(\frac{n_0}{|{\bf B}_0|} + \tilde\sigma_\perp \right)\epsilon_{ab}$, and the corresponding resistivity matrix as $\rho_{ab} \equiv (\sigma^{-1})_{ab} = \rho_\perp \delta_{ab} + \tilde\rho_\perp\, \epsilon_{ab}$, which defines $\rho_\perp$ and~$\tilde\rho_\perp$. Using these definitions, the two-point functions of the transverse currents $J^x$, $J^y$ give the transverse resistivities,
\begin{align}
\label{eq:Kubo-r1b}
  &  \coeff{1}{\omega} {\rm Im}\, G_{J^x J^x}(\omega,{\bf k}{=}0) = \omega^2 \rho_\perp \frac{w_0(w_0-\MO_{,\mu}B_0^2)}{B_0^4}  \,,\\
\label{eq:Kubo-r1c}
  &  \coeff{1}{\omega}{\rm Im}\, G_{J^x J^y}(\omega,{\bf k}{=}0) = 
     \frac{n_0}{B_0} - \omega^2 \tilde\rho_\perp \frac{w_0(w_0-\MO_{,\mu}B_0^2)}{B_0^4}\, {\rm sign}(B_0)\, ,
\end{align}
\end{subequations}	
Alternatively, the transverse resistivities can be found from correlation functions of momentum density,
\begin{subequations}
\label{eq:Kubo-r2}
\begin{align}
  & \coeff{1}{\omega} {\rm Im}\, G_{T^{tx} T^{tx}} (\omega, {\bf k}{=}0) = \rho_\perp \frac{w_0(w_0-\MO_{,\mu}B_0^2)}{B_0^2} \,,\\[5pt]
  & \coeff{1}{\omega} {\rm Im}\, G_{T^{tx} T^{ty}} (\omega, {\bf k}{=}0) = -\tilde \rho_\perp {\rm sign}(B_0) \frac{w_0(w_0-\MO_{,\mu}B_0^2)}{B_0^2}\,.
\end{align}
\end{subequations}
where $O_3=\frac{1}{2}(T^{xx} - T^{yy})$. The ``bulk'' viscosities may be expressed as
\begin{subequations}
\begin{align}\label{eq:Kubo-Tmne}
  & \coeff{1}{\omega}\delta_{ij} {\rm Im}\, G_{T^{ij} O_1}(\omega,{\bf k}{=}0) = 3 \zeta_1+\cdots\,,\\ \label{eq:Kubo-Tmnf}
  & \coeff{1}{3\omega} \delta_{ij} \delta_{kl}\, {\rm Im}\, G_{T^{ij} T^{kl}}(\omega,{\bf k}{=}0) = 3\zeta_1 + \zeta_2 +\cdots\,,\\ \label{eq:Kubo-Tmng}
  & \coeff{1}{\omega} {\rm Im}\, G_{O_1 O_1}(\omega,{\bf k}{=}0) = \zeta_1 - \coeff23 \eta_1+\cdots\,,\\ \label{eq:Kubo-Tmnh}
  & \coeff{1}{\omega} {\rm Im}\, G_{O_2 O_2}(\omega,{\bf k}{=}0) = 2\eta_2+\cdots\,,
\end{align}
where $O_1 = \frac12(T^{xx}+T^{yy})$, and $O_2 = T^{zz} - \frac12(T^{xx}+T^{yy})$. The $\delta^{ij}$ is the projector onto the spatial coordinates, i.e. $i=x,y,z$. The ellipsis denote terms that vanish when $M_1=M_3=M_4=0$, or when $B_0 \ll T_0^2$. 
The shear viscosities are given by\footnote{For parity preserving systems the $c_i$ coefficients vanish and the Kubo formulas are identical to those in~\cite{Hernandez:2017mch}.}

\label{eq:Kubo-Tmn}
\begin{align}
  & \coeff{1}{\omega}{\rm Im}\, G_{T^{xy} T^{xy}}(\omega,{\bf k}{=}0) =  \eta_\perp \,, \label{eq:etaPerpKubo}\\ 
  & \coeff{1}{\omega}{\rm Im}\, G_{T^{xy} O_3}(\omega,{\bf k}{=}0) =  \tilde\eta_\perp\, {\rm sign}(B_0) \,,\label{eq:tildeEtaPerpKubo}\\
 & \coeff{1}{\omega}{\rm Im}\, G_{T^{xz} T^{xz}}(\omega,{\bf k}{=}0) =  \eta_\parallel + (c_8 c_{15} - c_{10} \bar{c}_{17}) \rho_\perp - (c_8 \bar{c}_{17} + c_{10} c_{15} ) \tilde{\rho}_\perp \,,\\
  & \coeff{1}{\omega}{\rm Im}\, G_{T^{yz} T^{xz}}(\omega,{\bf k}{=}0) = \left( \tilde\eta_\parallel  + (c_8 \bar{c}_{17} + c_{10} c_{15} ) \rho_\perp +  (c_8 c_{15} - c_{10} \bar{c}_{17}) \tilde{\rho}_\perp \right) {\rm sign}(B_0)\,,
\end{align}
\end{subequations}
Using relation~\eqref{eq:Ons} yields the Onsager relations for the parity preserving transport coefficients
\begin{equation}
\label{eq:ons}
    3\zeta_2-6\eta_1-2\eta_2=0\,.
\end{equation}
In addition, the parity constraints coming from relation~\eqref{constraint:parity} imply that the tilded transport coefficients $\tilde{\rho}_\perp$, $\tilde{\eta}_\perp$ and $\tilde{\eta}_\parallel$ are odd functions of the chemical potential, while the untilded $\sigma_\parallel$, $\rho_\perp$, $\eta_\perp$, $\eta_\parallel$, $\zeta_1$, $\zeta_2$, $\eta_1$ and $\eta_2$ are even functions of the chemical potential.\footnote{These behaviours can be derived using ${\cal C}$ instead of ${\cal P}$ for vector gauge fields instead of axial gauge fields.}

The Kubo formulas for the parity violating non-equilibrium coefficients appearing in~\eqref{eq:PV-non-eq} are given by 
\begin{subequations}
\label{eq:Kubo-pv}
\begin{align}
  & \coeff{1}{\omega}{\rm Im}\, G_{T^{tx} T^{xz}}(\omega,{\bf k}{=}0) = - \frac{w_0-\MO_{,\mu}B_0^2}{B_0}(c_8 \tilde{\rho}_\perp + c_{10}\rho_\perp) \,,\\[5pt]  
   & \coeff{1}{\omega}{\rm Im}\, G_{T^{tx} T^{yz}}(\omega,{\bf k}{=}0) =- \frac{w_0-\MO_{,\mu}B_0^2}{|B_0|}(c_8 \rho_\perp - c_{10}\tilde{\rho}_\perp) \,,\\[5pt]
  & \coeff{1}{\omega}{\rm Im}\, G_{ T^{xz}T^{tx}}(\omega,{\bf k}{=}0) =  \frac{w_0}{B_0}(c_{15} \tilde{\rho}_\perp + \bar{c}_{17}\rho_\perp) \,,\\[5pt]
  & \coeff{1}{\omega}{\rm Im}\, G_{T^{yz}T^{tx}}(\omega,{\bf k}{=}0) =  -\frac{w_0}{|B_0|}(c_{15} \rho_\perp - \bar{c}_{17}\tilde{\rho}_\perp) \,,\\[5pt]
  & \coeff{1}{\omega}{\rm Im}\, G_{J^{z} O_1}(\omega,{\bf k}{=}0) = -c_4 \,{\rm sign}(B_0) + \cdots \,,\\[5pt]
  & \coeff{1}{\omega}{\rm Im}\, G_{J^{z} O_2}(\omega,{\bf k}{=}0) =  -c_5 \, {\rm sign}(B_0) +\cdots \,\\[5pt]
    & \coeff{1}{\omega}\delta_{ij}{\rm Im}\, G_{T^{ij}J^{z}}(\omega,{\bf k}{=}0) =  3 c_3 \,{\rm sign}(B_0) + \cdots \,,\\[5pt]
  & \coeff{1}{\omega}{\rm Im}\, G_{O_2 J^{z}}(\omega,{\bf k}{=}0) =  2 c_{14} \, {\rm sign}(B_0) +\cdots\,,
\end{align}
\end{subequations}
where once again the terms 
in the ellipsis vanish for $M_1=M_3=M_4=0$ or $B_0 \ll T^2_0$. The rest of the Kubo formulas in the previous section~\eqref{eq:Kubo-r1} \eqref{eq:Kubo-r2} and \eqref{eq:Kubo-Tmn} remain valid when the gauge fields are axial. As mentioned in section~\ref{sec:onsager}, the Onsager relations give constraints on the transport coefficients. The constraints on the parity-violating coefficients can be derived using $\Theta = {\cal T}$ in~\eqref{eq:Ons}. Here, ${\cal T}$ refers to time-reversal. In addition to~\eqref{eq:ons}, these are
\begin{equation}
\label{eq:c-onsager}
c_3 = -c_4 - \coeff13 c_5 \,,\quad c_{14} = - \coeff12 c_5\,, \quad c_{15} = - \frac{w_0-\MO_{,\mu} B^2_0}{w_0} c_8\,,\quad \bar{c}_{17} =-  \frac{w_0-\MO_{,\mu} B^2_0}{w_0} c_{10}\,.
\end{equation}
In addition, the parity constraints~\eqref{constraint:parity} imply that $c_{10}$ and $c_{17}$ are even functions of the chemical potential, while $c_3$, $c_4$, $c_5$, $\bar{c}_8$, $c_{14}$ and $c_{15}$ are odd functions of the chemical potential.

For a microscopic theory with a chiral anomaly, the Kubo formulas for the parallel shear viscosities are slightly modified
\begin{subequations}
\begin{align}\label{eq:KuboTT1an2}
     &\coeff{1}{\omega}{\rm Im}\, G_{T^{xz} T^{xz}}(\omega,{\bf k}{=}0) =  \eta_\parallel +  (c_8 c_{15} - \tilde{c}_{10} \tilde{c}_{17}) \rho_\perp  - ( c_8 \tilde{c}_{17} + \tilde{c}_{10} c_{15})  \tilde{\rho}_\perp \,, \\ 
   &\coeff{1}{\omega}{\rm Im}\, G_{T^{yz} T^{xz}}(\omega,{\bf k}{=}0) = \left( \tilde\eta_\parallel  + ( c_8 \tilde{c}_{17} + \tilde{c}_{10} c_{15}) \rho_\perp +  ( c_8 c_{15} - \tilde{c}_{10} \tilde{c}_{17} ) \tilde{\rho}_\perp \right) {\rm sign}(B_0)\,, \label{eq:KuboTT1an2Hall}  
\end{align}
\end{subequations}
where
$$\tilde{c}_{10} = c_{10} -\xi_{TB}\,,\quad \tilde{c}_{17} = \bar{c}_{17} +\xi_{TB} = c_{17} + B_0^2 M_{2,\mu}+\xi_{TB} \,.
$$
Recall that $\xi_{TB} = \frac12C\mu^2 +c_1T^2+2c_2 T\mu $. The Kubo formulas for parity violating non-equilibrium coefficients that are modified are\footnote{To isolate $c_8$ and $\tilde{c}_{10}$, we invert eqs.~\eqref{eq:KuboTT2an1} and \eqref{eq:KuboTT1an2}, 
then, using eqs.~\eqref{eq:Kubo-r2} we find
\begin{equation}
    \begin{aligned}
    \tilde{c}_{10} &= -\frac{w_0}{B_0} \frac{{\rm Im} G_{T^{tx} T^{tx}}\, {\rm Im} \, G_{T^{tx} T^{xz}} + {\rm Im} \, G_{T^{tx} T^{ty}}\, {\rm Im} \, G_{T^{tx} T^{yz}}}{\left( {\rm Im} \, G_{T^{tx} T^{tx}}\right)^2 + \left({\rm Im} \, G_{T^{tx} T^{ty}} \right)^2}\,,\label{eq:c10tildeIsolated}\\
    c_{8} &= \frac{w_0}{|B_0|} \frac{{\rm Im}\, G_{T^{tx} T^{ty}} \, {\rm Im}\, G_{T^{tx} T^{xz}} - {\rm Im}\, G_{T^{tx} T^{tx}} \,  {\rm Im}\, G_{T^{tx} T^{yz}}}{\left( {\rm Im}\, G_{T^{tx} T^{tx}}\right)^2 + \left({\rm Im}\, G_{T^{tx} T^{ty}} \right)^2}\,.
    \end{aligned}
\end{equation}
We may isolate $c_{15}$ and $c_{17}$ in a similar way. 
}

\begin{subequations}
\begin{align}\label{eq:KuboTT2an1}
  & \coeff{1}{\omega}{\rm Im}\, G_{T^{tx} T^{xz}}(\omega,{\bf k}{=}0) = - \frac{w_0-\MO_{,\mu}B_0^2}{B_0}(c_8 \tilde{\rho}_\perp + \tilde{c}_{10}\rho_\perp) \,,\\[5pt]  \label{eq:KuboTT2an2}
   & \coeff{1}{\omega}{\rm Im}\, G_{T^{tx} T^{yz}}(\omega,{\bf k}{=}0) = - \frac{w_0-\MO_{,\mu}B_0^2}{|B_0|}(c_8 \rho_\perp - \tilde{c}_{10}\tilde{\rho}_\perp) \,,\\[5pt] \label{eq:KuboTT2an3}
  & \coeff{1}{\omega}{\rm Im}\, G_{ T^{xz}T^{tx}}(\omega,{\bf k}{=}0) = \frac{w_0}{B_0}(c_{15} \tilde{\rho}_\perp + \tilde{c}_{17}\rho_\perp) \,,\\[5pt] \label{eq:KuboTT2an4}
  & \coeff{1}{\omega}{\rm Im}\, G_{T^{yz}T^{tx}}(\omega,{\bf k}{=}0) = - \frac{w_0}{|B_0|}(c_{15} \rho_\perp - \tilde{c}_{17}\tilde{\rho}_\perp) \,.
\end{align}
\end{subequations}

In addition, we need to specify what currents we use in the correlation functions. From the on-shell expressions $T^{\mu\nu}_{\rm on-shell}[A,g]\,,\ J^\mu_{cov,\, \rm on-shell}[g,A]$, every gauge field variation introduces a consistent current, that is
\begin{equation}
\label{eq:cons-var}
    G_{T^{\mu\nu} J_{cons}^\alpha} = \frac{\delta}{\delta A_{\alpha}}   T^{\mu\nu}_\textrm{on-shell}[A,g]  \,,
  \quad  G_{J_{cov}^\mu J_{cons}^\alpha} = \frac{\delta}{\delta A_{\alpha}}  J^{\mu}_{cov,\,\textrm{on-shell}}[A,g] \,.
\end{equation}
Let us rewrite~\eqref{eq:Kubo-r1} and the rest of~\eqref{eq:Kubo-pv} with the explicit labels for these currents\footnote{We use here the covariant-consistent correlation functions for our Kubo formulas. In appendix~\ref{app:jconsjcons}, we write the Kubo formulas in terms of the consistent-consistent correlation functions instead.}
\begin{subequations}
\begin{align}
\label{eq:KuboJJ}
  &\coeff{1}{\omega} {\rm Im}\, G_{J^z_{cov} J^z_{cons}}(\omega,{\bf k}{=}0) =  \sigma_\parallel + \cdots\,, \\
  &  \coeff{1}{\omega} {\rm Im}\, G_{J_{cov}^x J^x_{cons}}(\omega,{\bf k}{=}0) = \omega^2 \rho_\perp \frac{w_0(w_0-\MO_{,\mu}B_0^2)}{B_0^4}  \,, \\
  &  \coeff{1}{\omega}{\rm Im}\, G_{J_{cov}^x J_{cons}^y}(\omega,{\bf k}{=}0) = 
     \frac{n_0}{B_0} - \omega^2 \tilde\rho_\perp \frac{w_0(w_0-\MO_{,\mu}B_0^2)}{B_0^4}\, {\rm sign}(B_0)\,, \\
     & \coeff{1}{\omega}{\rm Im}\, G_{J_{cov}^{z} O_1}(\omega,{\bf k}{=}0) = - c_4 \,{\rm sign}(B_0)+\cdots \,,\\[5pt]
  & \coeff{1}{\omega}{\rm Im}\, G_{J_{cov}^{z} O_2}(\omega,{\bf k}{=}0) =  - c_5\, {\rm sign}(B_0)+ \cdots \,\\[5pt]
    & \coeff{1}{\omega}\delta_{ij}{\rm Im}\, G_{T^{ij}J_{cons}^{z}}(\omega,{\bf k}{=}0) =  3 c_3 \,{\rm sign}(B_0) + \cdots \,,\\[5pt]
  & \coeff{1}{\omega}{\rm Im}\, G_{O_2 J_{cons}^{z}}(\omega,{\bf k}{=}0) =  2 c_{14} \, {\rm sign}(B_0) + \cdots \,,
\end{align}
\end{subequations}
where $O_1$ and $O_2$ are defined below~\eqref{eq:Kubo-Tmn}. The terms omitted vanish for $B_0 \ll T_0^2$ or when $M_1=M_3=M_4=0$. Note that the Bardeen-Zumino polynomial $J^\mu_{BZ}$ is proportional to $1/\sqrt{-g}$ so that $\sqrt{-g} J^\mu_{BZ}$ is independent of the metric and therefore $G_{J_{cov}^\mu T^{\nu\rho}} = G_{J_{cons}^\mu T^{\nu\rho}}$. This is important for using time reversal covariance to derive the Onsager constraints by~\eqref{eq:Ons}. The modified Onsager relations are
\begin{equation}
\label{eq:ons-pv2}
c_3 = -c_4 - \coeff13 c_5 \,,\quad c_{14} = - \coeff12 c_5\,, \quad c_{15} = - \frac{w_0-\MO_{,\mu} B^2_0}{w_0} c_8\,,\quad \tilde{c}_{17} =-  \frac{w_0-\MO_{,\mu} B^2_0}{w_0} \tilde{c}_{10}\,.
\end{equation}

\subsubsection{A comment on frequency-dependent transport coefficients} \label{sec:frequencyDependence}
Note that we may also compute frequency-dependent transport  coefficients and find Kubo relations for them. A common example are the AC electric conductivities defined in electrodynamics. 
Generally, one can define any frequency-dependent thermodynamic transport coefficient, $\kappa$ as 
\begin{equation}
 \kappa_{\text{thermo}}(\omega) = \lim\limits_{k\to 0} 
 \frac{G(\omega,k)-G(\omega,k=0)}{-i k} \, ,
\end{equation}
with the Green's function for the appropriate operator. 
Similarly, any frequency-dependent hydrodynamic transport coefficient can be defined as
\begin{equation}
 \kappa_{\text{hydro}}(\omega) = 
 \frac{G(\omega,k=0)}{-i \omega} \, \qquad\text{(no limit on $\omega$ is implied)} \, . 
\end{equation}
In this work, however, we are not going to consider such frequency-dependent transport coefficients, and instead leave this as a future task.

\subsection{Entropy constraints}
To find constraints on the transport coefficients, one method is to impose a local version of the second law of thermodynamics: the existence of a local entropy current with positive semi-definite divergence for every non-equilibrium configuration consistent with the hydrodynamic equations. As was shown in~\cite{Bhattacharyya:2013lha, Bhattacharyya:2014bha}\footnote{
This was demonstrated in the example of (2+1)-dimensional parity-violating hydrodynamics to first order in derivatives before~\cite{Jensen:2011xb}.
}, the constraints on transport coefficients derived from the entropy current are the same as those derived from the equilibrium generating functional, plus the inequality constraints on dissipative transport coefficients. We take the entropy current to be
$$
  S^\mu = S^\mu_{\rm canon} + S^\mu_{\rm eq.}\,,
$$
where the canonical part of the entropy current is
\begin{equation}
\label{eq:S-canon}
S^\mu_{\rm canon} = \frac{1}{T} \left(p u^\mu - T^{\mu\nu}_A u_\nu - \mu J_{cov}^\mu \right)\,,
\end{equation}
and $S^\mu_{\rm eq.}$ is found from the equilibrium partition function, as described in ~\cite{Bhattacharyya:2013lha, Bhattacharyya:2014bha}. The constraints on transport coefficients follow by demanding $\nabla_{\!\mu} S^\mu \geqslant0$.
Using the hydrodynamic equations \eqref{eq:CovViol}, the divergence of the modified canonical entropy current is
$$
\nabla_{\!\mu} S^\mu_{\rm canon} = \nabla_{\!\mu} \left(\frac{p}{T}u^\mu\right) - T_{A}^{\mu\nu}\nabla_{\!\mu} \frac{u_\nu}{T} + J_{cov}^\mu\left(\frac{E_\mu}{T}-\partial_\mu \frac{\mu}{T}\right) -\frac{\mu}{T} C B{\cdot}E\,.
$$
The $S^\mu_{\rm eq.}$ part of the entropy current is explicitly built to cancel out the part of $\nabla_{\!\mu} S^\mu_{\rm canon}$ that arises from the equilibrium terms in the constitutive relations, i.e.~the terms in $T^{\mu\nu}$ and~$J^\mu$ derived from the equilibrium generating functional. These include the anomalous term $\frac{\mu}{T} C B{\cdot}E$. We thus focus on non-equilibrium terms, and write the thermodynamic frame constitutive relations as $T_A^{\mu\nu} = T^{\mu\nu}_{\rm eq.} + T^{\mu\nu}_\textrm{non-eq.}$ and $J^{\mu}_{cov} = J^{\mu}_{\rm eq.} + J^{\mu}_\textrm{non-eq.}$. The divergence of the entropy current is then
\begin{align*}
  \nabla_{\!\mu} S^\mu 
  & = \frac{1}{T}J^\mu_\textrm{non-eq.}\left( E_\mu-T\partial_\mu \frac{\mu}{T}\right) 
    - T^{\mu\nu}_\textrm{non-eq.} \nabla_{\!\mu} \frac{u_\nu}{T}\\[5pt]
  & = \frac{1}{T}\left( \ell^\mu_{\perp \textrm{non-eq.}} + \frac{B^\mu}{B}\ell_\textrm{non-eq.}\right) V_{\mu} -\frac{1}{T} f_\textrm{non-eq.} \nabla{\cdot} u 
    - \frac{1}{2T} t^{\mu\nu}_\textrm{non-eq.} \sigma_{\mu\nu} \,.
\end{align*}
Using the constitutive relations (\ref{eq:TTF}), (\ref{eq:JTF}), this leads to
\begin{align}
\label{eq:DS2}
  T \nabla_{\!\mu} S^\mu 
  & = \coeff12 \eta_\perp (\sigma^{\mu\nu}_\perp)^2  
    + \sigma_\perp V_\perp^2 
    +  \eta_\parallel \Sigma^2 
    + \left(c_8 - c_{15}\right) \Sigma{\cdot}V_\perp \nonumber\\[5pt]
  & + (\zeta_1 - \coeff23 \eta_1) S_3^2
    + 2\eta_2 S_4^2 + \sigma_\parallel S_5^2 + (2\eta_1 + \zeta_2 -\coeff23 \eta_2) S_3 S_4 \nonumber \\[5pt]
  & + \left( c_4 - c_3 +\coeff23 c_{14}\right) S_3 S_5 + \left( c_5 - 2 c_{14}\right) S_4 S_5\,,
\end{align}
where $S_3\equiv \nabla{\cdot}u$, $S_4 \equiv b^\mu b^\nu \nabla_{\!\mu}u_\nu$ and $S_5 \equiv b{\cdot}V$.
Demanding $\nabla_{\!\mu}S^\mu\geqslant0$ now gives $\eta_\perp \geqslant 0$ together with the condition that the quadratic forms made out of $V_\perp$, and $\Sigma$ and $S_3$, $S_4$ and $S_5$ are non-negative, which implies
\begin{equation}
\label{eq:entropy-constraints-b}
\begin{aligned}
  & \sigma_\perp \geqslant 0\,, \quad \eta_\parallel \geqslant 0\, \quad \eta_2 \geqslant 0\,,\quad \zeta_1 -\coeff23 \eta_1 \geqslant 0\,, \quad \sigma_\parallel \geqslant 0\,, \\[5pt]
  & \sigma_\perp \eta_\parallel \geqslant \coeff14 \left(c_8 - c_{15} \right)^2\,, \quad  2\eta_2 (\zeta_1 - \coeff23 \eta_1)
    \geqslant \coeff14
    (2\eta_1 + \zeta_2 -\coeff23 \eta_2)^2\,, \\[5pt]
    & \sigma_\parallel (\zeta_1 -\coeff23 \eta_1) \geqslant \coeff14 (c_4 - c_3 + \coeff23 c_{14})^2\,, \quad 2 \eta_2 \sigma_\parallel \geqslant \coeff14 (c_5 -2 c_{14})^2\,, \\[5pt]
    & {\rm det}(M) \geqslant 0\,,
\end{aligned}
\end{equation}
where 
\begin{equation}
\label{eq:M}
    M = 
\begin{bmatrix}
\zeta_1 - \coeff23 \eta_1 & \eta_1+\coeff12 \zeta_2 -\coeff13 \eta_2 & \coeff12 c_4 -\coeff12 c_3 + \coeff13 c_{14} \\
\eta_1+\coeff12 \zeta_2 -\coeff13 \eta_2  & 2\eta_2 & \coeff12 c_5 - c_{14} \\
\coeff12 c_4 -\coeff12 c_3 + \coeff13 c_{14} & \coeff12 c_5 - c_{14} & \sigma_\parallel
\end{bmatrix}\,.
\end{equation}

The coefficients $\tilde\eta_\perp$, $\tilde\eta_\parallel$, $\tilde\sigma$, $c_{10}$ and $c_{17}$ do not contribute to entropy production, and are not constrained by the above analysis. Thus, $\tilde\eta_\perp$, $\tilde\eta_\parallel$, $\tilde\sigma$, $c_{10}$ and $c_{17}$ are non-equilibrium non-dissipative coefficients.  Note that using the Onsager relations~\eqref{eq:ons} and \eqref{eq:ons-pv2} these constraints reduce to the linear constraints
\begin{equation}
  \eta_\perp \geqslant 0\,, \quad \sigma_\perp \geqslant 0\,, \quad \eta_\parallel \geqslant 0\, \quad \eta_2 \geqslant 0\,,\quad \zeta_1 -\coeff23 \eta_1 \geqslant 0\,, \quad \sigma_\parallel \geqslant 0\,,
\end{equation}
the quadratic constraints  
\begin{equation}
\begin{aligned}
  & \sigma_\perp \eta_\parallel \geqslant (1-\frac{M_{5,\mu} B_0^2}{2})^2 c_8^2\,, \quad \quad   2\eta_2 (\zeta_1 - \coeff23 \eta_1)
    \geqslant 4 \eta_1^2 \,,\\ 
  &\sigma_\parallel (\zeta_1 -\coeff23 \eta_1) \geqslant c_4^2\,, \quad \quad \quad \quad \quad 2 \eta_2 \sigma_\parallel \geqslant c_5^2\,,
\end{aligned}
\end{equation} 
and the qubic constraint
\begin{equation}
    {\rm det}(M) \geqslant 0
\end{equation}
 where now
 $$M = 
\begin{bmatrix}
\zeta_1 - \coeff23 \eta_1 & 2\eta_1 & c_4 \\
2\eta_1 & 2\eta_2 & c_5\\
c_4 &  c_5 & \sigma_\parallel
\end{bmatrix}.
$$

\subsection{Eigenmodes}
\label{sec:eigenmodes-1}
From the hydrodynamic equations~\eqref{eq:CovViol} together with the constitutive relations~\eqref{eq:TTF}, \eqref{eq:JTF}, \eqref{eq:DeltaT}, \eqref{eq:Jcov}, one can study the eigenmodes of small oscillations about the thermal equilibrium state. We begin by including only the anomaly induced transport coefficients from the parity violating sector, i.e. $\xi,\,\xi_{B},\, \xi_{T},\,\xi_{TB}$, and set $c_3 = c_4 = c_5 = c_8 = c_{10} = c_{14} = c_{15} = c_{17} = 0$.\footnote{For $c_i\neq 0$, there are some corrections to the hydrodynamic dispersion relations.}
We also keep the $CPT$ violating constant $c_2=0$, and begin by ignoring the $c_1$ term that is related to the gravitational anomaly coefficient. At the end of the section we comment on the changes due to keeping $c_1\neq 0$.
We set the external sources to zero, and linearize the hydrodynamic equations near the flat-space equilibrium state with constant $T=T_0$, $\mu=\mu_0$, $u^\alpha=(1,{\bf 0})$, and $B^\alpha=(0,0,0,B_0)$. Taking the fluctuating hydrodynamic variables proportional to $\exp(-i\omega t+ i{\bf k}{\cdot}{\bf x})$, the source-free system admits five eigenmodes, two gapped ($\omega({\bf k}{\to}0)\neq0$), and three gapless ($\omega({\bf k}{\to}0)=0$). The frequencies of the gapped eigenmodes are\footnote{All dispersion relations in this section are exact to the order in momentum shown. There is one potential exception which we discuss separately below.}
\begin{align}
\label{eq:omega-cyclo}
  \omega = \pm\frac{B_0^2}{w_0} \sigma_{12} - \frac{i B_0^2}{w_0} \sigma_{11} + v_{gap \pm} k\, \cos\theta  - i D_c(\theta) k^2\,,
\end{align}
where $w_0 \equiv \epsilon_0 + p_0$ is the equilibrium enthalpy density, and we have taken $\aBB B_0^2 \ll w_0$, ${\MO}_{,\mu} B_0^2 \ll w_0$ in the hydrodynamic regime $B_0\ll T_0^2$. The $2\times2$ conductivity matrix in the plane transverse to ${\bf B}_0$ was defined in section~\ref{sec:kubo}. 
We repeat it here for the reader's convenience: $\sigma_{ab} \equiv \sigma_\perp \delta_{ab} + \left(\frac{n_0}{|{\bf B}_0|} + \tilde\sigma \right)\epsilon_{ab}$. The corresponding resistivity matrix is $\rho_{ab} \equiv (\sigma^{-1})_{ab} = \rho_\perp \delta_{ab} + \tilde\rho_\perp\, \epsilon_{ab}$, which defines $\rho_\perp$ and~$\tilde\rho_\perp$. Stability of these eigenmodes requires $\sigma_{11} = \sigma_\perp >0$, which is a direct consequence of the entropy production argument~\eqref{eq:entropy-constraints-b}. The analogous mode in 2+1 dimensional hydrodynamics was christened the hydrodynamic cyclotron mode in~\cite{Hartnoll:2007ih}, which also explored its implications for transport near two-dimensional quantum critical points. The gapped mode velocity
\begin{equation}\label{eq:vGap}
    v_{gap \pm} =  \frac{B^2_0 C \mu_0^3}{3 w_0^2}(\sigma_{12} \pm i \sigma_{11})
\end{equation}
is unique to systems in the presence of anomalies.

The coefficient $D_c(\theta)$ in the cyclotron mode eigenfrequency (\ref{eq:omega-cyclo}) at small $B_0$ is\footnote{Note that the Hall viscosities and the Hall conductivity show up in this coefficient and not in any of the others. Only the Hall conductivity contributes to the gap in eq.~\eqref{eq:vGap}, while both Hall viscosities and the Hall conductivity contribute to the diffusion coefficient~\eqref{eq:cyclotron}.}
\begin{equation}
\begin{aligned}\label{eq:cyclotron}
  D_c(\theta) & =   \left(  \frac{3\zeta_1{+}6\eta_\perp{-}2 \eta_1+ 6i \tilde{\eta}_\perp}{6w_0}   + \frac{(n_0^2 \chi_{11} {-} w_0^2 \chi_{33})w_0}{2 n_0^2 \det(\chi)}(\sigma_{11} \mp i \tilde{\sigma}) \pm\frac{i v_s^2 w_0}{2 n_0 B_0} \right) \sin^2\theta \\
  &+ \left(\frac{\eta_\perp}{w_0} \pm i \left(\frac{\tilde{\eta}_\parallel}{w_0} + \frac{C \mu_0^3}{3} \frac{v_{gap\,\pm}}{w_0} \right)\right)  \cos^2\theta + O(B_0^2)\,,
\end{aligned}
\end{equation}
where $\theta$ is the angle between $\mathbf{k}$ and $\mathbf{B}_0$. The nonzero elements of the $3\times 3$ susceptibility matrix are $\chi_{11} = T (\partial \epsilon/\partial T)_{\mu/T}$, $\chi_{13} = \chi_{31} = (\partial \epsilon /\partial\mu)_{T}$, $\chi_{33} = (\partial n/\partial\mu)_{T}$, and $\chi_{22}=w_0$, with derivatives evaluated at constant $B^2$ in equilibrium. The speed of sound $v_s$ in eq.~\eqref{eq:cyclotron} is given by
\begin{equation}\label{eq:speedOfSound}
  v_s^2 = \frac{n_0^2 \chi_{11} + w_0^2 \chi_{33} -2n_0 w_0 \chi_{13}}{ \det(\chi)}\,.
\end{equation}
For the gapped modes, the limits $\theta \to 0$ and $\mathbf{k}\to 0$ as well as $\theta \to \pi/2$ and $\mathbf{k}\to 0$ commute.

For momenta ${\bf k}\parallel {\bf B}_0$, the three gapless eigenmodes are the two ``sound'' waves, and the chiral magnetic wave~\cite{Kharzeev:2010gd}. The eigenfrequencies in the small momentum limit are
\begin{subequations}
\label{eq:omega-par}
\begin{align}
  & \omega = k v_\pm - i\frac{\Gamma_{\parallel\pm}}{2} k^2\,,  \\
  & \omega = k v_0 -iD_\parallel k^2\,.
\end{align}
\end{subequations}
 
The velocities $v_0\,,\ v_+$ and $v_-$ can be expressed in terms of the speed of sound as well as the following expressions
\begin{equation}
\alpha = \frac{(s_0 T_0)^2}{{\rm det}(\chi)}\,, \quad \gamma  = \frac{\mu_0^2}{w_0}-2n_0 \mu_0 \frac{\chi_{11}-\mu_0\chi_{13}}{{\rm det}(\chi)}+w_0 \frac{\chi_{11}-\mu_0 \chi_{33}}{{\rm det}(\chi)}\,,
\end{equation}
from which we find
\begin{equation}
\begin{aligned}
    v_0 &= B_0 C \frac{\alpha}{v_s^2}+\cdots\,,\\
    v_\pm & = \pm v_s + B_0 C \frac{\gamma v_s^2 - \alpha}{2 v_s^2} +\cdots\,,
\end{aligned}
\end{equation}
where we have omitted terms of order $B_0^2 C^2$ and higher.

The damping coefficient is 
\begin{equation}\label{eq:Gamma-parallel}
\begin{aligned}
  \Gamma_{s,\parallel} &= \frac{3 \zeta_1 + 10 \eta_1 + 6 \eta_2}{3W_0} +  \frac{v_s^2 \chi_{11}-w_0}{\det(\chi)}  \frac{ w_0}{v_s^2} \sigma_\parallel \\
  &+ C B_0 \left( \Sigma_\parallel \sigma_\parallel + \Sigma_\perp \sigma_\perp \right) + \cdots
  \,,
\end{aligned}
\end{equation}
where
\begin{equation}
    \frac{w_0}{W_0} = 1 + {\cal O}\left(C B_0\right) \,.
\end{equation}
The ${\cal O}(CB_0)$ part of $\frac{1}{W_0}$ as well as the ${\cal O}(1)$ part of $\Sigma_\perp$ can be found in appendix~\ref{app:eigenmodes}. We have omitted higher order terms in $C B_0$ in eq.~\eqref{eq:Gamma-parallel}.

The longitudinal diffusion constant is 
\begin{equation}\label{eq:D-parallel}
  D_\parallel = \frac{w_0^2 \sigma_\parallel}{v_s^2\det(\chi)}+ {\cal O}\left(B_0^2 C^2\right) \,,
\end{equation}
where once again, the ${\cal O}\left(C^2B_0^2\right)$ terms can be found in appendix~\ref{app:eigenmodes}. The positivity of the diffusion constant implies $\sigma_\parallel >0$. 

For modes propagating at an angle $\theta \neq \pi/2$ with respect to ${\bf B}_0$, the velocities (and damping coefficients) of the ``sound'' waves and the chiral magnetic wave depend on the angle of propagation. For a fixed value of $\theta$, the small-momentum eigenfrequencies are $\omega =  k v_\pm \cos\theta -\frac{i}{2}\Gamma_s(\theta) k^2$, and $\omega= k v_0 \cos\theta -iD(\theta)k^2$, where 
\begin{eqnarray}\label{eq:theta-general}
  D(\theta) &=& D_\parallel \cos^2\theta +\left( \frac{n_0^2\, w_0^2\, \rho_\perp}{B_0^2 v_s^2 \det(\chi)^2} +{\cal O} (B_0C)\right)\sin^2\theta\,,\\
  \Gamma_{s}(\theta) &=& \Gamma_{s,\parallel} \cos^2\theta + \left( \frac{\eta_\parallel}{w_0} + \frac{(n_0 \chi_{13} - w_0 \chi_{33})^2w_0^3}{B_0^2\, \det(\chi)^2} \rho_\perp + {\cal O} (B_0C)\right) \sin^2\theta\,.
\end{eqnarray}

The limits $\theta\to \pi/2$ and $k\to0$ in the gapless eigenfrequencies do not commute. For momenta ${\bf k}\perp {\bf B}_0$, the three gapless eigenmodes include two diffusive modes, and one ``subdiffusive'' mode with a quartic dispersion relation,\footnote{One might worry that the quartic relation in eq.~\eqref{eq:quartic} could be affected by ${\cal  O}\left(\partial^2\right)$ terms in the constitutive relations. However, we verified that the only term that could modify eq.~\eqref{eq:quartic} is a term in $\nabla_\mu J^\mu \sim k_y^3$. But $\nabla_\mu J^\mu$ is a scalar equation and there are no scalar terms which contain $k_y^3$.}
\begin{subequations}
\label{eq:omega-perp}
\begin{align}
  & \omega = -i D_{\perp\,\pm} k^2\,,\\[5pt]
  & \omega = -i \frac{\eta_\perp k^4}{B_0^2\, \chi_{33}}\,.\label{eq:quartic}
\end{align}
\end{subequations}
 
The transverse diffusion constant is given by
\begin{equation}
\begin{aligned}\label{eq:Dperp-pm}
  D_{\perp\pm} = D_\perp \pm \sqrt{D_\perp^2 - \frac{w_0^2 \chi_{33} \eta_{\parallel} \rho_\perp}{B_0^2 \left(\det(\chi)-B_0^2 C^2 T_0^2 \mu_0^2 \frac{ds}{dT} \right)}}\,,
\end{aligned}
\end{equation}
where
\begin{equation}
\begin{aligned}\label{eq:Dperp}
D_\perp =& \frac{w_0^4 \chi_{33} + B_0^2 C^2 \mu_0^2 \left(w_0^2 \mu_0 T_0 \frac{dn}{dT} + \coeff14 \mu_0^2 \det(\chi)- w_0^3 \right) }{w_0  B_0^2\left( \det(\chi) - B_0^2 C^2 T_0^2 \mu_0^2 \frac{ds}{dT} \right)}\frac{\rho_\perp}{2}\\
&+\frac{\det(\chi)}{w_0\left(\det(\chi) - B_0^2 C^2 T_0^2 \mu_0^2 \frac{ds}{dT}\right)} \frac{\eta_{\parallel}}{2} \,.
\end{aligned}
\end{equation}
again using ${\MO}_{,\mu} B_0^2 \ll w_0$.
Stability of the equilibrium state requires $\eta_\perp >0$, $\eta_\parallel >0$, $\rho_\perp >0$, which is ensured by the entropy production argument~\eqref{eq:entropy-constraints-b}.

The $c_1$ constant related to the gravitational anomaly modifies the dispersion relations in a similar way than the $C$ gauge anomaly coefficient. For example, the combination in the denominators in the transverse diffusion constants $D_{\perp \pm}$ in eq.~\eqref{eq:omega-perp} with definitions given in eqs.~\eqref{eq:Dperp-pm} and \eqref{eq:Dperp} get modified
\begin{equation}
T_0^2 \mu_0^2 C^2 \frac{ds}{dT}\to T_0^2 \mu_0^2 C^2 \frac{ds}{dT} + 4 c_1 C \frac{dn}{dT} T_0^3 \mu_0 - 4 c_1^2 T_0^4 \chi_{33}\,,   
\end{equation}
and the terms in the gap velocity $v_{gap}$ and the cyclotron frequency $D_c(\theta)$ in eq.~\eqref{eq:omega-cyclo} which include the gauge anomaly coefficient appear in the following combination
\begin{equation}
    \frac13 C \mu_0^3 \to \frac13 C \mu_0^3 + 2 c_1 T_0^2 \mu_0 = \xi_T\,,
\end{equation}
with $\xi_T$ in eq.~\eqref{eq:xis}.

Lastly, we mention that taking $C=c_1 = 0$, and therefore $\xi = \xi_B = \xi_T = \xi_{TB}=0$, the results of this section agree with those found in section 3.5 of~\cite{Hernandez:2017mch}. In turn, those results reduce to the standard results for $B_0\to 0$. 
Including the other coefficients $c_n$ complicates this eigenmode analysis considerably, and is a excellent direction for future investigation. 

The equations~\eqref{eq:cyclotron}, \eqref{eq:quartic}, \eqref{eq:Dperp-pm}, \eqref{eq:Gamma-parallel}, \eqref{eq:D-parallel}, \eqref{eq:theta-general} may be regarded as Einstein relations. They are relating several transport coefficients to each other, in analogy to the simple examples of the shear diffusion $D_\eta=\eta/(\epsilon+P)$, the charge diffusion $D_\sigma = \sigma/\chi$ (with the charge susceptibility $\chi$), and sound attenuation $\Gamma=(\zeta + 4\eta/3)/(\epsilon+P)$ (with bulk viscosity $\zeta$) in the uncharged isotropic system.

\subsection{Interpretation of transport coefficients}
\label{sec:transportInterpretation}
With the systematic approach applied in this section, 22 
independent transport coefficients have been identified. Some of them have a standard interpretation with a new twist, some are novel and will be given a first interpretation here.\footnote{The pressure $p$ in our formulation acts as a generating functional for equilibrium $n$-point functions~\cite{Jensen:2011xb,Jensen:2012jh,Banerjee:2012iz,Kovtun:2016lfw}. It is not counted as a transport coefficient. Susceptibilities such as the $U(1)_A$ charge susceptibility $\chi_{33}$ are derivatives of the pressure and are not counted as individual transport coefficients.} 

\subsubsection{Discussion of all transport coefficients}
\label{sec:allTransportCoeffs}
If not specified otherwise, in the examples here we assume a flat metric $\eta_{\mu\nu} = \text{diag}(-1,1,1,1)$, and the equilibrium fluid velocity $u^\mu=(1,0,0,0)$. 
\paragraph{\bf The perpendicular magnetic vorticity susceptibility $\boldsymbol{M_2}$. 
} 
In order to interpret $M_2$, we may consider how it arises in various expressions originating from the generating functional. 

First, one may interpret $M_2$ with the help of the {vorticity of the magnetic field}, $\Omega_B^\mu$. This quantity appears in the most prominent terms in the equilibrium constitutive equations~\eqref{eq:TTF-eq}, which contain $M_2$, namely
\begin{equation}
    \mathcal{E}_{\text{eq}}\sim \mathcal{P}_{\text{eq}}\sim M_2  B {\cdot} \Omega_B \, .
\end{equation}
Here, in analogy to the vorticity of the fluid velocity, $\Omega^\mu=\epsilon^{\mu\nu\rho\sigma}u_\nu \nabla_\rho u_\sigma$, we define the vorticity of the magnetic field\footnote{Both, $u$ and $B$ are vector fields, however, $u$ is a dynamical field, while $B$ is a source, i.e.~an external field.} as 
$\Omega_B^\mu=\epsilon^{\mu\nu\rho\sigma} u_\nu \nabla_\rho B_\sigma$. 
In this sense, $M_2$ measures the response of energy or pressure to the vorticity of $B$ parallel to $B$. 
In other words, $M_2$ measures the response to that component of the curl of the magnetic field, which is perpendicular to both the magnetic field itself and to the fluid velocity. \\
\uline{Example:} Thermodynamic (time-independent): Consider a time-independent uncharged equilibrium state with an inhomogeneous anisotropic background magnetic field $\vec{B}=(B_x(y),B_y(x),B_z)$. Here, $B_z$ is a constant in space and time, however, $B_x$ depends on $y$, and $B_y$ depends on $x$. This leads to 
$T^{00}\sim M_2 \epsilon^{\mu\nu\rho\sigma}u_\mu B_\nu \nabla_\rho B_\sigma \sim 
M_2 \epsilon^{t z y x}u_t B_z \nabla_y B_x(y) \sim - M_2 B_z \partial_i \times B_j \sim - M_2 B_z \partial_y B_x(y)$. 
So in such an equilibrium state, the energy density (and pressure) receive a contribution from that part of the curl of the magnetic field, which is perpendicular to itself. $M_2$ measures how large that contribution is.
    
Alternatively, $M_2$ also appears in the magnetization measuring response to the curl of the magnetic field perpendicular to the fluid flow (in the fluid rest frame), or as response to the temperature gradient perpendicular to the magnetic field and the fluid flow (in the fluid rest frame):
\begin{equation}\label{eq:magneticNernst}
\mathfrak{m}^\mu \sim  M_2 (2 
\Omega^\mu_B
-\epsilon^{\mu\nu\kappa\sigma}u_{\nu}B_{\kappa}\partial_{\sigma}T ) \, .
\end{equation}
The second term may be interpreted as a magentic version of the Nernst effect as both are a response to the same tensor structure $\epsilon^{\mu\nu\kappa\sigma}u_{\nu}B_{\kappa}\partial_{\sigma}T$. However, here the response occurs in the magnetization, whereas the original Nernst effect has a response in the electric field. One may think of $-M_2$ as a magnetization Nernst coefficient.

Second, one may interpret $M_2$ in terms of the Poynting vector. Consider a 
setup with $B = B_0 \hat{z}$, $E= E_0 \hat{y}$, $T=T_0$ and $\mu = E_0 y$ on $|y|\leq R$ where $E_0 = {\cal O}(\partial)$. Then there is a nonzero shear term $T^{zx} = -M_{2,\mu} B^z \epsilon^{xtyz} u_t E_y B_z \sim M_{2,\mu} B_0^2 E_0$ due to $M_2$. That is, $M_{2,\mu}$ measures the response of the shear tensor to an external Poynting vector $S^\mu = \epsilon^{\mu\nu\rho\sigma} u_\nu E_\rho B_\sigma$ in the plane spanned by the magnetic field $B^\mu$ and the Poynting vector $S^\mu$. In the same setup, $M_{2,\mu}$ also gives the response of the magnetization to the external Poynting vector $\mathfrak{m}^\mu \sim -M_{2,\mu} S^\mu$.

In a theory which microscopically preserves parity invariance, $M_2$ can be non-vanishing in states in which parity is broken through an axial ($U(1)_A$) chemical potential.\footnote{Although a vector ($U(1)_V$) magnetic field breaks parity, the only scalar that can be formed from it to appear in the equilibrium generating function is the parity even $B^2$.} 
It can also be nonzero if parity is broken microscopically through a chiral anomaly.

\paragraph{\bf The magneto-thermal susceptibility $\boldsymbol{M_1}$.} It appears in the constitutive relation of the equilibrium energy momentum tensor, \eqref{eq:TTF-eq}, as a response of the energy density to the gradient of the dimensionless ratio $B^2/T^4$. This gradient is parallel to the magnetic field
\begin{equation}
    \mathcal{E}_{\text{eq}}\sim M_1 B^\mu \partial_\mu \left ( \frac{B^2}{T^4} \right ) \, .
\end{equation}
\uline{Example:} Consider a spatially modulated magnetic field in the $z$-direction, $B_z$, which is constant in time but depends on the $z$-coordinate, e.g. $B_z(z)=B_0\, \text{sin}(k z)$ with the wave vector of the modulation, $k$. Assume the temperature is constant. This leads to a spatially modulated energy density $T^{00}\sim M_1 B_z(z) (\partial_z B_z(z)^2)/T^4 = M_1\frac{2 k B_0^3}{T^4} \text{sin}^2(k z)\, \text{cos}(k z)$. Alternatively, the temperature can be spatially modulated. A pure time-modulation would not lead to any response, because the gradient $\partial_\mu$ needs to be aligned with the (spatial) magnetic field $B$. 

Alternatively, $M_1$ measures the response of the magnetization to the part of the gradient of $B^2/T^4$ which is perpendicular to the fluid velocity
\begin{equation}
\mathfrak{m}^\mu \sim M_1 \Delta^{\mu\nu} \partial_\nu \frac{B^2}{T^4} \, .
\end{equation}

\paragraph{\bf The magneto-acceleration susceptibility $\boldsymbol{M_3}$.} It multiplies $B{\cdot} a$ in the generating functional, eq.~\eqref{eq:Ws}. In the energy momentum constitutive relation, \eqref{eq:TTF-eq}, one finds the term
\begin{equation}
    \mathcal{E}_{\text{eq}}\sim \mathcal{P}_{\text{eq}}\sim M_{3,B^2} \, B{\cdot} a  \, .
\end{equation}
Hence, the thermodynamic derivative of $M_3$ with respect to $B^2$ measures the response of equilibrium energy and pressure to a magnetic field aligned with the acceleration of the fluid in any of the spatial directions. 
In the magnetization, $M_3$ appears directly measuring the response to the acceleration of the fluid 
\begin{equation}
   \mathfrak{m}^\mu \sim M_3 \, a^\mu  \, .
\end{equation}
\uline{Example:} A fluid which is accelerated in the $x$-direction 
gets magnetized along that direction. Its magnetization is proportional to the acceleration.

The magneto-acceleration susceptibility $M_3$ vanishes in conformal field theories regardless of the state breaking conformal invariance.

\paragraph{\bf The magneto-electric susceptibility $\boldsymbol{M_4}$.} This susceptibility has been discussed previously with both electric and magnetic field being strong  ($B\sim\mathcal{O}(1)$)~\cite{Kovtun:2016lfw}.\footnote{In~\cite{Kovtun:2016lfw} the magneto-electric susceptibility $M_4$ was named $\chi_{EB}$.} 
There and in our case, this susceptibility multiplies $B{\cdot} E$ in the generating functional, eq.~\eqref{eq:Ws}. For our case
\begin{equation}
    \mathcal{E}_{\text{eq}}\sim  M_{4,T} \, B{\cdot} E \, , \qquad {\cal P}_{\rm eq} \sim M_{4,B^2}\,  B{\cdot} E   \, .
\end{equation}
The magneto-electric susceptibility $M_4$ measures the response of the energy density or pressure to an electric field projected onto the direction of the magnetic field. Hence this response is proportional to $|B|\, |E| \text{cos}\,\theta$ with the angle, $\theta$, between the two fields.  
In the magnetization, $M_4$ measures the response to an electric field 
\begin{equation}
\mathfrak{m}^\mu \sim M_4 \, E^\mu \, ,
\end{equation}
and the only leading contribution to the polarization is given by
\begin{equation}
    \mathfrak{p}^\mu = M_4\, B^\mu\,.
\end{equation}
These two last equations highlight that the term $M_4 B{\cdot} E$ generates a response (polarization and magnetization) symmetric under exchange of $B$ and $E$.

\paragraph{\bf Novel ``expansion-induced longitudinal conductivities'' $\boldsymbol{c_{4}}$  and $\boldsymbol{c_{5}}$.} 
These transport coefficients appear in the constitutive relation for the current 
\begin{equation}\label{eq:expansionJ}
    \mathcal{J}^\mu \sim b^\mu (c_4 \nabla{\cdot} u + c_5 b^\alpha b^\beta \nabla_\alpha u_\beta)\, .
\end{equation}
There is similarity between these two terms and the longitudinal viscosity terms contributing to the energy-momentum tensor $T^{\mu\nu}\sim b^{\langle \mu \nu\rangle}(\eta_1 \nabla\cdot u+\eta_2 b^\alpha b^\beta \nabla_\alpha u_\beta)$ in eq.~\eqref{eq:TTF}. The latter is a symmetric traceless two-tensor contribution aligned with the magnetic field, and eq.~\eqref{eq:expansionJ} is a current aligned with the magnetic field. 
We refer to $c_4$ and $c_5$ as conductivities instead of viscosities since they appear in the charge current. 
In this sense they  are longitudinal conductivities in analogy to $\eta_1,\, \eta_2$ being longitudinal viscosities.  
Both, $c_4$ and $\eta_1$ measure the response of the respective currents to divergence of the velocity field, $\nabla{\cdot} u$. 
In a similar way, the coefficient $c_5$ appears in analogy to $\eta_2$ 
as a response to the gradient of the velocity field along the magnetic field projected onto the magnetic field direction, $b^\mu b^\nu \nabla_\mu u_\nu$. The latter can be thought of as an expansion of the fluid along the magnetic field.

\paragraph{Novel shear-induced conductivity \bf $\boldsymbol{c_{8}}$ and shear-induced Hall conductivity $\boldsymbol{c_{10}}$ (the latter is dissipationless).} 
Both transport coefficients measure the response to the shear within a particular plane, in that sense both are shear-induced. But contrary to the standard shear viscosities, $c_8$ and $c_{10}$ measure the response within that plane in which the shear occurs. Hence, we refer to $c_8$ and $c_{10}$ as being transverse. In order to stress that both measure a response of the current we refer to them as conductivities rather than viscosities. 
One example for the interpretation of $c_8$ and $c_{10}$ can be based on the constitutive relations containing:
\begin{eqnarray}
    \mathcal{J}^\mu &=& \mathcal{J}_\text{eq.}^\mu + \dots + c_8 \Sigma^\mu + c_{10} \tilde\Sigma^\mu\, ,\\
    \Sigma^\mu &=& (\Delta^{\mu\nu}-b^\mu b^\nu) \sigma_{\nu\rho} b^\rho \, , \label{eq:SigmaMuApx}\\
    \tilde\Sigma^\mu &=& \epsilon^{\mu\nu\alpha\beta} u_\nu b_\alpha \Sigma_\beta \, .
\end{eqnarray}
\uline{Example:} 
    Start by choice with the background magnetic field $B^\mu=(0,0,0,B_z)$, which implies $b^\mu=(0,0,0,1)$, which implies $(\Delta^{\mu\nu}-b^\mu b^\nu) = \text{diag}(0,1,1,0)$, which is the projector onto the two directions perpendicular to the background magnetic field, $B_z$, and the fluid velocity simultaneously. Working out eq.~\eqref{eq:SigmaMuApx}, we find that $c_8$ measures the response of the spatial current components, e.g.~$\mathcal{J}_x$ ($\mathcal{J}_y$), to a shear of the fluid in the plane of that current and the magnetic field, e.g.~$(x,z)$-plane ($(y,z)$-plane) for the response in the $x$-direction ($y$-direction):
    \begin{equation}
        \mathcal{J}_x\sim c_8 (\partial_x u_z+\partial_z u_x)\, , \quad 
        \mathcal{J}_y\sim c_8 (\partial_y u_z+\partial_z u_y)\, ,
    \end{equation}
    at linear order in derivatives. Equivalently, $c_{10}$ measures the (Hall-like) response of the current $\mathcal{J}$ to a shear in the plane of the magnetic field and perpendicular to the current response, e.g.~$(y,z)$-plane for the response in $x$-direction (and equivalently for the $y$-direction):
    \begin{equation}
        \mathcal{J}_x\sim c_{10} (\partial_y u_z+\partial_z u_y)\, , \quad 
        \mathcal{J}_y\sim c_{10} (\partial_x u_z+\partial_z u_x)\, .
    \end{equation}

\paragraph{The transverse Hall viscosity
$\boldsymbol{\tilde\eta_\perp}$, and the novel longitudinal Hall viscosity $\boldsymbol{\tilde\eta_{||}}$.} 
 Hall viscosities were first discovered in (2+1)-dimensional systems~\cite{Avron:1995fg,Avron:1997}. The (3+1)-dimensional counterparts have been discussed in~\cite{Hernandez:2017mch}. In (2+1)dimensions, the relevant term in the energy momentum tensor constitutive relation takes a form which is our $\tilde\sigma_\perp^{\mu\nu}$ projected onto the plane perpendicular to the magnetic field $B^\mu$. When simplified, this reads $T^{ij}\sim \eta_H (\epsilon^{ikl} u_k\sigma_l^j+\epsilon^{jkl} u_k\sigma_l^i)$ with the (2+1)-dimensional traceless symmetric stress $\sigma_l^j$ defined in analogy to our $\sigma_\mu^\nu$. 

In our system, the transport in the plane perpendicular to the magnetic field is associated with $\tilde \eta_{\perp}$. The relevant contribution to the energy-momentum tensor constitutive equation, eq.~\eqref{eq:TTF}, is given by
\begin{equation}
    T^{\mu\nu} \sim \tilde\eta_\perp \tilde\sigma_\perp^{\mu\nu}\, .
\end{equation}
One may imagine that in that plane the tensor structure giving rise to Hall viscosity is simply that of a (2+1)-dimensional system, given in the previous paragraph. It may be interpreted in the same way as in~\cite{Avron:1995fg,Avron:1997}, namely as the response of the energy-momentum tensor's diagonal components to a shear in the plane perpendicular to the magnetic field.\\
{\underline{Example:}} 
Considering the $(x,y)$-plane, one finds for example~$T^{xx}\sim\tilde\eta_\perp \sigma^{xy}$, if the magnetic field is chosen along the $z$-direction. 

On the contrary, the Hall viscosity in the plane along the magnetic field, $\tilde \eta_{||}$, is novel. It measures the response of the energy-momentum tensor off-diagonal components to a shear in the plane aligned with the magnetic field and one of the other spatial directions as seen in the constitutive relation, eq.~\eqref{eq:TTF},
\begin{equation}
   T^{\mu\nu} \sim \tilde\eta_{||} (b^\mu \tilde\Sigma^\nu +b^\nu \tilde \Sigma^\mu) \, .
\end{equation}
{\underline{Example}:} If the magnetic field is aligned with the $z$-direction, then we have $T^{xz}\sim \tilde\eta_{||} \sigma^{yz}$. 

\paragraph{The Hall conductivity $\boldsymbol{\tilde\sigma_\perp}$.} 
Hall transport is dissipationless and only occurs in the plane perpendicular to the magnetic field. 
For example, consider a magnetic field along the $z$-direction and an electric potential $V^x = E^x - T \partial^x \frac{\mu}{T}$ along the $x$-direction. This configuration induces a current in the $y$-direction proportional to $\tilde\sigma_\perp$. An equivalent Hall response in the longitudinal $(x,z)$- or $(y,z)$-plane does not exist. Note that this can also be related to the parity anomaly in the absence of any magnetic field from a (2+1)-dimensional point of view as discussed below, see section~\ref{sec:relationTo2DHydro}. 

\bigskip

The remaining 12 
transport coefficients have been identified previously, either at vanishing magnetic field, at vanishing charge density, and/or without a chiral anomaly. Our study generalizes these previous results.

\paragraph{\bf The magneto-vortical susceptibility $\boldsymbol{M_5}$.} 
This coefficient was first found, named, and interpreted in~\cite{Hernandez:2017mch}.\footnote{This coefficient in~\cite{Hernandez:2017mch} is designated by $M_\Omega$.}  
It appears as the response of energy/pressure to the magnetic field along the vorticity:
\begin{equation}
    T_{tt}\sim T_{zz}\sim M_5 B{\cdot}\Omega \, .
\end{equation}
Heuristically, one may imagine vortices of charged fluid (due to nonzero vorticity) distributed over the system. Each charged vortex acts like an elementary magnet. Depending on its charge and orientation with respect to the magnetic field, it either increases or decreases the energy of the fluid. The fluid can respond like a diamagnet or paramagnet, depending on the sign of $M_5$, which is a microscopic property of the system and has to be measured. 
$M_5$ also measures the response of the magnetization to vorticity in the fluid
\begin{equation}
\mathfrak{m}^\mu \sim M_5 \Omega^\mu + M_2 (2 \epsilon^{\mu\nu\rho\sigma} u_\nu \nabla_\rho B_\sigma -\epsilon^{\mu\nu\kappa\sigma}u_{\nu}B_{\kappa}\partial_{\sigma}T ) \,,
\end{equation}
and was interpreted to measure the angular momentum generated by an external magnetic field due to non-vanishing surface currents (see discussion of~\cite{Hernandez:2017mch} for more details). Note that $M_5$ requires a nonzero $\mu$ in order not to vanish, indicating that the fluid must be charged for these effects to take place. These terms capture the intuitive effects of a charged fluid with nonzero vorticity producing a magnetization, and a charged fluid subject to a magnetic field acquiring some angular momentum in response. 

Additionally, the magneto-vortical susceptibility induces a Nernst effect in the energy current
\begin{equation}\label{eq:QNernst}
{\cal Q}^\mu \sim \left(2 M_5 - T M_{5,T} - \mu M_{5,\mu}\right)\epsilon^{\mu\nu\rho\sigma} u_\nu B_\rho \partial_\sigma T/T
\end{equation}
from which one can identify $2 M_5/T - M_{5,T} - \mu M_{5,\mu}/T$ as a momentum Nernst coefficient.

\paragraph{\bf Chiral vortical, chiral magnetic, chiral thermal conductivities, $\xi=\xi_{TB},\, \xi_B,\, \xi_T$.}  As expected, these dissipationless chiral conductivities are given analytically as functions of thermodynamic quantities and the chiral anomaly coefficient of the microscopic theory, see eq.~\eqref{eq:xis}. Therefore, we confirm validity of these expressions in states with a strong magnetic field.

\paragraph{\bf The Nernst effect.} The thermodynamic constitutive relations encode the Nernst effect in the spatial current ${\cal J}^\mu$\footnote{In deriving eq.~\eqref{eq:Nernst}, we separated the electric field contribution from the temperature gradient contribution as a source to the equilibrium current. That is, we chose $T\Delta^{\mu\nu}\partial_\nu\frac{\mu}{T}\sim E^\mu$ and $\Delta^{\mu\nu}\partial_\nu T$ as our independent first order equilibrium scalars.}
\begin{equation}\label{eq:Nernst}
{\cal J^\mu} \sim -\left(\chi_{B,T} +\mu \chi_{B,\mu}/T \right) \epsilon^{\mu\nu\rho\sigma} u_\nu B_\rho \partial_\sigma T\,.
\end{equation}
The Nernst coefficient can therefore be identified with $- \chi_{B,T}- \mu \chi_{B,\mu}/T $. The magnetic susceptibility $\chi_B =2\, p_{,B^2}$ was defined in section~\ref{subsec:Kubo}. Furthermore, it has been shown that the conformal anomaly gives rise to a Nernst effect~\cite{Chernodub:2017jcp} with a Nernst coefficient proportional to the conformal anomaly coefficient. This can be seen in our setup by considering that, in the absence of a magnetic field, the conformal anomaly vanishes and thus $T^\mu_\mu = 4 p - T p{,T} - \mu p_{,\mu} - 2\chi_B B^2 = 0$ at $B=0$. Taylor expanding $T^\mu_\mu$ at small $B^2$ then gives the leading conformal anomaly coefficient $c_A$
\begin{equation}
    T^\mu_\mu  = c_A F_{\mu\nu} F^{\mu\nu} = 2 c_A B^2\,,
\end{equation}
where $c_A = -T \chi_{B,T} /2-\mu \chi_{B,\mu}/2$. This leads to the relation $N_{Nernst}=  2c_A /T$. This relation agrees with the result of~\cite{Chernodub:2017jcp} except for the numerical prefactor. The latter should depend on which charges the fermions in the one-loop diagram carry, that determines the conformal anomaly coefficient.

\paragraph{\bf Very well known transport coefficients.} It should be noted first that while the transport coefficients to be discussed here are well known, they have not yet been discussed for a strong magnetic field associated with an axial $U(1)_A$. This aspect is novel in our work. 

The remaining transport coefficients are all hydrodynamic. 
Due to the anisotropy caused by the magnetic field, there are two shear viscosities, $\eta_\perp$ for transport perpendicular,
and $\eta_{||}$ for transport longitudinal to the magnetic field.\footnote{We will see in the holography section that $\eta_{||}$  need not take on the value $s/(4\pi)$ because it does not satisfy the equation of motion of a minimally coupled scalar in asymptotically $AdS$ spacetime.} 
See e.g.~\cite{Rebhan:2009vc,Huang:2011dc,Critelli:2014kra,Hernandez:2017mch,Grozdanov:2016tdf,Grozdanov:2017kyl, Garbiso:2020puw} for shear viscosities in anisotropic systems. 
The bulk viscosities $\zeta_1$, $\zeta_2$, and $\eta_1$ (and the linearly dependent $\eta_2$) have been discussed in~\cite{Huang:2011dc,Critelli:2014kra,Hernandez:2017mch,Grozdanov:2016tdf,Grozdanov:2017kyl}. The same is true for the perpendicular and longitudinal conductivities $\sigma_\perp$, $\sigma_{||}$, and the associated resistivities. 

\paragraph{Remarks on the origin of the transport effects:} %
\begin{itemize}
    \item The coefficient $M_5$ can be nonzero if the chemical potential is nonzero, even if there is no anomaly. 
    \item We note that the susceptibilities $M_1,\, M_2,\, M_3,\, M_4$ were already considered in~\cite{Hernandez:2017mch} for a vector magnetic field associated with a $U(1)_V$. In that case, they have to vanish in a microscopic theory preserving parity.
    However, in our case with an axial magnetic field present, these coefficients can be nonzero even if the microscopic theory is parity preserving.
    In other words, for the thermodynamic transport coefficients $M_1,\, M_2,\, M_3,\, M_4$ one source of parity-violation suffices in order for them not to vanish. This parity violation may stem from a chiral anomaly in the microscopic theory or alternately from an 
    external axial $U(1)_A$ chemical potential in a parity-preserving microscopic theory. Hence, these coefficients are not exclusively caused by the anomaly.  
    \item While $M_{1,3,4}$ show up in constitutive equations multiplying a first order scalar $s_n$, $M_2$ and $M_5$ stand out from the crowd as they multiply other tensor structures at first order in derivatives. Consequently, $M_2$ and $M_5$ are the one of these transport coefficients which still contribute to the constitutive equations if all $s_{1,2,3,4,5}=0$.  
    \item All transport coefficients $c_i$ and $\tilde c_i$ are nonzero only if the system is chiral. This chirality can be caused by an anomaly, or a $U(1)_A$ chemical potential. 
    Hence, these coefficients are not exclusively caused by the anomaly.
\end{itemize}

\subsubsection{Relation to hydrodynamics in 2+1 dimensional fluids}
\label{sec:relationTo2DHydro}
It is theoretically and experimentally motivated to consider slicing a $(3+1)$-dimensional material into $(2+1)$-dimensional planes and suppressing the interactions between such slices. For example, graphene or the high temperature superconducting cuprates show a layered structure where transport along the layers is different from transport perpendicular to the layers. 

Parity-violating hydrodynamics in 2+1 dimensions has been constructed and discussed in~\cite{Jensen:2011xb}. While there is no chiral anomaly, here the parity anomaly manifests in the transport effects. In order to relate the transport in that lower dimensional system, we may think of our magnetic field as defining (2+1)-dimensional planes perpendicular to it. One can think of the (3+1)-dimensional hydrodynamics described in this section as being projected onto hypersurfaces perpendicular to the magnetic field. For the purpose of constructing constitutive relations for hydrodynamic transport on those (2+1)-dimensional hyperplanes, this simply means that we could project all (3+1)-dimensional tensor structures onto those hyperplanes. For example, the magnetic field may point along the $z$-direction. Then, itself is defined by the projection of $B^\mu= \frac12\, \epsilon^{\mu\nu\rho\sigma} u_\nu F_{\rho\sigma}$ onto the $z$-direction: $B_z= \frac12\, \epsilon_z{}^{ijk} u_i F_{jk}$, where $i,\, j,\, k \in \{t,\, x,\, y\}$. On the $(t,x,y)$-hyperplane $B_z$ transforms like a pseudoscalar if $B^\mu$ is associated with a vector $U(1)_V$, and like a scalar for an axial $U(1)_A$. Another example is the vorticity which becomes a pseudoscalar $\Omega_z=  \epsilon_z{}^{ijk} u_i \partial_j u_k$. These are the definitions of the (pseudo)scalar magnetic field and vorticity in~\cite{Jensen:2011xb}. 
This projection procedure can also be applied to the other tensor structures we used to construct the (3+1)-dimensional hydrodynamic constitutive relations, eqs.~\eqref{eq:TTF} and~\eqref{eq:JTF}. Of course, for a comparison, we need to take into account that weak magnetic fields, $B\sim\mathcal{O}(\partial)$ are considered in~\cite{Jensen:2011xb}. 
So similarities will generally be more obvious in the non-equilibrium part of the constitutive relations. 

Among the equilibrium quantities, $s_1,\, s_2,\, s_3,\, s_4$ have derivatives or other vectors pointing in the $z$-direction, which have no counterpart in (2+1) dimensions. However, $s_5$ has a trivial counterpart $\Omega B$, 
with $B\sim\mathcal{O}(1)$, $\Omega\sim\mathcal{O}(\partial)$ in our counting and $\Omega, B\sim\mathcal{O}(\partial)$ in the counting of~\cite{Jensen:2011xb}. 
The generating functional may depend on the pseudoscalars $B$ and $\Omega$ discussed above, as well as on the temperature $T$. 

It turns out that the perpendicular Hall viscosity of the (3+1)-dimensional hydrodynamics after projection of constitutive relations eq.~\eqref{eq:TTF} is identified with the Hall viscosity in the (2+1)-dimensional hydrodynamic constitutive relation. To see this, consider that the term $\tilde \eta_\perp \tilde\sigma_\perp^{\mu\nu}$ projected onto the $(t,x,y)$-hyperplane becomes $\tilde \eta_\perp \tilde\sigma_\perp^{ij}$ in analogy to the same tensor structure defined in~\cite{Jensen:2011xb}. Similarly, the perpendicular shear viscosity term $\eta_\perp \sigma^{\mu\nu}$ is projected onto the shear viscosity term in (2+1) dimensions. 

In the (2+1)-dimensional hydrodynamics the current constitutive relation analogous to eq.~\eqref{eq:JTF} contains the Hall conductivity term $(\tilde\sigma_\perp+\tilde\chi_E) \epsilon^{ijk} u_j E_k$. The first term can be thought of as a projection of $\tilde\sigma_\perp \tilde V^\mu$. 
The thermodynamic transport coefficient $\tilde\chi_E$ is entirely determined by thermodynamic quantities, and it does not vanish at zero magnetic field. 
The (3+1)-dimensional Hall conductivity also contains such a purely thermodynamic contribution to the current, namely $\epsilon^{\mu\nu\rho\sigma}u_\nu b_\rho E_\sigma$, which is first order in derivatives but we may choose these to be only spatial derivatives. 
Note that the parity anomaly leads to a Hall effect in absence of magnetic fields, as had been realized early by Haldane~\cite{Haldane:1988zza}.

Projecting either of the longitudinal shear or longitudinal Hall viscosity onto a $(t,x,y)$-hyperplane makes it vanish from the energy momentum tensor. Heuristically, this is clear because there can not be any shear or Hall response in the longitudinal $(x,z)$- or $(y,z)$-planes if there are no such planes in the (2+1)-dimensional system. We refer to such longitudinal transport effects as {\it out-of-plane transport} from the perspective of $(2+1)$-dimensional hyperplanes orthogonal to the anisotropy. The opposite to that is the {\it in-plane transport}. It turns out that  all of the novel\footnote{These are novel in that they are absent when the magnetic field is of linear or higher order in derivatives.} transport coefficients 
$\tilde\eta_{||}, \, \eta_{||}$, $\tilde\sigma_{||},\, \eta_1,\, \eta_2$, $\zeta_2,\, c_3,\, c_4,\, c_5,\, c_8,\, c_{10},\,c_{14}$, $c_{15},\, c_{17}$ describe out-of-plane transport effects as seen from the projection of the constitutive relations eq.~\eqref{eq:TTF} and~\eqref{eq:JTF}. 

The (2+1)-dimensional constitutive relations in the thermodynamic frame are given by
\begin{eqnarray}
   T^{ij}_{(2+1)D} & = & 
   (\epsilon_0 + \tilde\epsilon_1 B + \tilde \epsilon_2 \Omega) u^i u^j 
   + p_0 \Delta^{ij}
   +\tilde \phi_2 \tilde E^{(i} u^{j)} +\frac{\tilde\epsilon_2}{T} \epsilon^{(i k l} u_k \partial_l u^{j)} + \mathcal{O}(\partial^2)\, ,\\
    J^i_{(2+1)D} & = & (n_0 + \tilde\phi_1 B + \tilde\phi_2 \Omega) u^i 
    +\tilde\phi_1 \tilde E^i 
    +\frac{\tilde\epsilon_1}{T} \epsilon^{ijk} u_j \partial_k T  + \mathcal{O}(\partial^2)\, .
\end{eqnarray}
The identification between the thermodynamic transport coefficients for (2+1)-dimensional parity violating hydrodynamics and (3+1)-dimensional hydrodynamics with strong magnetic fields can be easily done by comparing our equilibrium constitutive relations to the results of~\cite{Jensen:2012jh}. Formally, the comparison becomes straightforward by expanding the generating functional in eq.~\eqref{eq:Ws} in small magnetic field and keeping only the terms which don't vanish under the assumption of no fluctuations parallel to the magnetic field
\begin{equation}
\label{eq:Ws-new}
W_s = \int d^4x \sqrt{-g}\left(p(T,\mu,B^2=0) +p_{,B^2}(T,\mu,B^2=0) B^2 + M_5(T,\mu,B^2=0) B^\alpha\Omega_\alpha +\cdots \right)\,.
\end{equation}
We can take this generating functional  
over a thin (2+1)-dimensional sheet orthogonal to the magnetic field as the generating functional for (2+1)-dimensional parity violating hydrodynamics 
and compare it to the generating functional
in \cite{Jensen:2012jh}
\begin{equation}
    W_{2+1} = \int d^3 x \sqrt{-\gamma}\left(p_{2+1}(T,\mu) + \tilde{\alpha}_1 B + \tilde{\alpha}_2 \Omega + \cdots \right)\,,
\end{equation}
where $\Omega = B^\mu\Omega_\mu/B$ with $B= \sqrt{B^2}$. Comparing the two expansions above leads to the identification of the thermodynamic transport coefficients in the following way
\begin{equation}
    p_{2+1}(T,\mu) = p(T,\mu,0)\,, \quad \tilde{\alpha}_1 = B\, p_{,B^2}(T,\mu,0) =0\,, \quad \tilde{\alpha}_2 = B\, M_5(T,\mu,0)=0\,.
\end{equation}
Note that because of the derivative counting $B\sim {\cal O}(\partial)$, the first order thermodynamic transport coefficients $\tilde{\alpha}_i$ must vanish. This is a direct consequence of the assumption that the 3+1 dimensional system behaves analytically at small $B^\mu$, that is $p= p(T,\mu,B^2)$. Indeed, upon making these identifications, the constitutive relations in eqs.~\eqref{eq:TTF-eq} and \eqref{eq:JTF-eq} are inconsistent with the results of~\cite{Jensen:2011xb,Jensen:2012jh} by a factor of 2 whenever $\tilde{\alpha}_1$ shows up. To derive the correct 2+1 equilibrium constitutive relations, we must start with a 3+1 system with a generating functional that is not analytic at small $B^\mu$, so that $p = p(T,\mu, B)$. Then a similar analysis leads instead to the identification
\begin{equation}
    p_{2+1}(T,\mu) = p(T,\mu,0)\,, \quad \tilde{\alpha}_1 = \, p_{,B}(T,\mu,0)\,, \quad \tilde{\alpha}_2 = B\, M_5(T,\mu,0)\,.
\end{equation}
The thermodynamic transport coefficients $\tilde{\alpha}_i$ can now be nonzero. Note that $\tilde{\alpha}_2 \neq 0$ requires $M_5$ to diverge at small magnetic field, which is allowed since we didn't assume $W_s$ was analytic at small $B^\mu$. The resulting equilibrium constitutive relations would then match precisely the results of~\cite{Jensen:2011xb,Jensen:2012jh} after truncating fluctuations along the magnetic field and higher order terms in the magnetic field.  

A different way of dimensional reduction could be defined for hydrodynamics on a (2+1)-dimensional hypersurface with normal vector $n^\mu$ pointing along the magnetic field, in a (3+1)-dimensional material. Using this vector $n^\mu$ to project all tensor structures onto the lower-dimensional surface, one obtains a (2+1)-dimensional hydrodynamic description of such a hypersurface \cite{Hernandez:2017xan}.

We are interested in the Hall response $(\tilde\chi_E+\tilde\sigma) \tilde E^\mu$ and its (3+1)-dimensional analog. 
Note that within the (2+1)-dimensional constitutive equations in Landau versus thermodynamic frame the following relations hold~\cite{Jensen:2012jh}:\footnote{Note that the charge density is denoted by $\rho$ in~\cite{Jensen:2011xb,Jensen:2012jh}, whereas we have used $n$ in this work. For the purpose of this comparison we simplify $\rho=n_{2+1}=n$. Similarly, the pressure is denoted $P=p_{2+1} = p$. Note also that the coefficients in~\cite{Jensen:2012jh} were related as $\tilde\phi_1=\tilde\delta_1,\, \tilde\phi_2=\tilde\gamma_1$.}
\begin{equation}
    \tilde\chi_E 
    = \tilde \phi_1 -\frac{n}{\epsilon+p}\tilde\phi_2 \, .
\end{equation}

There is one obvious contribution to the Hall response in the current, $J^\mu = \tilde\sigma_\perp \tilde E^\mu + ...$, but there is another contribution to the Hall response, coming from $\mathcal{Q}^\mu$, in the frame invariant combination 
\begin{equation}
    J^\mu - \frac{n}{\epsilon+p - \aBB B^2}\mathcal{Q}^\mu \sim \tilde E^\mu \left(\tilde\sigma_\perp +2B\, p_{,B^2, \mu}- \frac{n}{\epsilon+p - \aBB B^2} (M_{5,\mu} - 2 p_{,B^2}) \, B
    \right) \, .
\end{equation}  This should be Taylor-expanded in small $B$ in order to keep only terms linear in derivatives for comparison. In the (2+1)-dimensional frame-invariant from~\cite{Jensen:2012jh}, one finds 
\begin{equation}
  J_{2+1}^i - \frac{n}{\epsilon+p}\mathcal{Q}_{2+1}^i \sim \tilde E^i \left(\tilde\sigma_\perp + \tilde\phi_1 - \frac{n}{\epsilon+p}  \tilde\phi_2\right)  \, .   
\end{equation}
This implies that our relations projected on (2+1) dimensions and taking hydrodynamic frames into account give
\begin{equation}
    \tilde\chi_E = \tilde\phi_1 - \frac{n}{\epsilon+p}  \tilde\phi_2 = 2B\,n_{,B^2} +
    \frac{n}{\epsilon+p} (2 p_{,B^2}-M_{5,\mu})\, B \, ,
\end{equation}
where we have used that $n=p_{,\mu}$. Now this agrees with~\cite{Jensen:2011xb,Jensen:2012jh} if we 
recall that in (2+1) dimensions $\frac{\partial p}{\partial \Omega} = B\,M_5$, as is shown in~\cite{Jensen:2011xb}.\footnote{There, the coefficient equivalent to our $M_5$ is named $M_\Omega$.}


\section{Holography}
\label{sec:holography}
\noindent Our goal is to investigate an explicit example of a charged plasma in quantum field theory with a chiral anomaly at strong coupling and in presence of a (strong) magnetic background field. 
In particular, we determine the transport coefficients within that quantum field theory and show that most of the novel transport coefficients discussed in section \ref{sec:hydrodynamics} are indeed non-vanishing. 

For this purpose, we consider $\mathcal{N}=4$ Super-Yang-Mills~(SYM) theory with gauge group $SU(N_c)$. The field content of $\mathcal{N}=4$ SYM theory consists of the following: one vector, four left-handed Weyl fermions, and six real scalar fields. All the matter content is in the adjoint representation of the $SU(N_c)$ gauge group. $\mathcal{N}=4$ SYM theory is invariant under a global $SU(4)_R$ R-charge symmetry. 
In the following we consider a $U(1)$ subgroup of $SU(4)_R$, which we label $U(1)_A$. In $\mathcal{N}=4$ SYM theory the associated global current is axial and has non-vanishing divergence due to a chiral anomaly in the theory. Hence we refer to this as an axial $U(1)_A$ current. 
The fermions and scalars are charged under the anomalous $U(1)_A$ symmetry, while the vector field is uncharged. The anomalous current $J^\alpha$ associated with the $U(1)_A$ symmetry is coupled to an external axial gauge field $A_\alpha^{\text{ext}}$ via the interaction term\footnote{Note that indices referring to boundary field theory coordinates $\{t,\, x,\, y,\, z\}$ are represented by Greek indices such as $\alpha, \,\beta,\, \mu,\, \nu,\, \dots = 0,\,1,\,2,\,3$.}
\begin{equation}\label{eq:extgaugefieldSYM}
S= S_{\text{SYM}} +\int d^4 x \, J^\alpha A_\alpha^{\text{ext}} \, .
\end{equation}
We are specifically interested in the effects of a constant external magnetic field $B$ described by $F = \textrm{d} A^{\text{ext}} = B\, dx \wedge dy$. Moreover, we allow for a nonzero chemical potential $\mu$ associated with the $U(1)_A$ symmetry. We may think of $\mu$ and $B$ as an \textit{axial} chemical potential and a magnetic field associated with an axial $U(1)_A$ symmetry, respectively. 
This is an example for the $U(1)_A$ symmetry, associated gauge fields, chemical potential, and currents discussed in section~\ref{sec:hydrodynamics}.

In order to obtain results for the charged plasma state within this strongly coupled theory, we utilize the gauge/gravity correspondence 
in the large $N_c$ limit and at large 't Hooft coupling $\lambda$ limit \cite{Maldacena:1997re} (for textbooks see \cite{Ammon:2015wua, Nastase, Schalm}). Then we  
perform the relevant calculations in the dual gravitational theory, or a consistent truncation thereof, namely within Einstein-Maxwell-Chern-Simons theory subject to an external magnetic field. The relevant solutions dual to the desired 
equilibrium state at finite temperature, finite axial chemical potential, and subject to an external magnetic field, are charged magnetic black branes~\cite{D'Hoker:2009mm,Ammon:2016szz}. The stability of these black branes was investigated in \cite{Ammon:2017ded}  by computing the quasinormal modes.\footnote{Note that the linear stability of the dual gravitation theory is not guaranteed in the presence of a magnetic field. In fact, there are examples of holographic models in which a magnetic field induces a linear instability, see e.g. \cite{Ammon:2011je}.}

\subsection{Holographic Setup}\label{sec:setup}
The holographically dual gravitational theory mentioned above is 
five-dimensional Einstein-Maxwell-Chern-Simons theory with a negative cosmological constant $\Lambda = - 6 / L^ 2$ and the $AdS_5$ radius $L$. This theory is defined via\footnote{Note that we refer to five-dimensional coordinate indices with lower case Latin letters, such as $m, n = 0,\, 1,\, 2,\, 3,\, 4$. Recall that we use Greek indices such as $\mu,\nu=0, \dots, 3$ for field theory vectors, as well as tensors.
}
\begin{equation}\label{eq:actionS}
S_{\textrm{grav}}=\frac{1}{2\kappa^{2}}
\left[\, \int\limits_{\mathcal{M}}\! d^{5}x\, 
\sqrt{-g}\left(
R+\frac{12}{L^{2}}-\frac{1}{4}F_{mn}F^{mn}
\right)
-\frac{\gamma}{6}\int\limits_{\mathcal{M}} A\wedge F\wedge F\right]\,, 
\end{equation}
where $\mathcal{M}$ denotes the asymptotically $AdS_5$ spacetime, while $\partial \mathcal{M}$ denotes its conformal boundary. Furthermore, $g_{mn}$ is the five-dimensional metric and $F_{mn}=\partial_m A_n - \partial_n A_m$ is the five-dimensional $U(1)_A$ field strength tensor. Moreover, the parameter $\gamma$ is the Chern-Simons coupling which is related to the anomaly coefficient $C$ introduced in section \ref{sec:chiralanomalyhydro} by  $C=-\gamma.$ Let us specify the parameters $c_1$ and $c_2$ introduced in the consistent generating functional \eqref{eq:WA}: the coefficient $c_1$ is related to the mixed gauge-gravitational anomaly and is sub-leading in the large $N_c$ limit. Hence, in our case $c_1=0.$\footnote{In order to mimic a nonzero coefficient $c_1$ we may add by hand a mixed gauge-gravitational Chern-Simons
term  to the action $S_{\textrm{grav}}$. See \cite{Landsteiner:2011iq} for more details.} Moreover, since the dual $\mathcal{N}=4$ supersymmetric field theory is CPT-invariant, we also conclude that $c_2=0$. 

For $\gamma = 2/\sqrt{3}$, the action~\eqref{eq:actionS} is the bosonic part of minimal gauged supergravity in five spacetime dimensions and hence it is  a  consistent  truncation  of  the  most  general  class  of  type  IIB
supergravity in ten  dimensions  or  supergravity  in  eleven  dimensions  which  are  dual  to
$\mathcal{N}= 1$ superconformal field theories, see e.g. \cite{Buchel:2006gb,Gauntlett:2006ai,Gauntlett:2007ma,Colgain:2014pha}.  In this paper however, we restrict ourselves not just to this particular value of $\gamma$ but rather treat $\gamma$ as a free parameter and study the transport coefficients as a function thereof. In particular, we investigate the cases $\gamma=0$ and  $\gamma = 2/\sqrt{3}$. The latter value, we refer to as {\it supersymmetric}.

The action~\eqref{eq:actionS} has to be amended by boundary terms \cite{Henningson:1998gx, Balasubramanian:1999re,Taylor:2000xw,Sahoo:2010sp} of the form 
\begin{equation}\label{eq:actionSbdy}
S_{\text{bdy}}= \frac{1}{\kappa^2} \int\limits_{\partial\mathcal{M}}\! d^4x  \sqrt{-\hat{g}} \left( K - \frac{3}{L} - \frac{L}{4} \hat{R} +  \ln\left( \frac{\varrho}{L} \right) \left( \frac{L}{8} F_{\mu\nu} F^{\mu\nu} - \frac{L^3}{8} \hat{R}_{ \mu\nu} \hat{R}^{\mu\nu} +  \frac{L^3}{24} \hat{R}^{2} \right)\right) ,
\end{equation}
where $\varrho$ is the radial coordinate of $AdS_5$ in the Poincar\'{e} slicing. 
The metric $\hat{g}_{\mu\nu}$ is induced by $g_{mn}$ on the conformal boundary of $AdS_5$, while the extrinsic curvature is given by 
\begin{equation}\label{eq:extrinsicK}
K_{mn} = \mathcal{P}_m^{\ \, o} \,  \mathcal{P}_n^{\ \, p} \, \nabla_o n_p \, ,\qquad \mbox{with} \quad \mathcal{P}_m^{\ \, o} = \delta_m^{\ o} - n_m n^o \, .
\end{equation}
Here, $\nabla$ is the covariant derivative and $n_m$ are the components of the outward pointing normal vector of the boundary $\partial\mathcal{M}.$ Moreover, $K$ is the trace of the extrinsic curvature with respect to the metric at the boundary, $\hat{R}_{ \mu\nu}$ denotes the Ricci tensor associated with the metric $\hat{g}_{\mu\nu}$ and $\hat{R}$ is the corresponding Ricci scalar. For simplicity we are going to choose $L=1$ and $2\, \kappa^2 =16 \pi G_5 =1$ from now on.

The equations of motion associated with the action \eqref{eq:actionS} in terms of the Ricci tensor $R_{mn}$, metric $g_{mn}$, and field strength $F_{mn}$ read 
\begin{equation}\label{eq:EOM1}
R_{mn}=-4\,g_{mn}+\frac{1}{2}\left(F_{mo}\,F_{n}{}^{o}-\frac{1}{6}\,g_{mn}\,F_{op}F^{op}\right)
\end{equation}
from variation with respect to the metric, as well as 
\begin{equation}\label{eq:EOM2}
\nabla_m F^{m n} + \frac{\gamma}{8 \sqrt{-g}} \, \varepsilon^{nmopq} F_{mo} F_{pq} = 0
\end{equation}
from variation with respect to the gauge field. Here, $\varepsilon^{mnopq}$ is the totally antisymmetric Levi-Civita symbol in five spacetime dimensions with $\varepsilon^{txyz\varrho}=1$.

We are interested in describing the charged plasma state in the presence of an external (axial) magnetic background field, (axial) chemical potential and in the presence of the chiral anomaly. The appropriate ansatz for the metric and the field strength tensor $F$ in ingoing Eddington-Finkelstein coordinates reads\footnote{To keep the notation simple, we still use the variable $t$ for the null ingoing Eddington-Finkelstein time.}
\begin{eqnarray}\label{eq:ansatzmetric}
d s^{2}  &=&  \frac{1}{\varrho^{2}}\left[\left(-u(\varrho)+c(\varrho)^2\,w(\varrho)^2\right)\, d t^{2} -2 \, dt \, d \varrho + 2\, c(\varrho)\,w(\varrho)^2 \, d z\, d t \right. \nonumber \\
 &&\left. + v(\varrho)^{2}\, \left( d x^{2} +d y^{2}\right)  + w(\varrho)^{2}\,d z^{2}\right] \, , \\
\label{eq:ansatzF} 
F  &=&A_t'(\varrho)\,d \varrho \wedge d t +  B\,d x\wedge d y+ P'(\varrho)\,d \varrho \wedge d z \, ,
\end{eqnarray}
where the horizon of the black brane is located at $\varrho = 1$, while the conformal boundary is located at $\varrho = 0$. Moreover, prime denotes derivatives with respect to the radial coordinate $\varrho$. The field strength tensor~\eqref{eq:ansatzF} may be obtained from a gauge field $A$ of the form
\begin{equation}\label{eq:ansatzA}
A= A_t(\varrho)\, d t + \frac{B}{2} \left(- y \, d x + x \, d y \right) +P(\varrho)\, d z\, .
\end{equation}
Note that the ansatz for the metric and the field strength tensor preserve the $SO(2)$ rotational symmetry in the $(x, y)$-plane. 

The metric functions $u(\varrho), v(\varrho), w(\varrho)$ and $c(\varrho)$ are chosen 
such that the spacetime is asymptotically $AdS_5$ with a flat Minkowski metric $\hat{g}$ of the conformal boundary which is located at $\varrho=0$. In particular, we set $u(0)= v(0)= w(0)=1$ and $c(0)=0$. Moreover, the leading component of $A_t$ is identified with the axial chemical potential $\mu$, while there is no explicit source for $P(\varrho)$. 
The latter choice sets the source for the field theory current $J_z$ to zero. 

Moreover, we  impose the conditions $A_t(1)=0$ and $c(1)=u(1)=0$ at the horizon which is located at $\varrho=1$. The condition on $c(1)$ and $u(1)$ prevents a conical singularity at the horizon in the Euclideanized metric. 
The subleading coefficient of $u(\varrho)$ at the horizon is related to the temperature
\begin{equation}\label{eq:T}
T = \frac{|u'(1)|}{4\pi} \, .
\end{equation} 
The explicit form of functions close to the conformal boundary and near the horizon are shown in appendix \ref{sec:detailsapp}.

In order to find the functions $u(\varrho),\, c(\varrho),\, w(\varrho),\, v(\varrho),\, A_t(\varrho)$, and $P(\varrho)$ subject to the boundary conditions specified above we use  spectral methods to solve the ordinary differential equations (for more details see ~\cite{Ammon:2016szz}). We can also read off thermodynamic quantities such as the density of the partition function in the grand canonical ensemble $\Omega$, the entropy density $s$ as well as the one-point function of the energy-momentum tensor $T_{\mu\nu}$, the covariant current $J^\mu_{cov}$ and the consistent current $J^\mu_{cons}$ from the numerical solution.

The form of the expectation value of the energy-momentum tensor depends on the chosen action \eqref{eq:actionS} and on its boundary terms \eqref{eq:actionSbdy}. Given these boundary terms, which corresponds to a particular choice of the renormalization scheme, the energy-momentum tensor may be extracted from ~\cite{Balasubramanian:1999re}\footnote{If we do not set $2\kappa^2=1$, the expressions for the energy-momentum tensor and the current have to be multiplied by $2\kappa^2.$ In case of $\mathcal{N}=4$ SYM with gauge group $SU(N)$, this overall prefactor is given by $N^2/(8\pi^2)$ in terms of field theory quantities.}
\begin{eqnarray} \nonumber
\left\langle T_{\mu\nu} \right\rangle=\lim\limits_{\varrho\rightarrow 0}\frac{1}{\varrho^{2}} &\Big(&-2K_{\mu\nu}+2(K-3) \, g_{\mu\nu}+\ln(\varrho)\left(F_{\mu}^{\ \alpha}F_{\nu\alpha}
-\frac{1}{4} \, g_{\mu\nu}F^{\alpha\beta}F_{\alpha\beta}\right)\\
&&+ \hat{R}_{\mu\nu} - \frac{1}{2} \hat{R} \hat{g}_{\mu\nu} + 8 \ln(\varrho) \, h^{(4)}_{\mu\nu} \Big) \, \label{eq:Tmunu}
\end{eqnarray}
with
\begin{equation}
h^{(4)}_{\mu\nu} =    \frac{1}{8} \hat{R}_{\mu\nu\rho\sigma} \hat{R}^{\rho\sigma} + \hat{R} \hat{\nabla}_\mu \hat{\nabla}_\nu  \hat{R} - \frac{1}{16} \hat{\nabla}^2 \hat{R}_{\mu\nu} - \frac{1}{24} \hat{R} \hat{R}_{\mu\nu} + \frac{1}{96} \left( \hat{\nabla}^2 \hat{R} + \hat{R}^2 - 3 \hat{R}_{\rho\sigma} \hat{R}^{\rho\sigma}  \right) \hat{g}_{\mu\nu}
\end{equation}
Here, $\hat{g}_{\mu\nu}$ is the metric on the conformal boundary of $AdS_5$; $\hat{\nabla}_\mu$, $\hat{R}_{\mu\nu\rho\sigma}$, $\hat{R}_{\mu\nu}$ and $\hat{R}$ are the covariant derivative, the Riemann curvature tensor, the Ricci tensor and the Ricci scalar of the boundary metric $\hat{g}$. Also, $h^{(4)}_{\mu\nu}$ is proportional to the Bach tensor, which spoils the power series expansion of the metric in the Fefferman-Graham expansion~\cite{FG,Fefferman:2007rka} for $d=4$, introducing a term that is logarithmic in the radial coordinate~\cite{Alvarez:2002jy}. Moreover, $K_{\mu\nu}$ is the (projected) extrinsic curvature \eqref{eq:extrinsicK}.  

Let us now turn to the expectation value of the consistent and covariant form of the current $\left\langle J^{\mu} \right\rangle.$ The consistent current is given in terms of a variation of the consistent generating functional $W_{cons}$ with respect to the boundary gauge field $A^{\textrm{ext}}_\mu$ (see eq. \eqref{eq:extgaugefieldSYM} for the definition of the external gauge field). The consistent generating functional is identified with the action \eqref{eq:actionS}  -- including its boundary terms \eqref{eq:actionSbdy} -- evaluated on-shell. Note that this proposal 
is adequate since both, $W_{cons}$ as well as the gravitational action $S_{\textrm{grav}}$ transforms in the same way under an infinitesimal $U(1)$ gauge transformation, namely as specified by equation \eqref{eq:WA-gauge}. Hence, the expectation value of the consistent current $J_{cons}^{\mu}$ is given by~\cite{D'Hoker:2009bc,Landsteiner:2011iq,Gynther:2010ed,Landsteiner:2016led,Landsteiner:2011tf,Landsteiner:2013aba}
\begin{equation}\label{eq:currentJcons}
\left\langle J_{cons}^{\mu} \right\rangle= \lim\limits_{\varrho\rightarrow 0}\left(
\sqrt{-g} n_a g^{a\nu}F_{\nu\sigma} g^{\sigma \mu}+\frac{\gamma}{6}\epsilon^{\alpha\beta\gamma\mu}A_{\alpha}F_{\beta\gamma} + \ln\varrho \, \sqrt{-\hat{g}} \, \hat{\nabla}_\nu F^{\nu\mu} \right) 
\, ,
\end{equation}
where $n_a$ is the unit normal vector orthogonal to the AdS-boundary. 

Let us turn to the expectation value of the covariant form of the current $J_{cov}^{\mu}$. The recipe is to drop the term in \eqref{eq:currentJcons} which arises from the Chern-Simons term, i.e. the term $\frac{\gamma}{6}\epsilon^{\alpha\beta\gamma\mu}A_{\alpha}F_{\beta\gamma}$. This is also in accordance with the proposal put forward in the previous paragraph: the covariant current is given in terms of a variation of the covariant generating functional $W_{cov}$ which differs from $W_{cons}$ by a Chern-Simons term -- see eq. \eqref{eq:WCS}. This is chosen in such a way that $W_{cov}$ is gauge invariant. 
In terms of the dual gravitational theory this means that we have to drop the Chern-Simons term from the action $S_{\textrm{grav}}$ before evaluating $S_{\textrm{grav}}$ and $S_{\textrm{bdy}}$ on-shell.\footnote{However, this does not imply that the Chern-Simons parameter is not relevant for computing $\left\langle J_{cov}^{\mu} \right\rangle$. We still have to solve the equation \eqref{eq:EOM2} for the gauge fields even though we have droped the Chern Simons term to determine the generating functional $W_{cov}.$} Note that this also fits nicely with the identification $C=-\gamma$ which we stated earlier. 

In summary, the expectation value of the covariant current reads 
\begin{equation}\label{eq:currentJcov}
\left\langle J_{cov}^{\mu} \right\rangle= \lim\limits_{\varrho\rightarrow 0}\left(
\sqrt{-g}n_a g^{a\nu}F_{\nu\sigma} g^{\sigma \mu} + \ln\varrho \, \sqrt{-\hat{g}} \, \hat{\nabla}_\nu F^{\nu\mu} \right) 
\,.
\end{equation}

In appendix \ref{sec:detailsapp} we evaluate the expectation value of the energy momentum tensor and the covariant current for the charged magnetic brane considered here and relate them to coefficients in the near-boundary expansion. A word of caution in order:  components of the one-point functions, e.g. the energy density $\epsilon=\langle T_{tt} \rangle,$ as well as the pressure $p$ may be scheme-dependent. Adding finite counter terms in $S_{\text{bdy}}$ (see eq. \eqref{eq:actionSbdy}) will change the renormalization scheme. In fact, here we use $1/L$ as our renormalization scale.\footnote{Note that due to setting $L=1$ the renormalization scale is not transparent in the explicit expressions.} Other physically  significant choices of the renormalization scale are discussed in \cite{Fuini:2015hba}.

Besides thermodynamics, we are also interested in thermodynamic and hydrodynamic transport coefficients given in terms of Kubo formulas in section~\ref{sec:kubo}. 
In order to compute the corresponding correlation functions, we consider fluctuations to linear order in the metric, $h_{mn}(\varrho, x^\mu),$ and of the gauge field, $a_m(\varrho, x^\mu)$, on top of the background discussed above. We perform a Fourier transformation along the field theory coordinates $x^\mu$ and solve the corresponding equations of motion \eqref{eq:EOM1} and \eqref{eq:EOM2} for the fluctuations $\tilde{h}_{mn}(\varrho, k^\mu)$ and $\tilde{a}_{m}(\varrho,k^\mu)$.\footnote{To keep the notation simple, we suppress the $k$-dependence of $\tilde{h}_{mn}(\varrho, k^\mu)$ and $\tilde{a}_{m}(\varrho,k^\mu)$ and neglect the tilde in the following.}

Since we either determine correlation functions at zero momentum, or we choose the momentum to be aligned with the magnetic field, (i.e. along the $z$-axis) we may classify the fluctuations according to the unbroken SO(2) symmetry corresponding to rotations around the $z$-axis (for more details see e.g. \cite{Ammon:2017ded}). In order to consider only the physical modes of the system, we have to fix the gauge freedom. To do so, we choose a modified radial gauge in which $a_\varrho = 0$ and $h_{m \varrho} = 0$ for $m\neq t$, as well as $h_{t\varrho} = 1/2 \, h_{tt}.$\footnote{Note these unusual gauge. However, setting $h_{t\varrho}$ to zero does not allow us to specify arbitrary sources for $h_{\mu\nu}(\varrho=0)$ at the conformal boundary.}

To determine the thermodynamic/transport coefficients we proceed as follows: we perturb the system by switching on a source term, e.g. $h_{\alpha\beta}(0)=\delta g_{\alpha\beta}$ and then study its linear response $\delta \langle T_{\mu\nu}\rangle$ and $\delta \langle J_{cov}^\mu \rangle$; using eq. \eqref{eq:corr-funcspr} we may read off the correlators $G_{T^{\mu\nu} T^{\alpha\beta}}(\omega, {\bf k})$ and $G_{J_{cov}^{\mu} T^{\alpha\beta}}(\omega, {\bf k}).$ Likewise we  may allow for non-vanishing sources $a_\alpha(0)$ from which we deduce $G_{J_{cov}^{\mu} J_{cons}^{\alpha}}(\omega, {\bf k})$ (see eq. \eqref{eq:cons-var}).
The naive way to compute correlators of the form
$\lim\limits_{k \rightarrow 0} \frac{1}{k} G_{\mathcal{O} \tilde{\mathcal{O}}}(\omega=0, k \hat{\bf{z}})$ and $\lim\limits_{\omega \rightarrow 0} \frac{1}{\omega} G_{\mathcal{O} \tilde{\mathcal{O}}}(\omega, \vec{k}=\vec{0})$ is to evaluate the Green's function at a small value of $k$ (or a small value of $\omega$), and then divide by that small value of $k$ (or $\omega$).  Not surprisingly, this leads to large numerical errors. In appendix \ref{sec:detailsapp} we outline a method which just determines the linear coefficient in a small $k$ (or $\omega$) expansion of the corresponding Green's function.

\subsection{Results: Thermodynamics and thermodynamic transport coefficients}
\subsubsection*{Partition function in the grand canonical ensemble and magnetization} 
First we determine  thermodynamic quantities of the dual field theory such as $\Omega(T, \mu, B)$ being the density of the partition function in the grand canonical ensemble which in turn gives the pressure $p=-\Omega.$ In fact, we study the system as a function of the dimensionless chemical potential $\tilde{\mu}=\mu/T$ and the dimensionless magnetic field $\tilde{B}=B/T^2.$ From here on out we will refer to these two quantities simply as the chemical potential $\tilde{\mu}$ and the magnetic field $\tilde{B}$ for brevity. 
Moreover, any observable discussed in the following sections will be rescaled by powers of temperature such that it is dimensionless. For example, the dimensionless density of the grand canonical potential $\Omega$ is given by $\Omega(T, \mu, B)/T^4$.

\begin{figure}
    \centering
    \includegraphics[width=0.45 \linewidth]{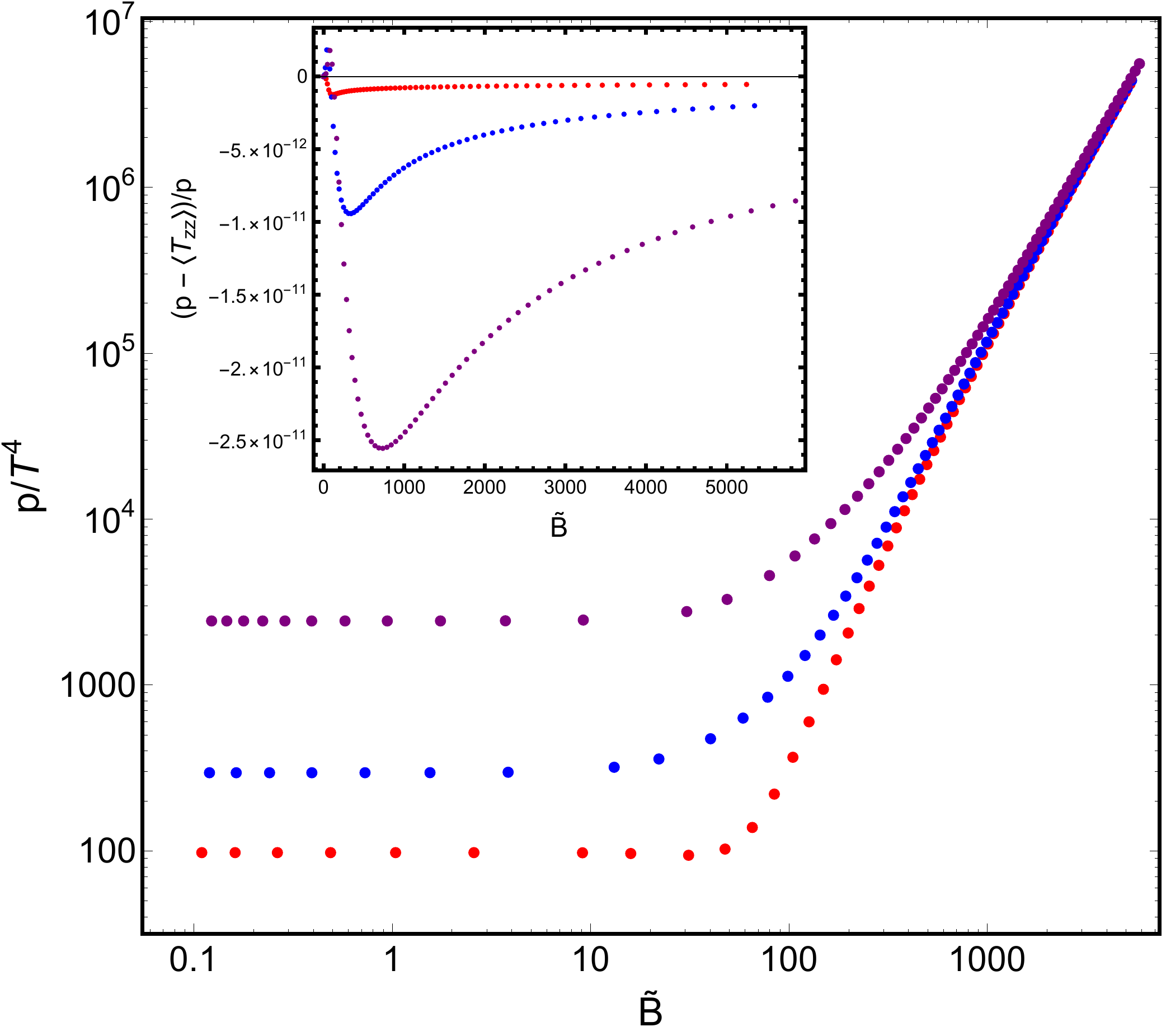}
    \caption{The dimensionless pressure $p/T^4$ as a function of the magnetic field $\tilde{B}=B/T^2$ for $\gamma = 2/\sqrt{3}$ and three different values of $\tilde{\mu}$, namely $\tilde{\mu}=0$ (red), $\tilde{\mu}=4$ (blue), $\tilde{\mu}=10$ (purple). Inset: Difference between $p$ and $\langle T_{zz} \rangle$ which serves as a check of the numerics.}\label{fig:pressure}
\end{figure}

In figure \ref{fig:pressure} we display the dimensionless pressure $p/T^4$ as a function of the  magnetic field $\tilde{B}$ for different values of the chemical potential $\tilde{\mu}.$ Since there is no major qualitative difference between the cases $\gamma=0$ and $\gamma=2/\sqrt{3}$ we only display the latter one. As a cross-check for our numerics we checked that the -- from the gravitational point of view non-trivial -- 
relation $p = \langle T_{zz} \rangle$ holds to high accuracy.\footnote{See \cite{Ammon:2012qs} or a discussion concerning thermodynamic potentials in the presence of magnetic fields and their relation to components of the energy-momentum tensor.} 

\begin{figure}
    \centering
    \includegraphics[width=0.45 \linewidth]{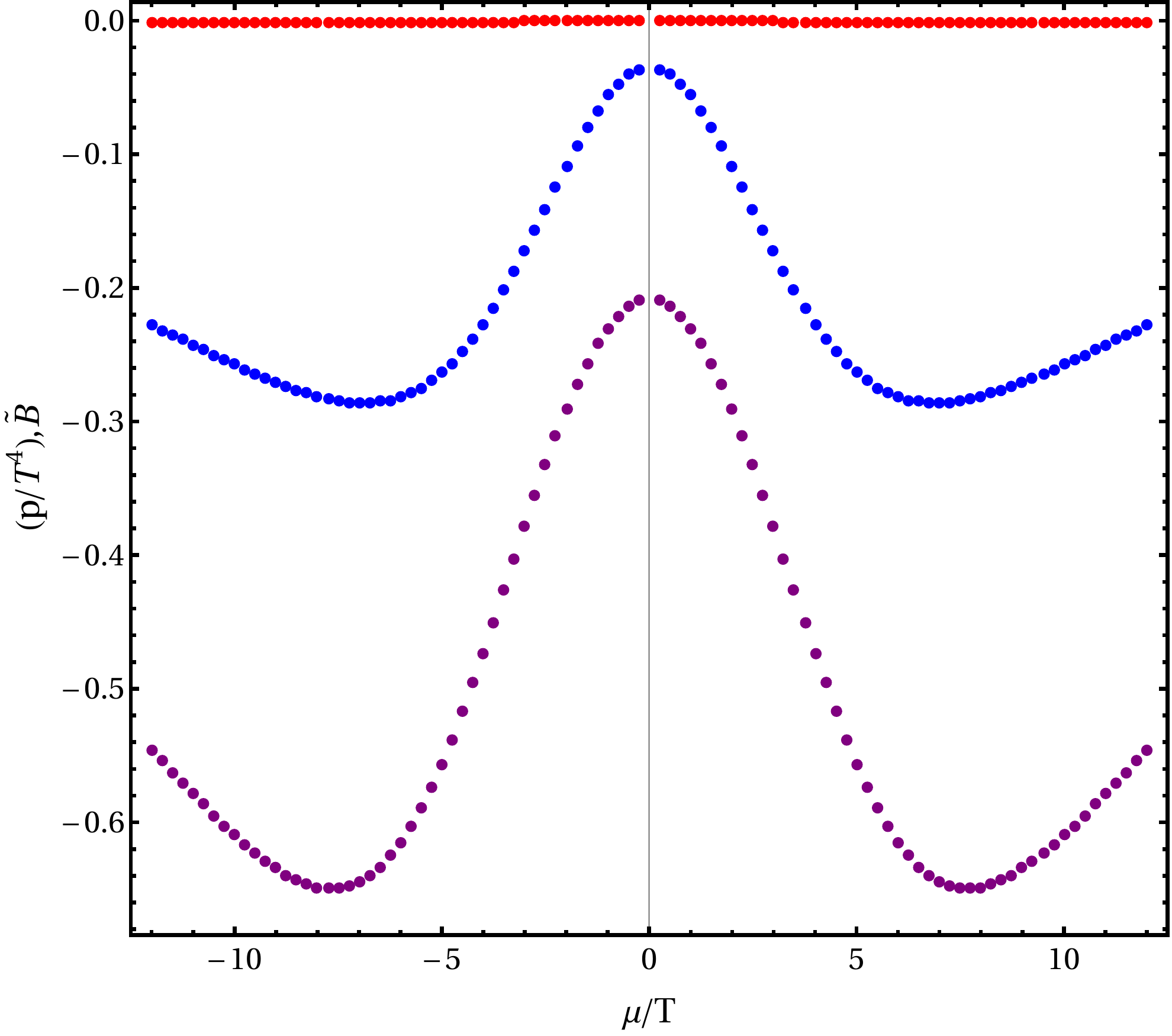}
    \hspace{0.5cm}
    \includegraphics[width=0.45 \linewidth]{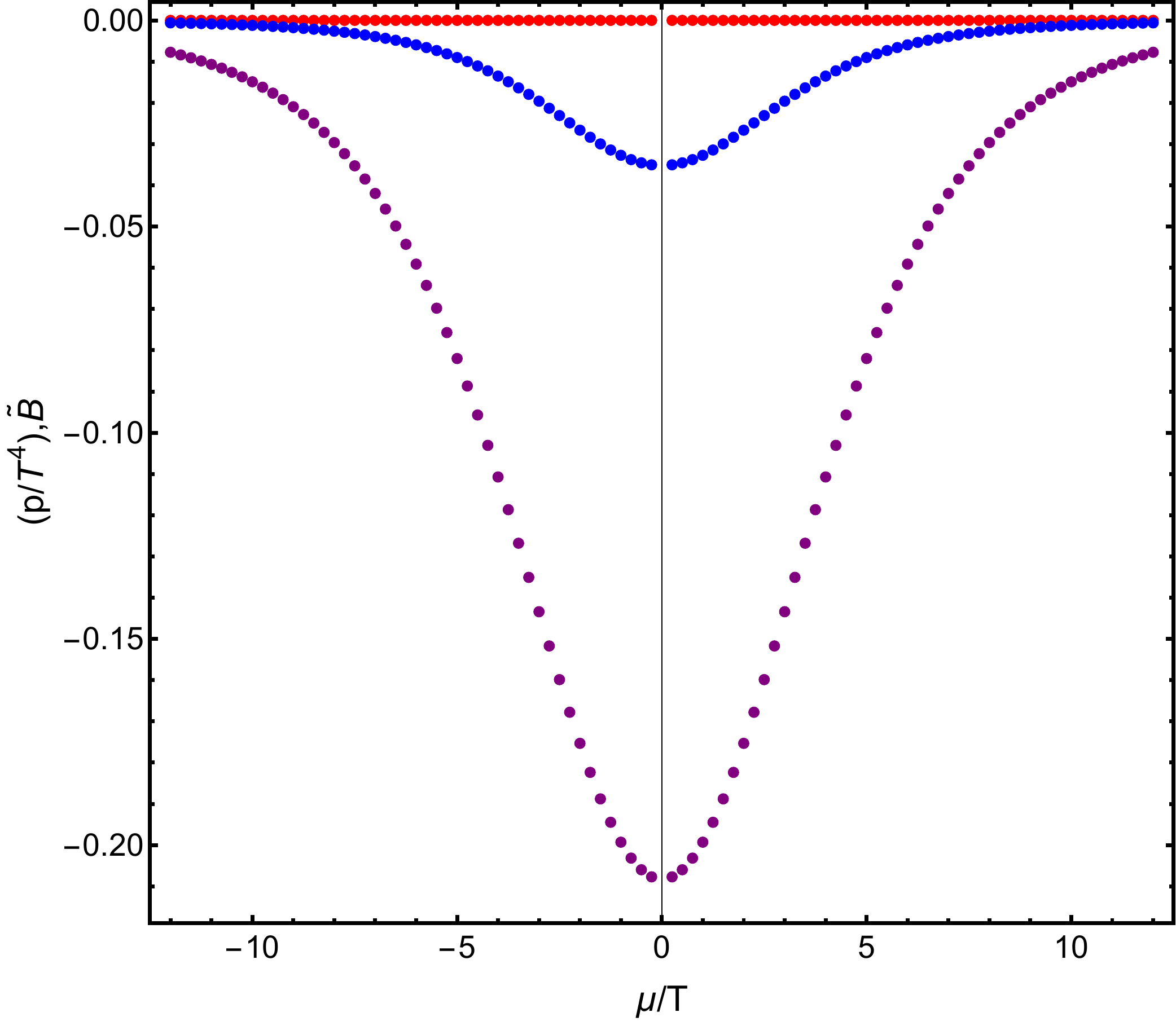}
    \caption{The dimensionless magnetization $(p/T^4)_{, \tilde{B}}$ as a function of $\tilde{\mu}=\mu/T$ for  $\gamma=2/\sqrt{3}$ (left figure) and $\gamma=0$ (right figure). The different curves correspond to magnetic fields $B/T^2=\{0.05,12.5,30\}$ (red, blue, purple).}\label{fig:magnetization}
\end{figure}

Next, we determine the magnetization $M$ which is defined by $M = - \left(\frac{\partial\Omega}{\partial B}\right)_{\mu, T}$ where $\mu$ and $T$ are kept constant. Note that we can express the dimensionless magnetization $M/T^2$ by $\partial \! \left(p/T^4\right)\!/\partial\tilde{B}|_{\tilde{\mu}}$ where we keep  $\tilde{\mu}$ constant.\footnote{In the following we abbreviate the derivative by $\left( p / T^4 \right)_{,\tilde{B}}$. We refer the reader to appendix \ref{app:dimderiv} for more details.} The results are shown in figure \ref{fig:magnetization}. The left figure shows the dimensionless magnetization for the supersymmetric case while the right figure displays the case $\gamma=0.$ In both cases, the magnetization is a symmetric function of $\tilde{\mu}$. Figure \ref{fig:magnetization} shows three different cases. Some remarks are in order: Firstly, the magnetization vanishes for $\tilde{B} \rightarrow 0$. In fact, the red curve in the left and right panel of figure \ref{fig:magnetization} corresponds to the smallest value of $\tilde{B}$ which we investigated, namely $\tilde{B}=0.05$. In both cases, the magnitude of the magnetization nearly vanishes (however, zooming into the red curve we see a behavior similar to the one with larger values of $\tilde{B}$). Secondly, the magnetization is an even function of the  
chemical potential $\tilde \mu$. Note that in the case $\gamma=0$ the magnetization increases as a function of $|\tilde{\mu}|,$ while for the supersymmetric value of $\gamma$, the magnetization as a function of $|\tilde{\mu}|$ first decreases and then again increases. 
Our numerics do not allow an exact statement, however, figure~\ref{fig:magnetization} suggests that in both cases the magnetization approaches zero for large values of $\tilde\mu$. 
Finally, the values for the dimensionless magnetization agree for both cases of $\gamma$ at zero chemical potential which is an important cross-check for our numerics. This is because at zero chemical potential our solutions reduce to magnetic black branes, which can be shown to receive no contribution from the Chern-Simons term, neither through the on-shell action, nor through the equations of motion. 

Next, we determine the thermodynamic transport coefficients $M_i$ with $i=1, \dots, 5$ in this holographic model using the Kubo formulas \eqref{eq:M2M5Kubo} and \eqref{eq:M1M3M4Kubo}. From the correlation functions within the Kubo formulas using dimensional analysis we then read off that $M_{1,3,5}$ have dimension of temperature (or chemical potential), while $M_2$ has dimension of inverse temperature. Finally, $M_4$ is dimensionless. 

\subsubsection*{Magneto-vortical susceptibility $M_5$} 
\begin{figure}
    \centering
    \includegraphics[width=0.45 \linewidth]{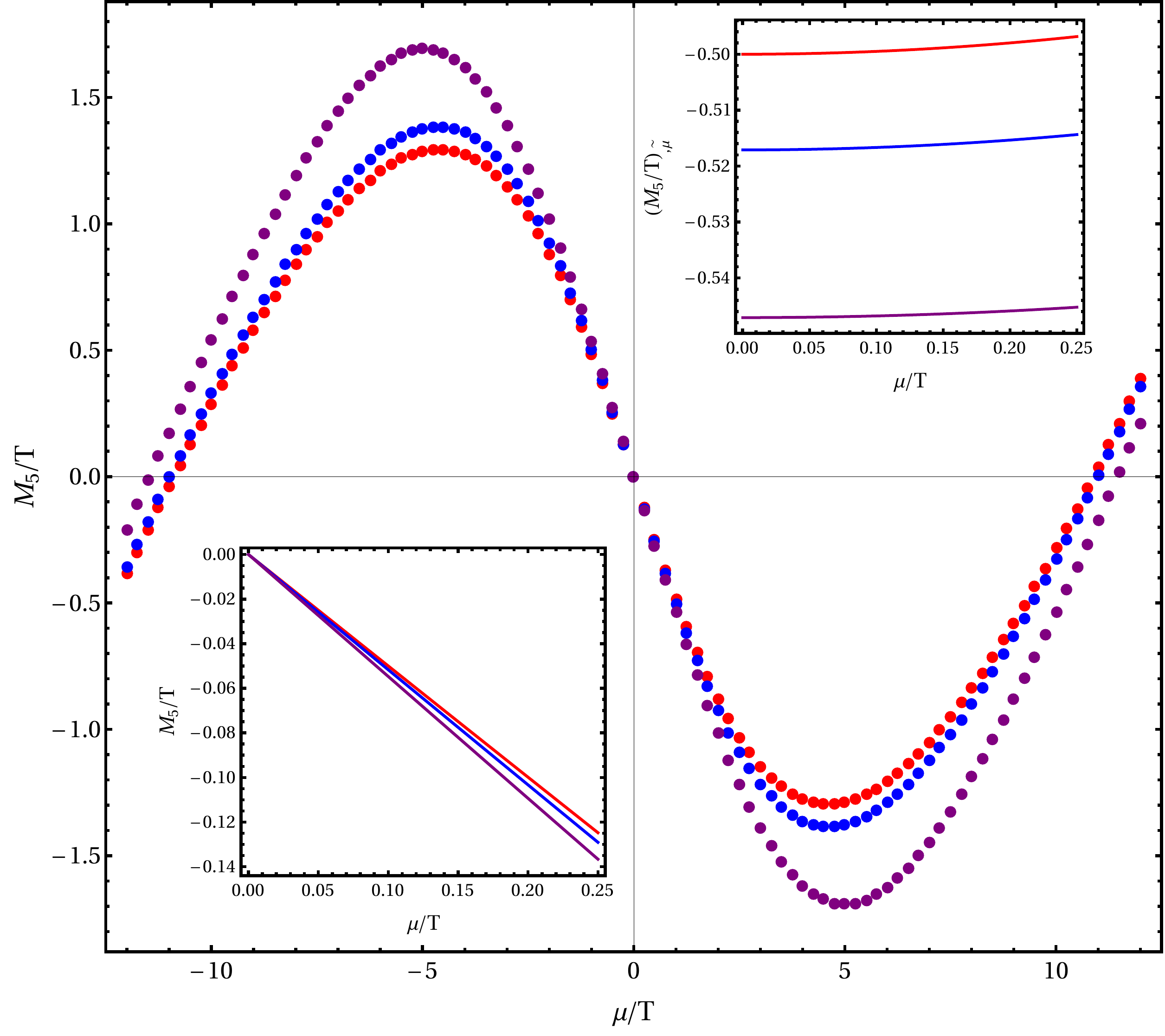}\hspace{0.5cm} \includegraphics[width=0.45 \linewidth]{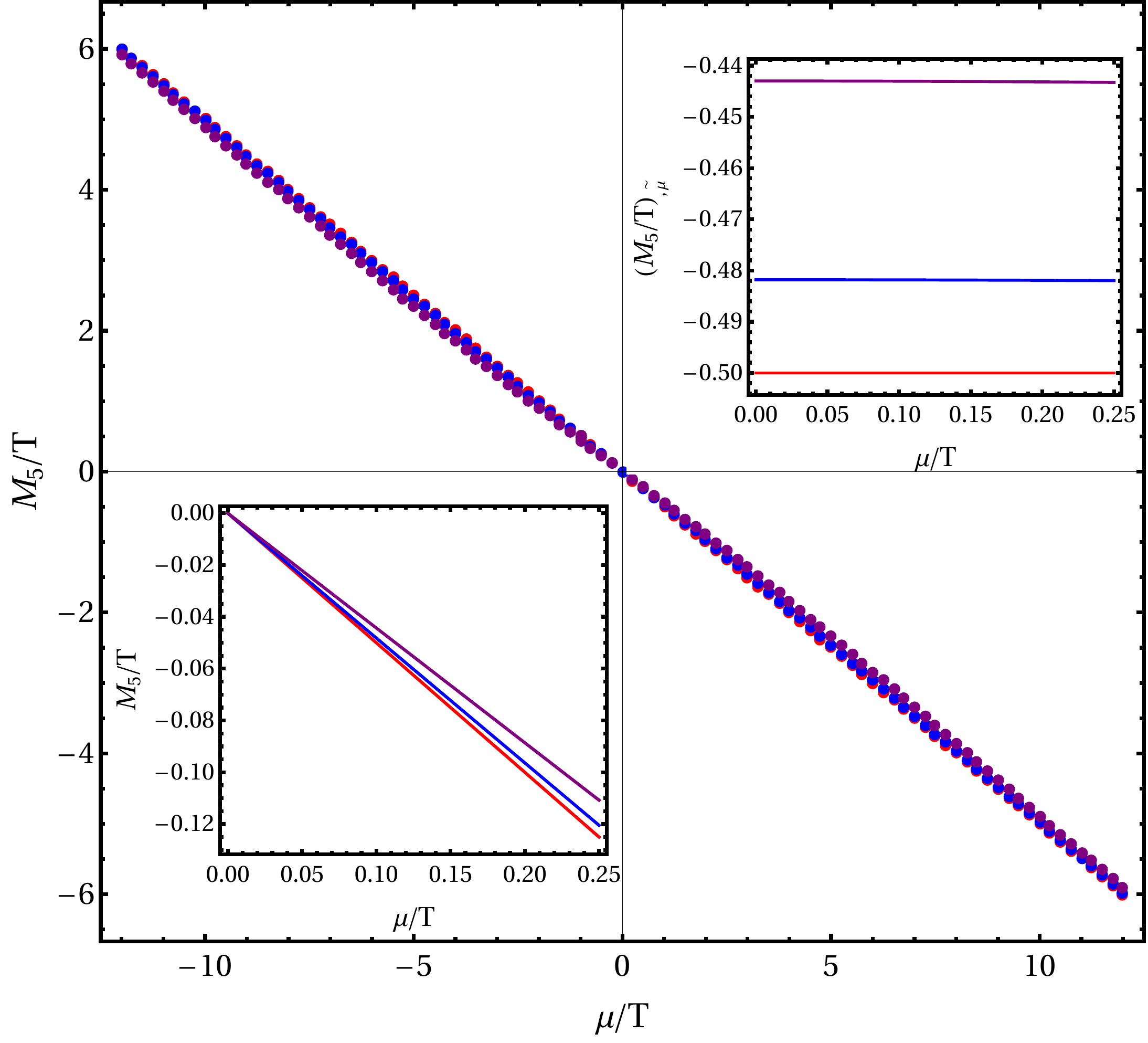}
    \caption{The dimensionless magneto-vortical susceptibility $M_5/T$ as a function of the  chemical potential $\mu/T$ for  $\gamma=2/\sqrt{3}$ (left subfigure) and $\gamma=0$ (right subfigure). The different curves correspond to magnetic fields $B/T^2=\{0.05,12.5,30\}$ (red, blue, purple).}\label{fig:M5}
\end{figure}

First, we study the thermodynamic coefficient $M_5$ which is also known as magneto-vortical susceptibility and has energy dimension one. We consider the dimensionless combination $M_5/T$. We numerically confirm that in both cases for $\gamma$, the magneto-vortical susceptibility $M_5/T$ is non-vanishing for nonzero magnetic field and/or nonzero chemical potential. Moreover, as expected and as evident from Figure \ref{fig:M5}, $M_5/T$ is an anti-symmetric function of the chemical potential $\tilde{\mu}$. 

Note that the left and right panel of Figure \ref{fig:M5} correspond to the case $\gamma=2/\sqrt{3}$ (left panel) and to the case $\gamma=0$ (right panel), respectively. We first discuss the case $\gamma=0.$  
For vanishing magnetic field $\tilde{B}=0$ we find a linear relation between $M_5/T$ and $\tilde{\mu}$ (at least for the values of the axial chemical potential displayed) of the form $M_5/T = - \tilde{\mu}/2$ which is in agreement with~\cite{Bu:2019qmd}. 
However for finite magnetic fields, $M_5/T$ is no longer directly proportional to $\mu/T$ as evident from the inset in the lower left corner. In fact, we numerically find that 
\begin{equation}\label{eq:M5Tch}
\frac{\partial(M_5/T)}{\partial \tilde{\mu}}\Big|_{\tilde{\mu}=0} = - \frac{1}{2} + c \tilde{B}^2
\end{equation}
where $c$ is a positive constant. 
If we evaluate $\partial(M_5/T)/\partial \tilde{\mu}$ at non-vanishing (but small) chemical potential  $\tilde{\mu}$ there are additional corrections to the right hand side of eq. \eqref{eq:M5Tch}.

In case of $\gamma=2/\sqrt{3}$ the thermodynamic coefficient $M_5/T$ transitions through zero at a finite value of $\tilde{\mu}$ as opposed to the case $\gamma=0$ where $M_5/T$ is decreasing monotonically with increasing chemical potential. In particular, for $\tilde{B}=0$, the relation $M_5/T = - \tilde{\mu}\,T/2$ does not longer hold for finite values of the chemical potential $\tilde{\mu}$. This can be seen from the inset in the lower left corner of the right panel of figure~\ref{fig:M5}. For finite but small magnetic fields, there will be again corrections  To be more precise we numerically verified the relation \eqref{eq:M5Tch}  for sufficiently small $\tilde{B}$. However, the constant $c$ is negative in contrast to the case in which $\gamma=0$.

\subsubsection*{Thermodynamic coefficient $M_2$} 
\label{sec:thermoCoeffsHolo}
\begin{figure}
    \centering
  \includegraphics[width=0.45 \linewidth]{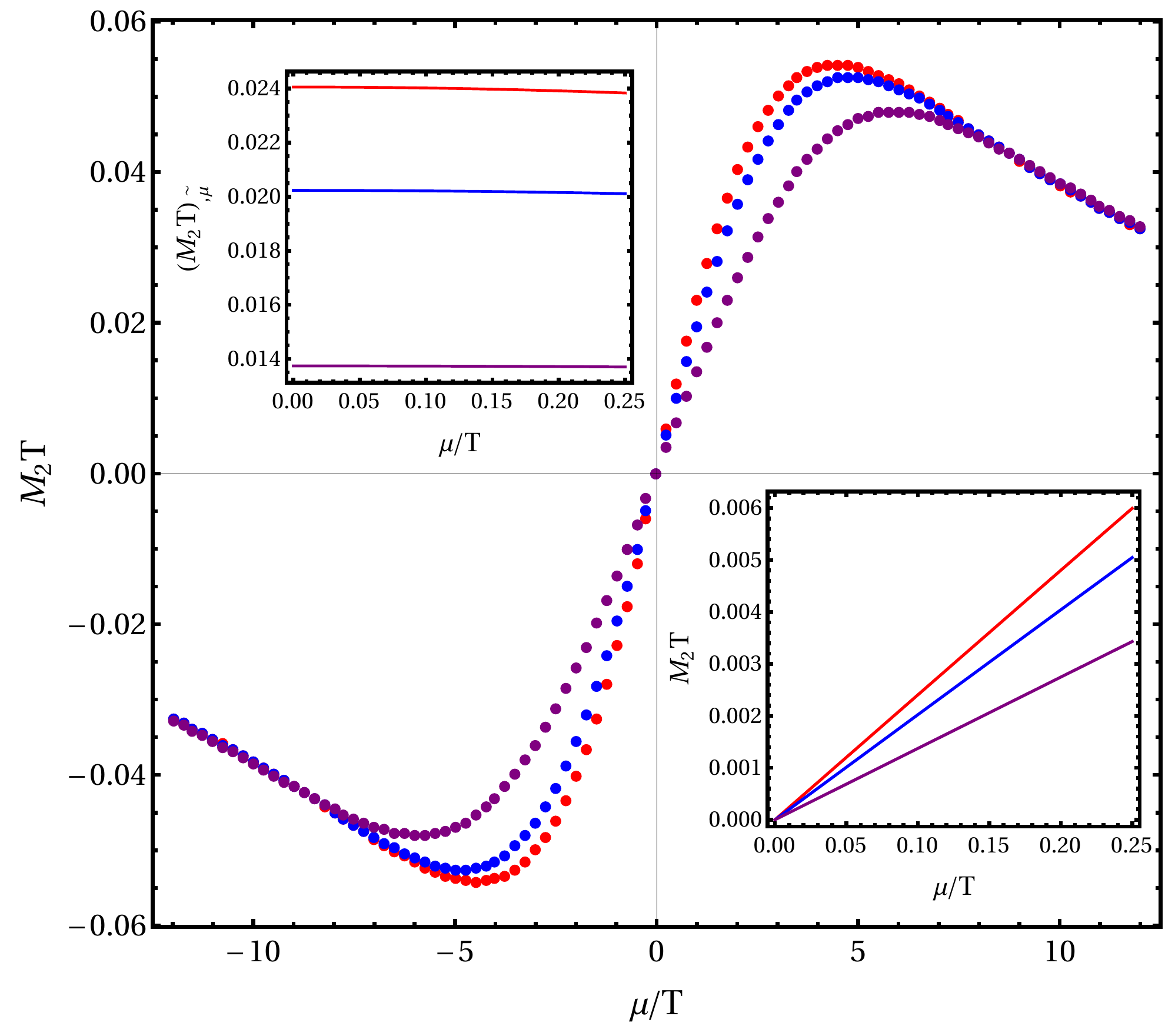}
    \caption{The dimensionless thermodynamic susceptibility $M_2 T$ as a function of the  chemical potential $\mu/T$ for  $\gamma=2/\sqrt{3}$. The different curves correspond to magnetic fields $B/T^2=\{0.05,12.5,30\}$ (red, blue, purple).}
    \label{fig:M2}
\end{figure}

Let us turn to the thermodynamic coefficient $M_2$ which has inverse energy dimension, hence we consider the dimensionless combination $M_2\cdot T$. We here determine this transport coefficient for the first time in holography. In fact, $M_2$ vanishes in the holographic Einstein-Maxwell model, i.e. for the case $\gamma=0$. Hence, in figure~\ref{fig:M2} we only present results for $\gamma=2/\sqrt{3}$.

As expected, $M_2 T$ is an odd function of the chemical potential $\tilde{\mu}.$ For small chemical potential $\tilde{\mu} < 0.25$ we approximately find a linear proportionality between $M_2 T$ and $\tilde{\mu}$ as evident from the inset in the lower right corner of figure~\ref{fig:M2}, implying $M_2 T = a \, \tilde{\mu}$, where the coefficient $a$ depends on $\tilde{B}$. 
In fact, the coefficient $a$ is given by $a\approx 0.024$ for vanishing magnetic field 
which is not related to the chiral anomaly coefficient $C=-\gamma$ in an obvious way. Finally, the value of $M_2 T$ seems to not be sensitive to $\tilde{B}$ for chemical potentials of order $\tilde{\mu}=7$ or larger -- at least for reasonably small magnetic fields $\tilde{B}$ displayed in figure~\ref{fig:M2}.

\subsubsection*{Remaining thermodynamic coefficient $M_1, M_3$ and $M_4$}
Next, we turn our attention to the remaining thermodynamic transport coefficients $M_1, M_3$ and $M_4.$ As discussed in the section on hydrodynamics there are no Kubo formulas available so far which give us directly one of the thermodynamic transport coefficients $M_i$ with $i=1,3,4.$ However, we can evaluate the Kubo relations \eqref{eq:M1M3M4Kubo} to numerically  determine $M_{4,\tilde{\mu}},$ $(M_1/T)_{,\tilde{\mu}}$ and $(M_3/T)_{,\tilde{\mu}}$    i.e. the derivative of the dimensionless thermodynamic coefficient $M_i$ with respect to the chemical potential $\tilde{\mu}.$ It turns out that all these three derivatives vanish numerically. Following the logic of section \ref{sec:commentKubo} we conclude that $M_4(\tilde{\mu},\tilde{B})=0$ in our holographic model. A similar conclusion cannot be made for the thermodynamic transport coefficients $M_1$ and $M_3$. $M_3$ is zero due to conformal invariance of theory. 

\subsection{Results: Dissipationless transport coefficients} \label{sec:dissipationlessTransportResults}
In this section we determine the dissipationless transport coefficients in our holographic model which are the Hall conductivity $\tilde\sigma_\perp$, the Hall viscosities $\tilde\eta_\perp$ and $\tilde\eta_\parallel$ as well as 
the coefficient $c_{10}$ (or alternatively $c_{17}$) which may be interpreted as a shear-induced Hall conductivity.

We first turn our attention to the  Hall viscosity $\tilde\eta_\perp$ which may be computed by the Kubo formula \eqref{eq:tildeEtaPerpKubo}. We find that this transport coefficient vanishes in our holographic model. In fact this is obvious from the gravity side since the relevant fluctuations of the metric, namely $h_{xy}$ and $h_{xx}-h_{yy}$, decouple. 

\subsubsection*{Hall conductivity $\tilde\sigma_\perp$}
\begin{figure}
\includegraphics[width=0.491 \linewidth]{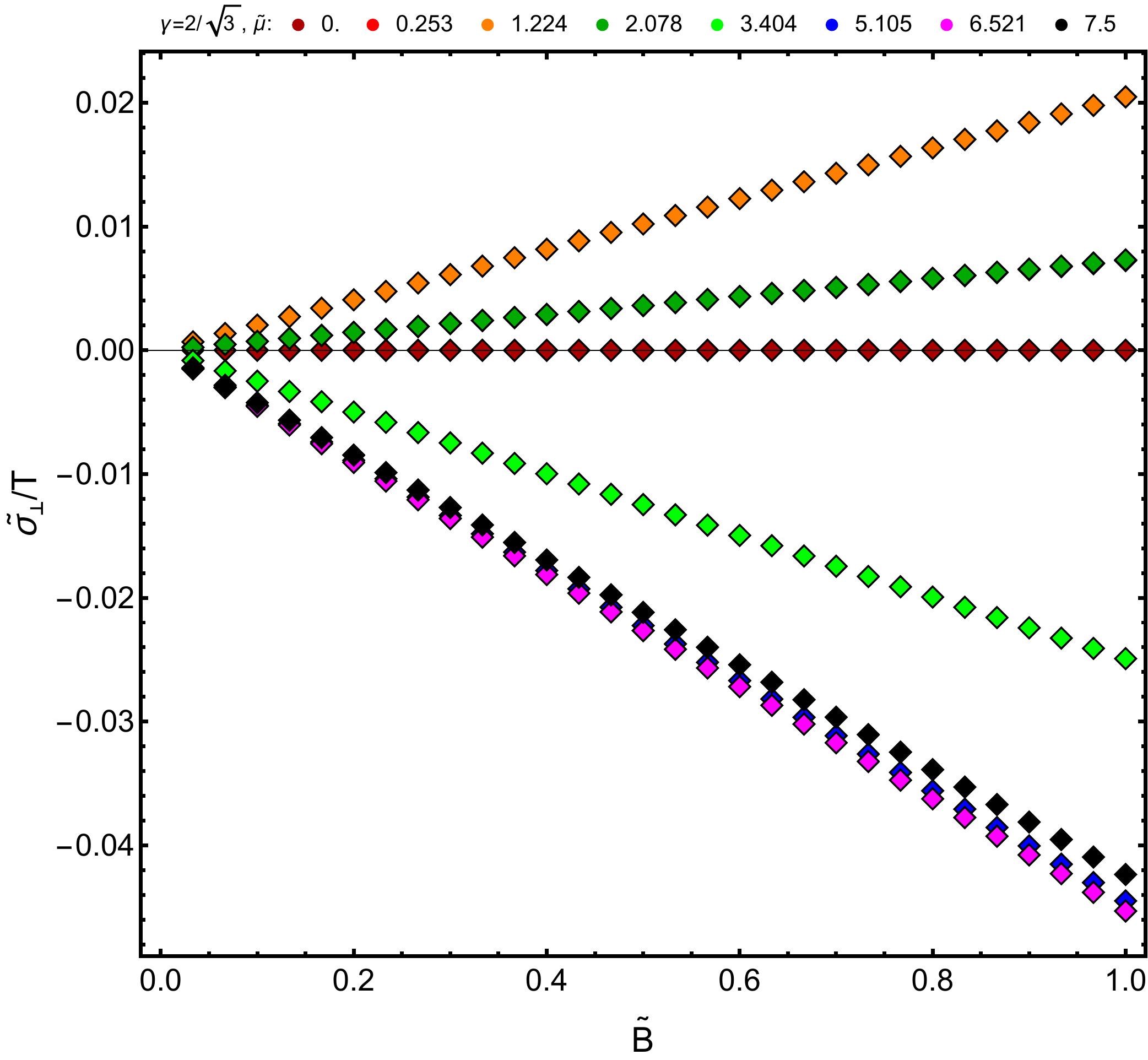}\ \includegraphics[width=0.48 \linewidth]{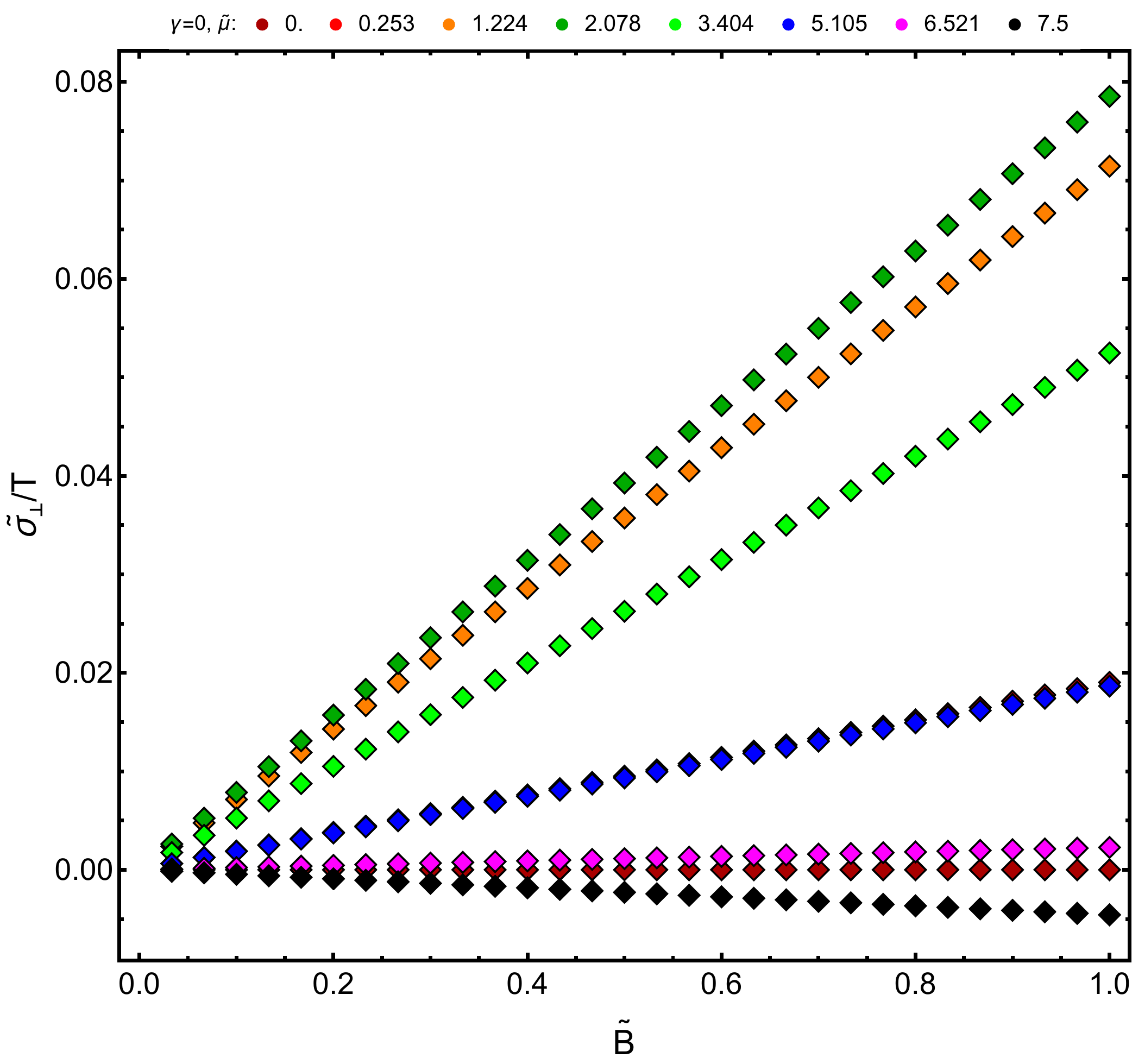}\caption{The dimensionless Hall conductivity $\tilde{\sigma}_\perp/T$ as a function of a small magnetic field $\tilde{B}$. The left panel shows the case $\gamma=2/\sqrt{3}$, while the right panel displays $\gamma=0$.  The different curves correspond to chemical potentials \mbox{$\tilde\mu\in\{0.016, 0.253, 1.224, 2.078, 3.404, 5.105, 6.521, 7.5\}$}. Note that the specific values of $\tilde \mu$ are part of a Chebychev grid in $\tilde \mu$ needed to compute the thermodynamic derivatives to high precision. For further reference 
see appendix~\ref{app:dimderiv}.
}
\label{fig:HallconsmallB}
\end{figure}

\begin{figure}
\includegraphics[width=0.48 \linewidth]{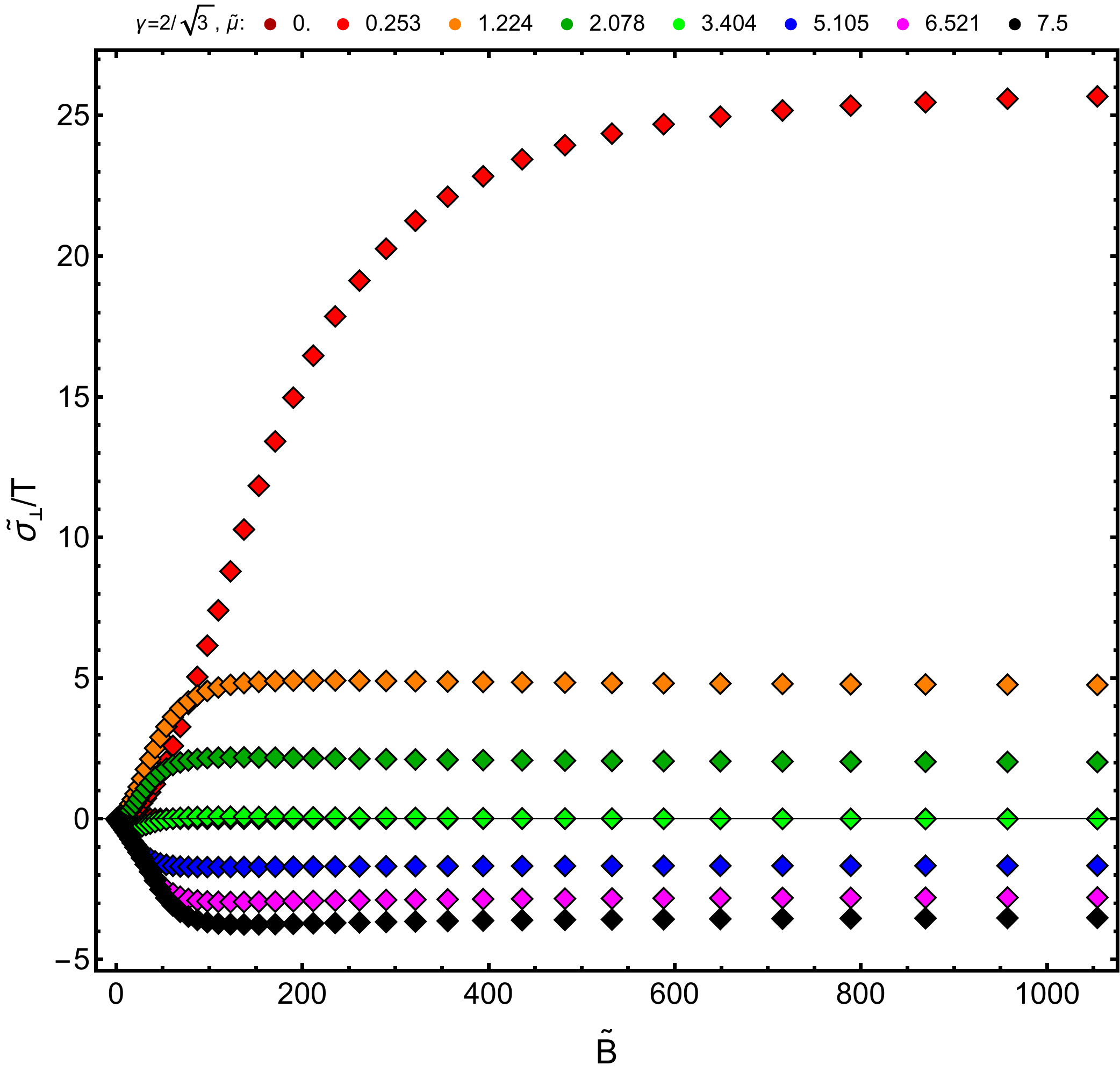}\ \includegraphics[width=0.48 \linewidth]{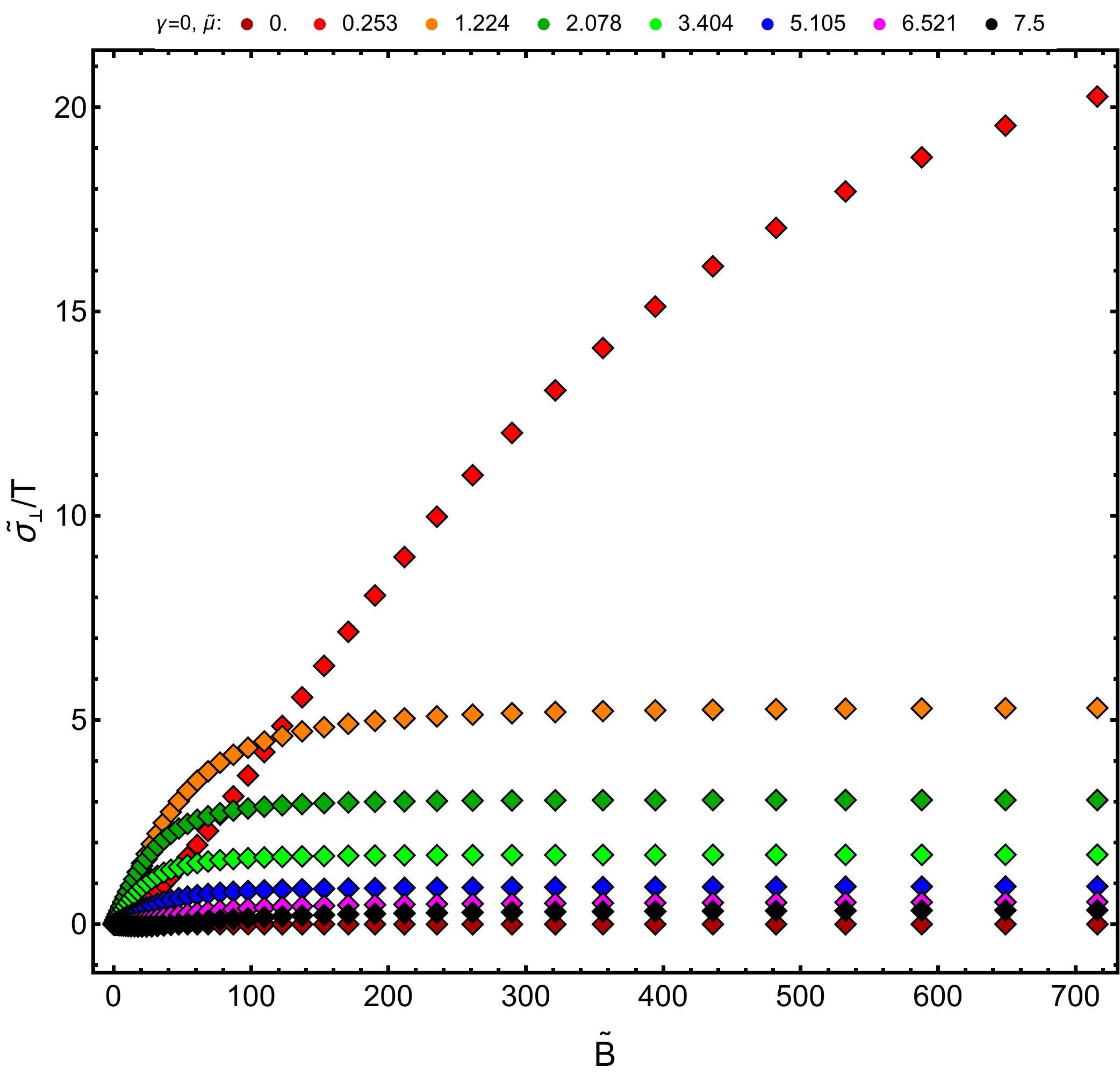}\caption{The dimensionless Hall conductivity $\tilde{\sigma}_\perp/T$ as a function of a large dimensionless magnetic field $\tilde{B}$. The left panel shows the case $\gamma=2/\sqrt{3}$, while the right panel displays $\gamma=0$. 
}
\label{fig:HallconlargeB}
\end{figure}

Next, we determine numerically the Hall conductivity $\tilde\sigma_\perp$. To do so we first employ the Kubo formulas \eqref{eq:Kubo-r2} for $\rho_\perp$ and $\tilde{\rho}_\perp$ and then invert the resistivity matrix to obtain the conductivity matrix which contains the Hall conductivity $\tilde{\sigma}_\perp$. The Hall conductivity $\tilde{\sigma}_\perp$ has dimension of temperature and hence we consider the dimensionless Hall conductivity  $\tilde{\sigma}_\perp/T$. Its behaviour for fixed chemical potential as a function of the  magnetic field is displayed in figures \ref{fig:HallconsmallB} and \ref{fig:HallconlargeB}. 
The Hall conductivity is nonzero even for vanishing chiral anomaly coefficient $C=-\gamma$ as evident from the right panels of those figures. Moreover, the Hall conductivity is only nonzero for nonzero chemical potential and nonzero magnetic field.  

We first focus on the behavior of the dimensionless Hall conductivity as a function of the magnetic field for fixed chemical potential. As evident from figure \ref{fig:HallconsmallB} the dimensionless Hall conductivity $\tilde{\sigma}_\perp/T$ is linear in $\tilde{B}$ for small magnetic fields $\tilde{B}$ and for fixed $\tilde{\mu}$, i.e. $\tilde{\sigma}_\perp/T \approx a(\tilde{\mu}) \, \tilde{B}.$ 
The proportionality constant $a(\tilde{\mu})$ displays an interesting behavior as a function of $\tilde{\mu}$ for both values of the chiral anomaly coefficient $C=-\gamma=0$ and $\gamma=2/\sqrt{3}$: first it monotonically increases with increasing chemical potential $\tilde{\mu}$, then turns around and decreases. Finally, the proportionality constant turns negative for even larger values of $\tilde{\mu}$. 

In addition, the behavior of the dimensionless Hall conductivity for large values of the magnetic field $\tilde{B}$ at fixed chemical potential $\tilde{\mu}$ is displayed in figure \ref{fig:HallconlargeB}. The dimensionless Hall conductivity approaches a $\tilde{\mu}$-dependent value for large magnetic fields. 
Moreover, we may investigate the ratio $\tilde{\sigma}_\perp/\sigma_\perp$ of the Hall conductivity and the perpendicular conductivity $\sigma_\perp$ (the latter one will be discussed in the next subsection) as a function of the magnetic field. For fixed chemical potential $\tilde{\mu},$ the ratio $\tilde{\sigma}_\perp/\sigma_\perp$ is directly proportional to the magnetic field. This linear relationship persists even for very large values of the magnetic field. The proportionality constant depends on $\tilde{\mu}$ and shows as a function of $\tilde{\mu}$ a behavior similar to that of the proportionality constant $a(\tilde{\mu})$ of the previous paragraph. Specifically, the similarities are that both proportionality constants vanish at $\tilde\mu=0$, both increase monotonically with increasing $\tilde\mu>0$, reach a maximum value, then decrease monotonically and then turn negative within the range covered by our data. 

\subsubsection*{Hall viscosity $\tilde\eta_\parallel$ and shear-induced Hall conductivity $c_{10}$}

Finally, we investigate the Hall viscosity  $\tilde\eta_\parallel$ and the coefficient $c_{10}$ which may be interpreted as a shear-induced Hall conductivity. Note that both transport coefficients turn out to be zero in case of vanishing chiral anomaly and hence these are novel transport coefficients which are determined here for the first time within a holographic model.

\begin{figure}
\includegraphics[width=0.49 \linewidth]{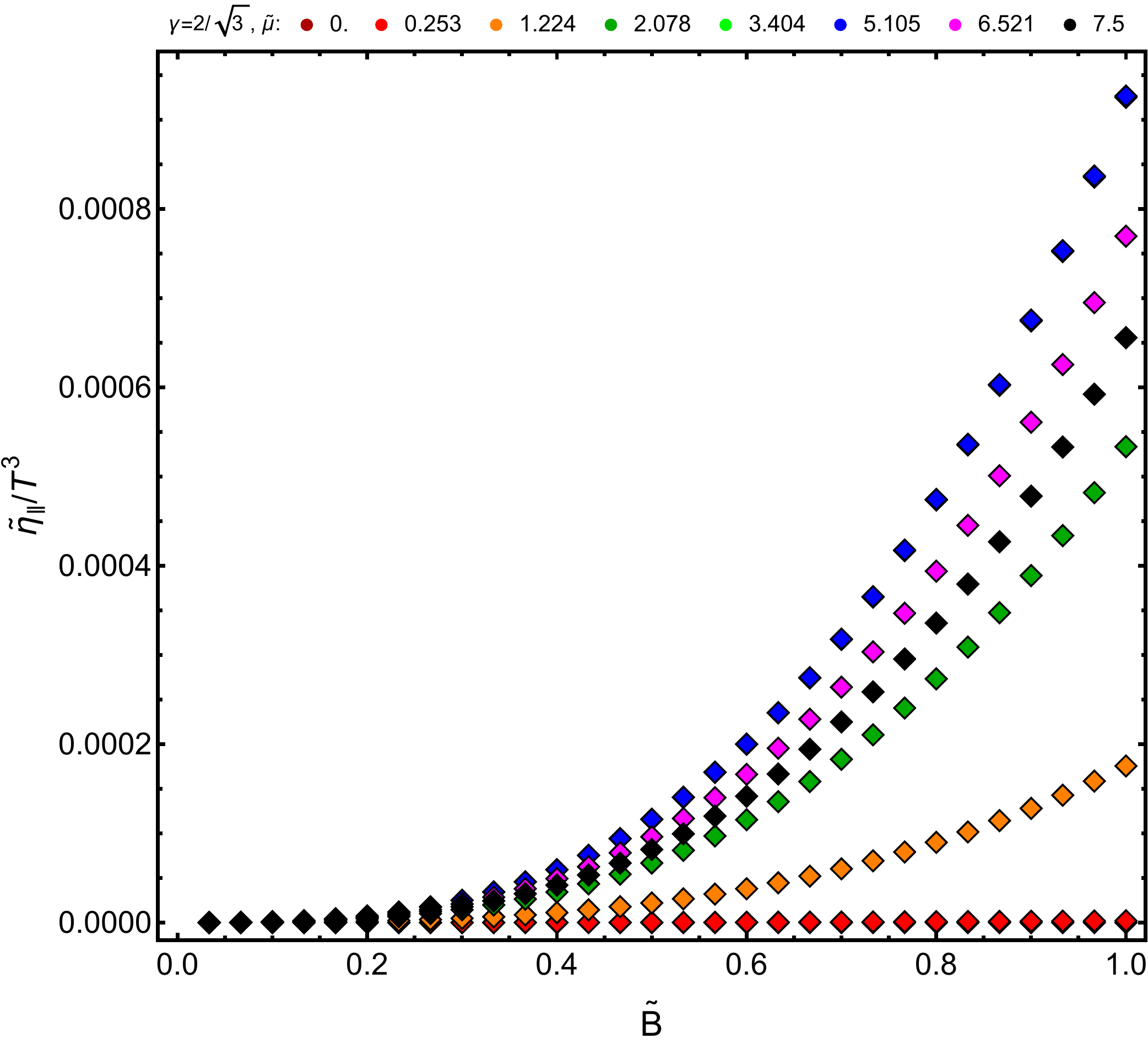}\ \includegraphics[width=0.48 \linewidth]{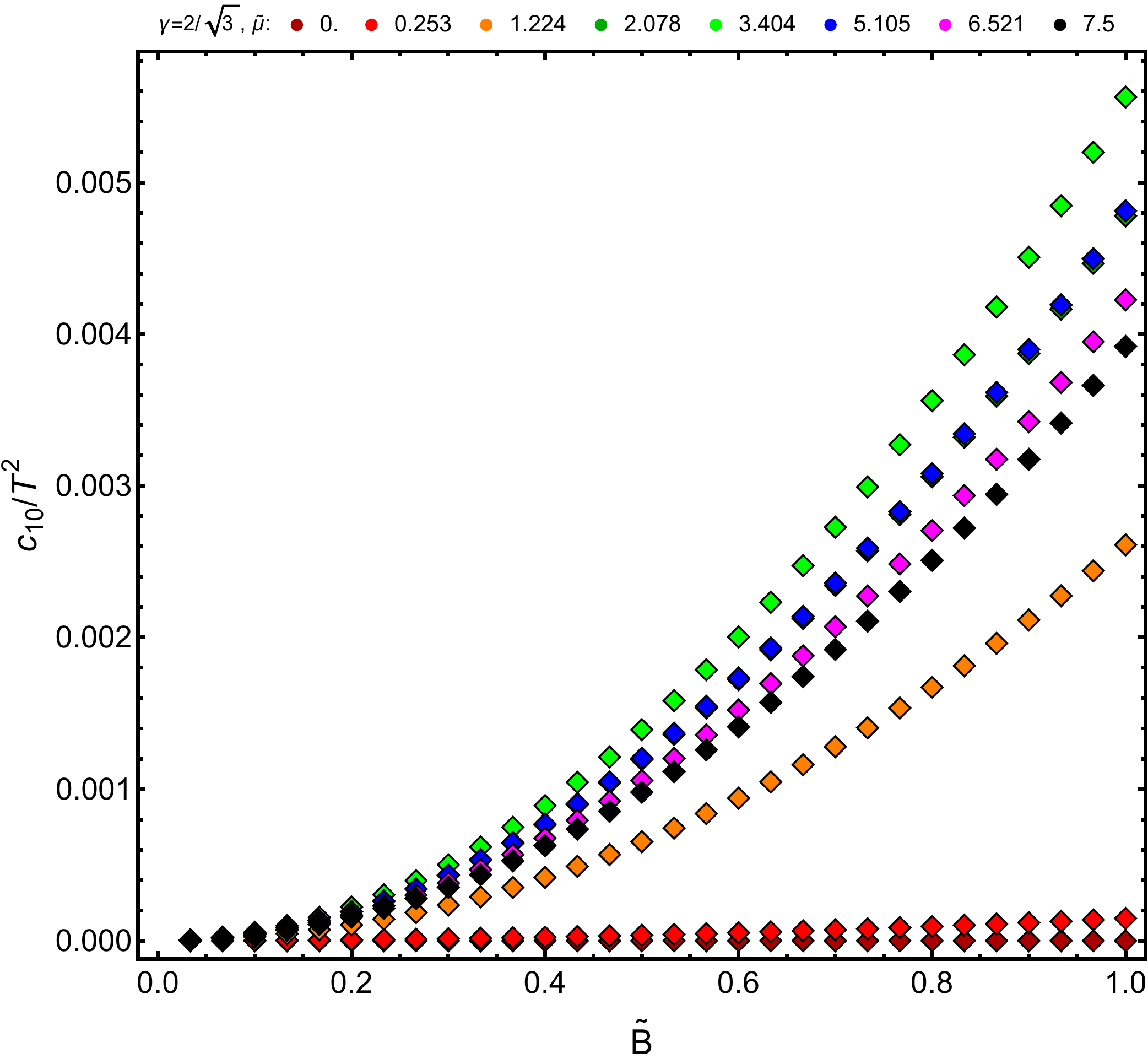}\caption{The dissipationless transport coefficients $\tilde{ \eta}_\parallel/T^3$ and $c_{10}/T^2$ in presence of the chiral anomaly $\gamma=2/\sqrt{3}$. Without chiral anomaly, both transport coefficients are zero. The green curve with $\tilde\mu=3.404$ is hidden under the blue curve.
}\label{fig:Hallvisc10smallB}
\end{figure}

\begin{figure}
\includegraphics[width=0.497 \linewidth]{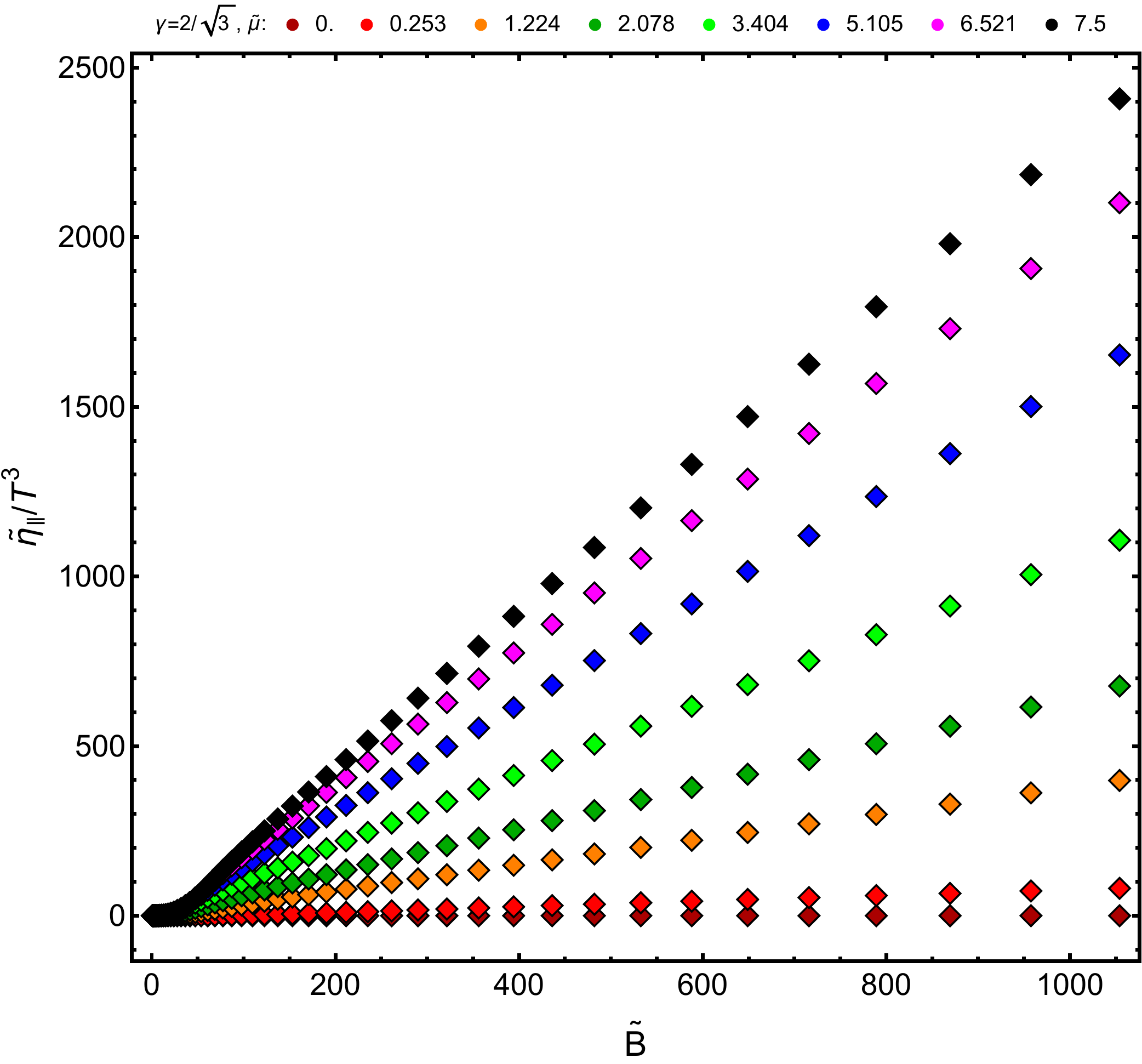}\ \includegraphics[width=0.48 \linewidth]{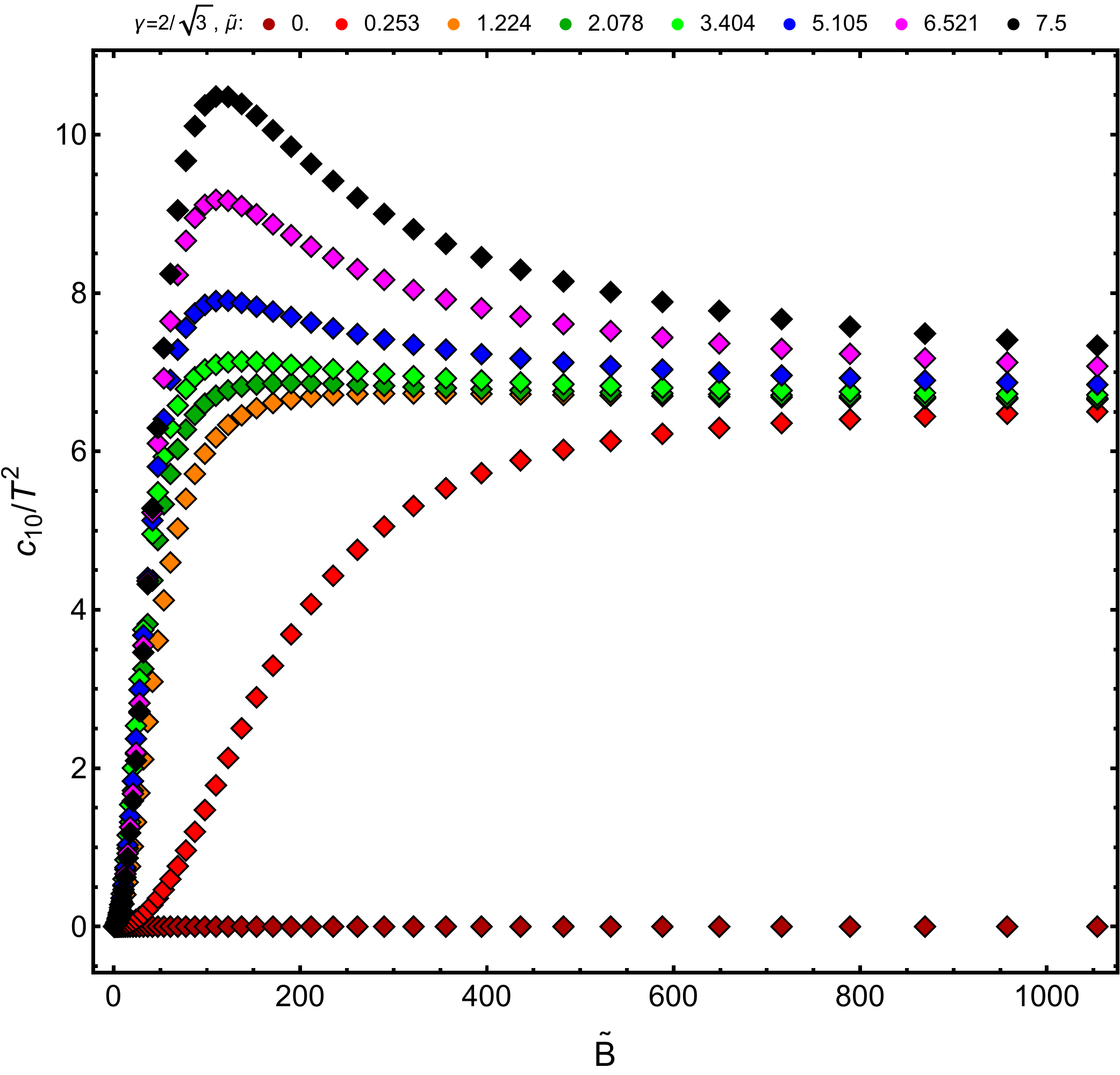}\caption{The dissipationless transport coefficients $\tilde{ \eta}_\parallel/T^3$ and $c_{10}/T^2$ in presence of the chiral anomaly $\gamma=2/\sqrt{3}$. Without chiral anomaly, both transport coefficients are zero.}\label{fig:Hallvisc10largeB}
\end{figure}

Figures \ref{fig:Hallvisc10smallB} and \ref{fig:Hallvisc10largeB} display dimensionless versions of the Hall viscosity $\tilde\eta_\parallel$ and of $c_{10}$ as a function of the magnetic field for fixed chemical potential. Note that the Hall viscosity  $\tilde\eta_\parallel$ has units of temperature $T^3$, and hence $\tilde\eta_\parallel/T^3$ is dimensionless.\footnote{We may instead plot the dimensionless quantity $\tilde\eta_\parallel/s$ where $s$ is the entropy density instead as it is usually done for viscosities.} Moreover, $c_{10}$ has the same units as $T^2$ implying that $c_{10}/T^2$ is dimensionless.

As evident from figures \ref{fig:Hallvisc10smallB} and \ref{fig:Hallvisc10largeB} both dimensionless quantities $\tilde\eta_\parallel/s$ and $c_{10}/T^2$ are only nonzero in the presence of both a non-vanishing magnetic field and a non-vanishing chemical potential. Moreover, note that both quantities are also positive (at least for the magnetic fields and chemical potentials investigated in this paper) even though these quantities are not constrained by the entropy positivity argument from hydrodynamics. 

We first investigate the behaviour of the Hall viscosity and the novel shear-induced Hall conductivity for small magnetic fields $\tilde{B}$. As indicated 
in figure \ref{fig:Hallvisc10smallB} the dimensionless quantities $\tilde\eta_\parallel/T^3$ and $c_{10}/T^2$ show the following scaling laws with the magnetic field for fixed chemical potential:
\begin{equation}
\label{eq:c10tildeEtaFits}
    \tilde\eta_\parallel/T^3 \sim \tilde{B}^3 \, , \qquad \textrm{and} \qquad c_{10}/T^2 \sim \tilde{B}^2 \, .
\end{equation}
We defer the plots with the fit curves superimposed, figure~\ref{fig:fitPlotAppendix}, to the appendix.  
The behavior of the corresponding $\tilde{\mu}$-dependent proportionality constants are not very illuminating; both proportionality constants monotonically increase for small chemical potential $\tilde{\mu}$, while they monotonically decrease for larger values of $\tilde{\mu}$.

Figure \ref{fig:Hallvisc10largeB} displays the dissipationless Hall viscosity and shear-induced Hall conductivity for fixed chemical potential and large magnetic fields. In particular, we find that the dissipationless Hall viscosity is linear in $\tilde{B}$ as shown in the left panel of figure \ref{fig:Hallvisc10largeB}.  In fact, we find that the proportionality constant between $\tilde{\eta}_\parallel / T^3$ and $\tilde{B}$ is linear in $\tilde{\mu}$, hence giving us the following universal result for the holographic model considered here:
\begin{equation}
\frac{\tilde{\eta}_\parallel}{T^3} \approx 0.305 \, \tilde{\mu} \, \tilde{B} \, . 
\end{equation}
Within the full parameter range covered by our numerical data, the above approximation is valid for sufficiently large $\tilde{\mu} \tilde{B}$.  
The shear-induced Hall conductivity for large magnetic fields $\tilde{B}$ is shown in the right panel of figure \ref{fig:Hallvisc10largeB}.  It is also tempting to speculate that the novel dimensionless shear-induced Hall conductivity, $c_{10}/T^2$, approaches a value for large magnetic fields which is independent of the chemical potential.

\subsection{Results: Dissipative hydrodynamic transport coefficients}
In this section we investigate the dissipative transport coefficients in our holographic model which may be grouped into 
\begin{itemize}
    \item components of the shear viscosity tensor, namely $\eta_\perp$, $\eta_\parallel$, $\eta_1$ and $\eta_2$, as well as the components of the bulk viscosity tensor, namely $\zeta_1$ and $\zeta_2$,
    \item dissipative components of the conductivity tensor, namely the longitudinal and perpendicular conductivities $\sigma_\parallel$ and $\sigma_\perp$,
    \item novel transport coefficients $c_4$ and $c_5$, as well as $c_8$ which in our model are only nonzero in the presence of a nonzero chiral anomaly coefficient $C=-\gamma \neq 0$, magnetic field $B \neq 0$, and chemical potential $\mu \neq 0$. (In our model $c_3=0$ for any values of $\gamma, \,B,\, \mu$.)
\end{itemize}
\subsubsection*{Viscosities}
We first focus on the various components of the shear viscosity tensor. The perpendicular shear viscosity $\eta_\perp$ satisfies $\eta_\perp/s = 1/(4\pi)$. 
Note that both $\eta_\perp/T^2$ and $s/T^2$ change as functions of the magnetic field. Remarkably, their functional dependence on $B$ is identical and cancels in the ratio $\eta_\perp/s$. 
This can be shown analytically even in the presence of the chiral anomaly as well as in the presence of $B$ and $\mu$, following~\cite{Critelli:2014kra}.
\begin{figure}
       \centering
    \includegraphics[width=7.2cm]{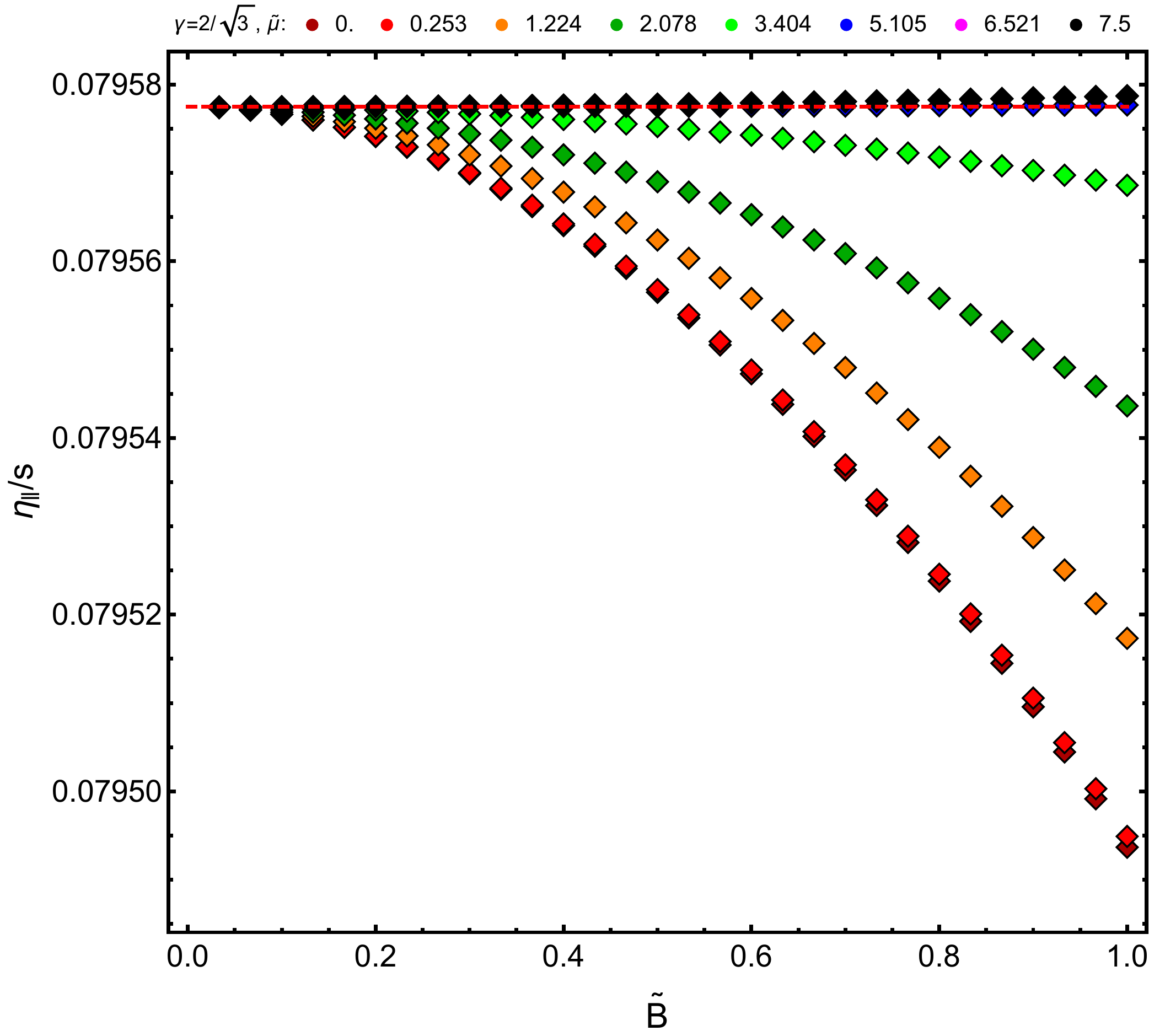}\hspace{0.5cm} \includegraphics[width=7.2cm]{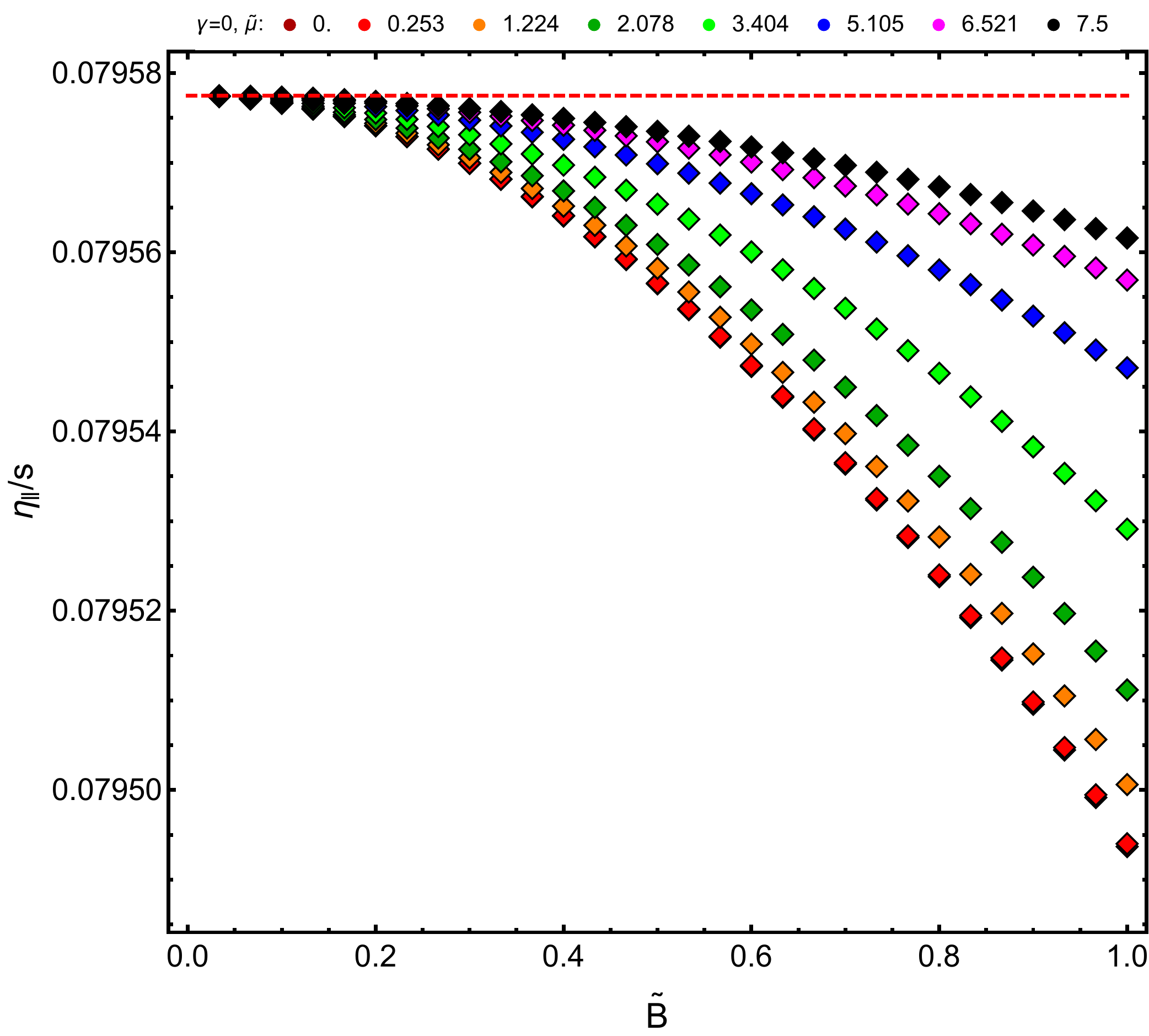}\vspace{-0.1cm}
    \caption{
    Parallel shear viscosity as function of $\tilde B$ for fixed values of $\tilde\mu$. The dashed red line indicates the well-known result for $\eta_\perp/s$, namely $1/(4\pi).$ Left: $\gamma=2/\sqrt{3}$. For this supersymmetric value of $\gamma$, the ratio of parallel shear viscosity increases with $\tilde B$ for $\tilde\mu\ge5$ within the range of parameters displayed here. 
    See figure~\ref{fig:sp1etaLarge} for a refined statement. 
    Right: $\gamma=0$. 
    }
  \label{fig:sp1etaSmall}
\end{figure}

\begin{figure}
    \centering
 \includegraphics[width=7.52cm]{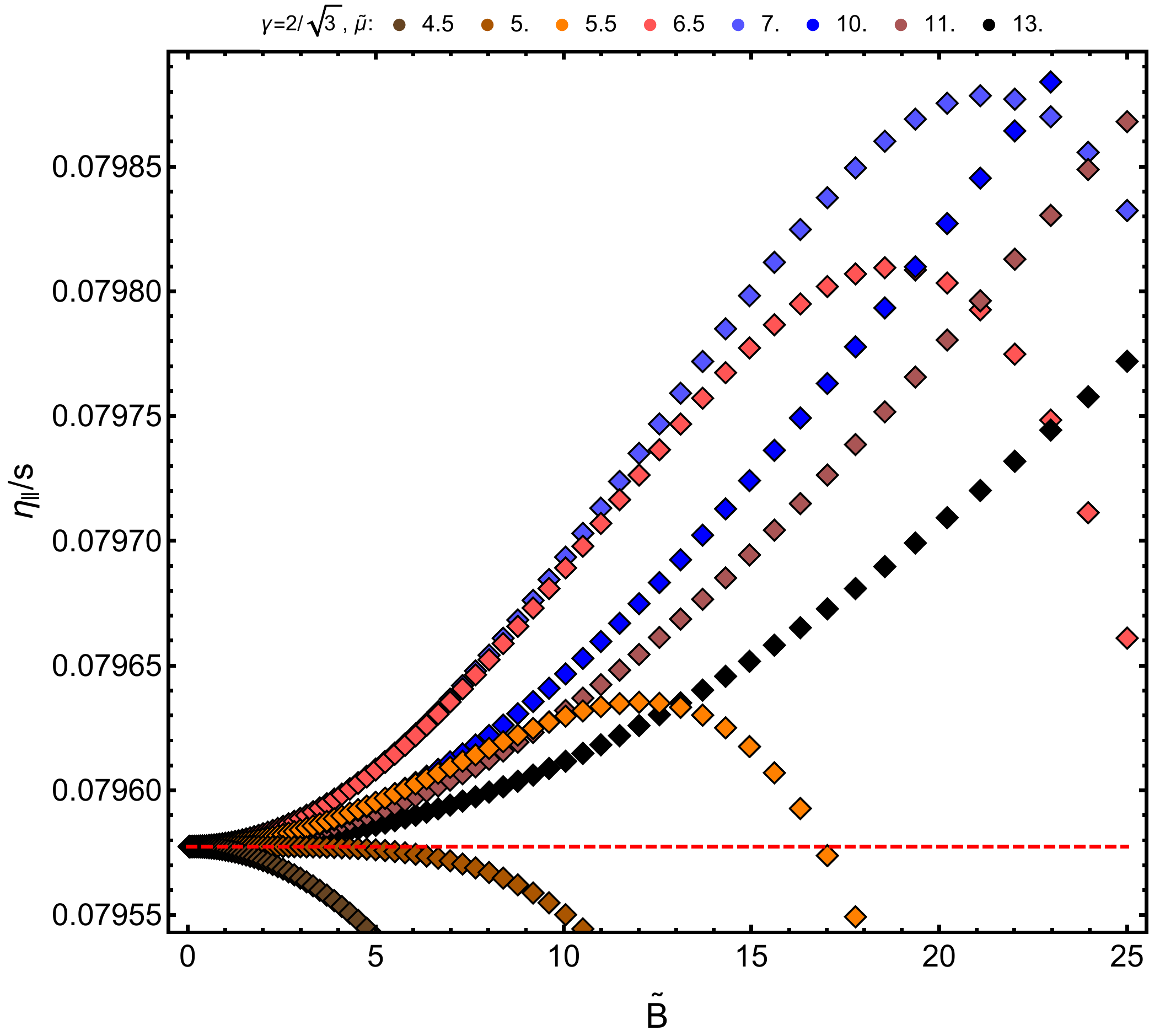}\hspace{0.1cm} \includegraphics[width=7.2cm]{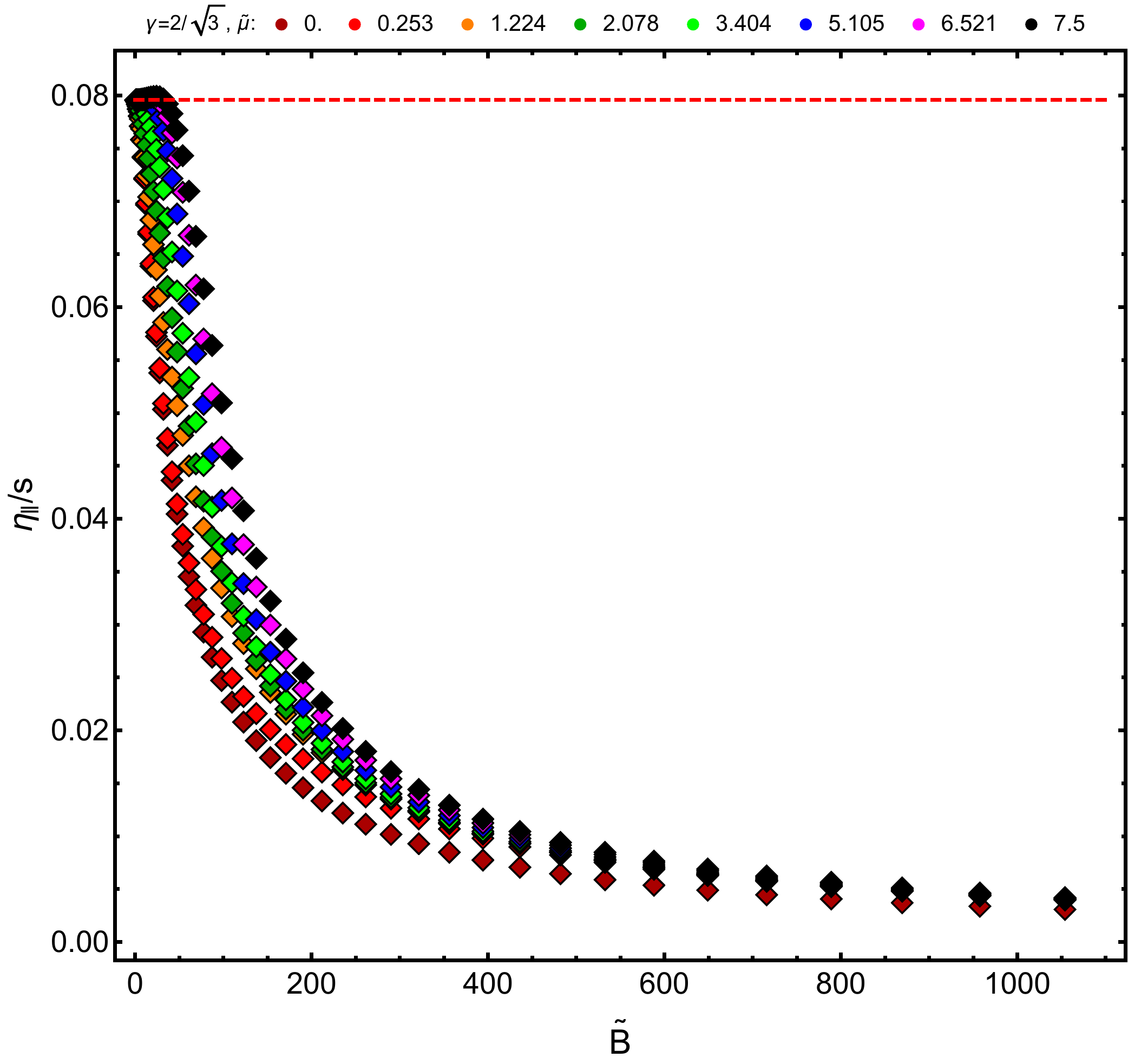}
    \caption{Ratio of parallel shear viscosity and entropy density as function of $\tilde B$ with $\gamma=2/\sqrt{3}$; 
    in the left plot the color scheme distinct from right since we use a different set of $\tilde\mu$. We observe that $\eta_\|/s$ increases initially for $\tilde\mu\ge5$. This behavior is only present with the chiral anomaly. Eventually, all curves tend to zero as displayed in the right figure. 
    }
    \label{fig:sp1etaLarge}
\end{figure}

The parallel shear viscosity $\eta_\parallel/s$ depends non-trivially on the magnetic field and the chemical potential. The dimensionless parallel shear viscosity $\eta_\parallel /s$ is shown in figure \ref{fig:sp1etaSmall} for small values of the magnetic field $\tilde{B}$  and in figure \ref{fig:sp1etaLarge} for medium and large magnetic fields. It is evident from both figures that $\eta_\parallel$ is positive 
as implied by the hydrodynamic stability analysis. Moreover, for vanishing magnetic field, $\eta_\parallel/s$ takes the value $1/(4\pi)$ as expected. Let us first investigate the behavior for small magnetic fields $\tilde{B}$ displayed in figure \ref{fig:sp1etaSmall}. The dimensionless ratio $\eta_\parallel/s$ deviates from $1/(4\pi)$ quadratically, i.e. 
\begin{equation}\label{eq:scalingetaparallels}
\frac{\eta_\parallel}{s} \sim \frac{1}{4\pi} - c(\tilde{\mu}) \tilde{B}^2 \, ,
\end{equation}
where $c(\tilde{\mu})$ is a model-dependent coefficient. In fact, $c(\tilde{\mu})$ is positive for vanishing chiral anomaly and monotonically decreases with increasing $\tilde{\mu}$; in contrast, for $\gamma=2/\sqrt{3}$ the coefficient $c(\tilde{\mu})$ turns negative for $\tilde{\mu}>5$. In other words, for those values of $\tilde{\mu}$ the ratio $\eta_\parallel/s$ increases for small and intermediate magnetic fields $\tilde{B}$, as evident from the right panel of figure \ref{fig:sp1etaLarge}. However, for large magnetic fields $\tilde{B}$ the dimensionless parallel shear viscosity $\eta_\parallel/s$ decreases monotonically to zero. This is true for both values of the chiral anomaly. 

We also investigated whether we can predict the transport coefficient $\eta_\parallel/s$ by using horizon data of the charge magnetic black brane. In fact, extending the analysis of~\cite{Critelli:2014kra} 
we were able to show that this is indeed the case for vanishing chiral anomaly coefficient and in the presence of a magnetic field and a chemical potential. However, in the case of non-vanishing chiral anomaly coefficient we were not able to find an horizon formula for $\eta_\parallel/s$. 

\begin{figure}
    \centering
    \includegraphics[width=7.2cm]{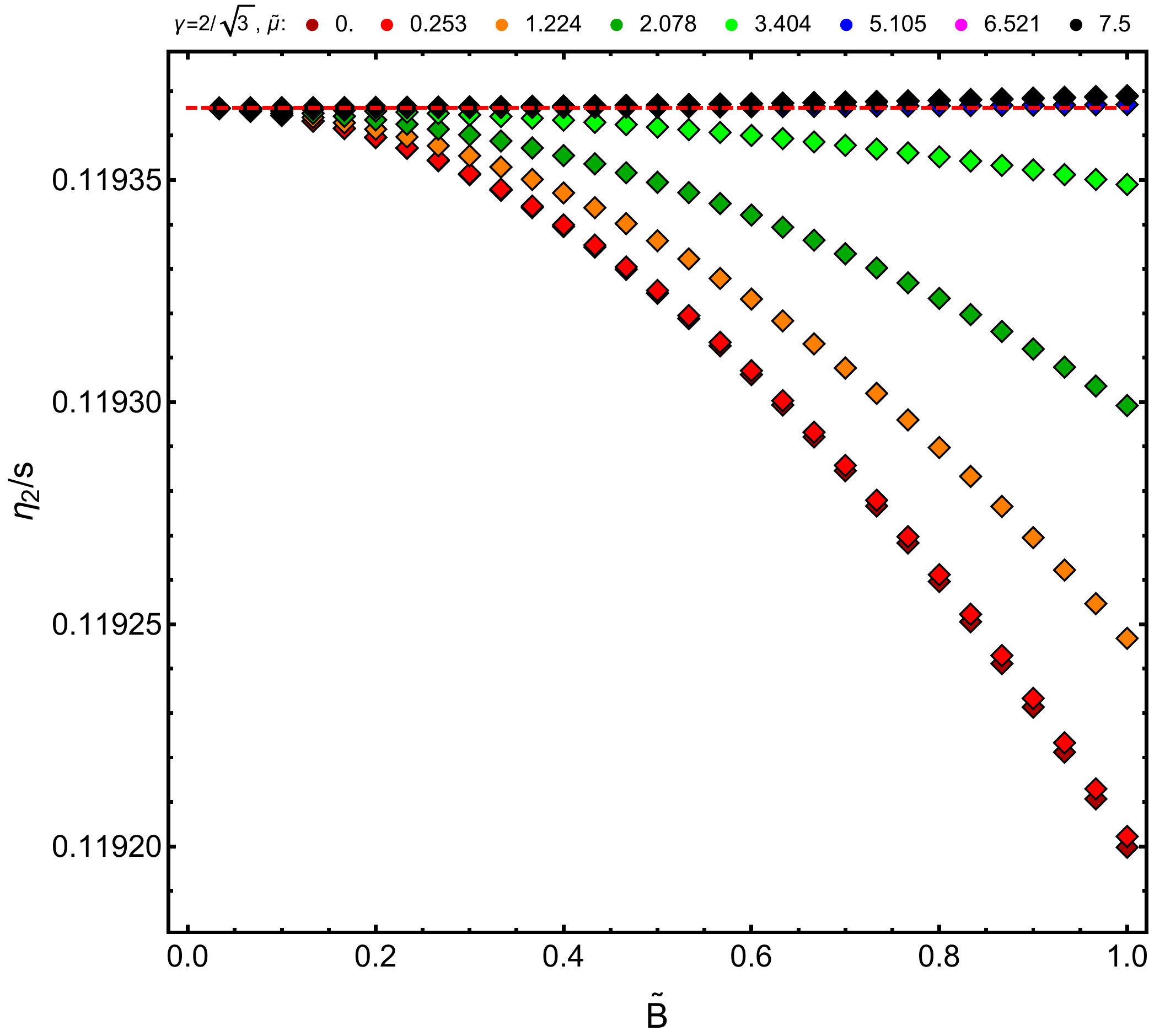}\hspace{0.5cm} \includegraphics[width=7.2cm]{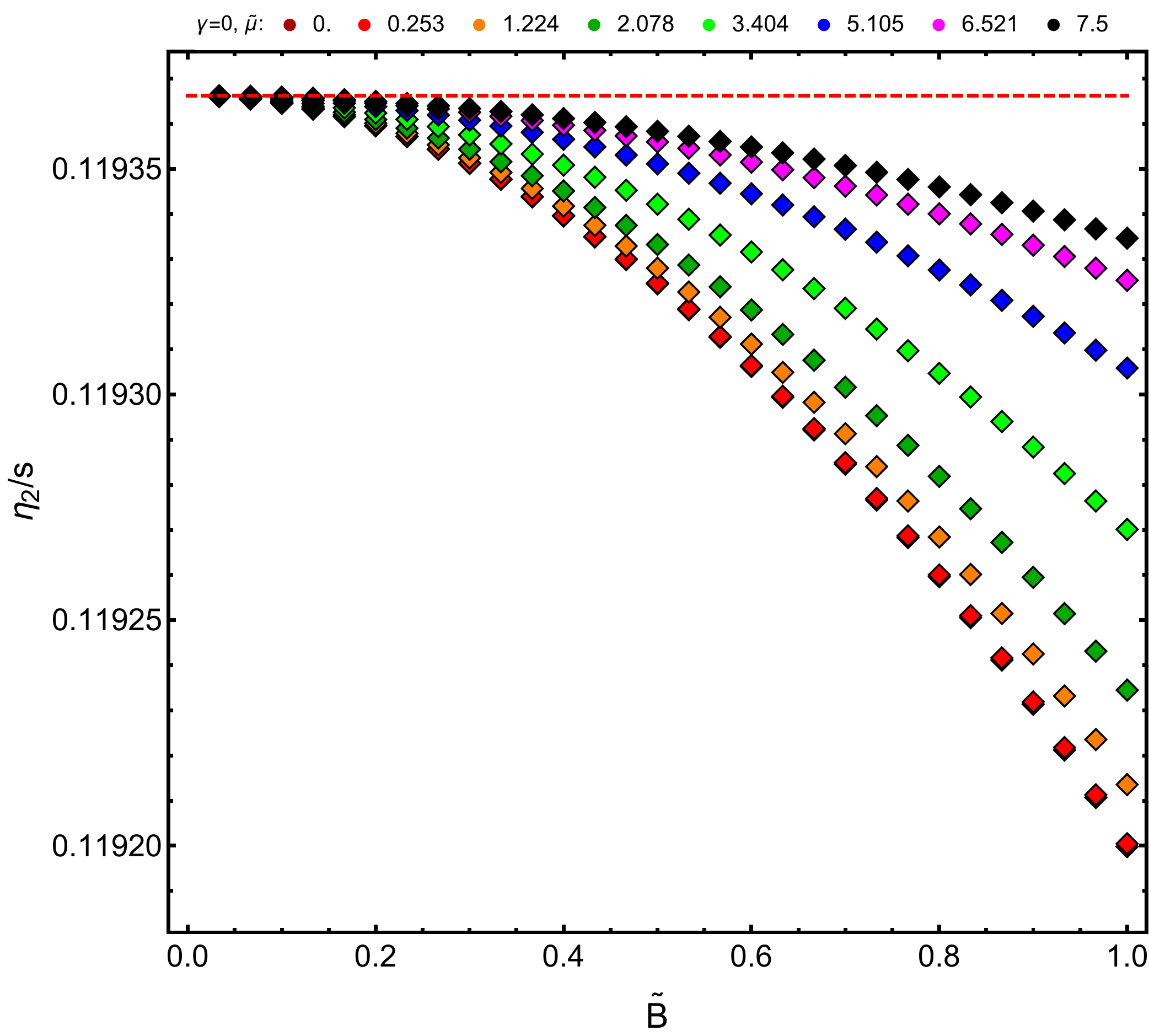}\vspace{-0.1cm}
    \caption{Dimensionless ratio of bulk viscosity $\eta_2$ and entropy density for fixed chemical potentials $\tilde\mu$. The red dashed line indicates the value for $\eta_2/s$ at zero magnetic field, namely $3/(8\pi).$ Note the behavior similar to $\eta_{\parallel}/s$, namely that for $\gamma=2/\sqrt{3}$ there is a distinctly different behavior from the case $\gamma=0$: for the largest three displayed chemical potentials, i.e. for $\tilde\mu\ge5$, the value of the transport coefficient $\eta_2/s$ increases with $\tilde B$, while it always decreases for $\gamma=0$.
}
  \label{fig:eta2small}
\end{figure}

Next, we turn to the dissipative components $\eta_1$ and $\eta_1$ of the shear viscosity tensor. It suffices to display only $\eta_2$ since the bulk viscosity $\zeta_2$ vanishes in our holographic model. Hence, the transport coefficients $\eta_1$ and $\eta_2$ are related by $\eta_2 = - \frac{1}{3} \eta_1$ due to the Onsager relation \eqref{eq:ons-pv2}. The behavior of the dimensionless ratio $\eta_2/s$ as a function of the magnetic field $\tilde{B}$ is depicted in figure \ref{fig:eta2small}.\footnote{Note that we do not display the transport coefficient $\eta_2/s$ for large magnetic fields since we expect contributions from the thermodynamic transport coefficients $M_1, M_3$ and $M_4$ which we cannot determine in our holographic model.} As expected, the ratio $\eta_2 /s =3/(8\pi)$ for vanishing magnetic field and quadratically deviates from that value for small magnetic fields up to $\tilde{B} = 1$. We obtain scaling law similar to the one for $\eta_\parallel/s,$ see equation  \eqref{eq:scalingetaparallels}. The analogous coefficient $c(\tilde{\mu})$ qualitatively 
shows the same behavior as in the case of $\eta_\parallel/s:$ it monotonically decreases for $\gamma=0$ but stays positive; in contrast for $\gamma=2/\sqrt{3}$ it is not monotonic for the whole range of chemical potentials $\tilde{\mu}$ and even turns negative for intermediate values of $\tilde{\mu}.$

Finally, we numerically checked that the bulk viscosities $\zeta_1$ and $\zeta_2$ vanish in our model.

\subsubsection*{Conductivities} 
Next, we investigate the dissipative components of the conductivity tensor, namely the longitudinal conductivity $\sigma_\parallel$ and the perpendicular component $\sigma_\perp$. Both quantities have units of temperature, implying that $\sigma_\perp/T$ and $\sigma_\parallel/T$ are dimensionless. Both dimensionless quantities are almost insensitive to the magnetic field and to $\gamma$. In fact, for both values of the chiral anomaly coefficient and within the range covered by our numerical data, that is $\tilde\mu$ between $0$ and $7.5$ $\sigma_{\parallel}/T$ has a relative variation of order $10^{-3}$ for $\tilde{B}<1.$

\begin{figure}
    \centering
 \includegraphics[width=7cm]{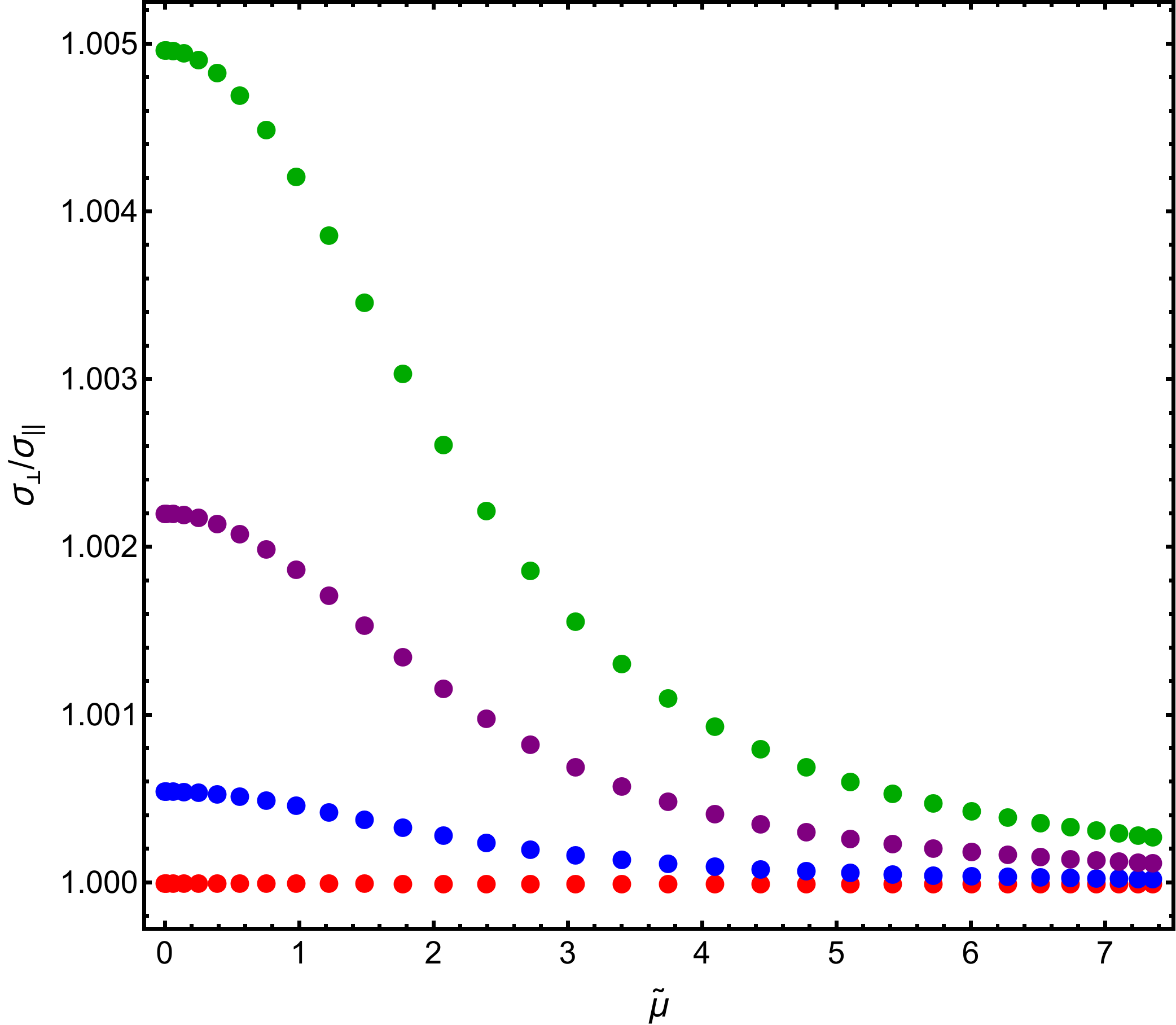}
 \hspace{0.5cm}\includegraphics[width=7cm]{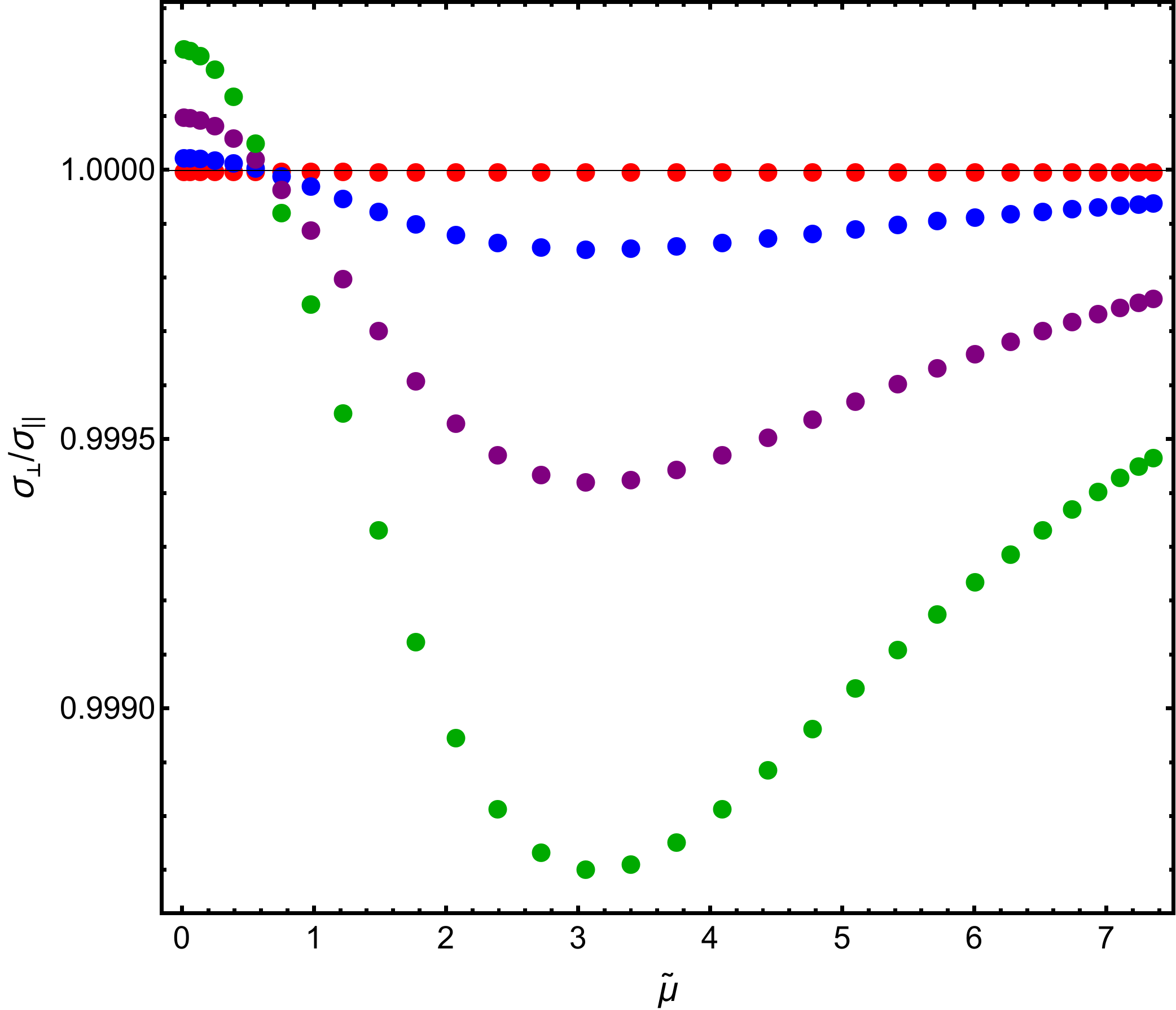}
    \caption{
    The ratio of longitudinal and perpendicular conductivity $\sigma_\perp/\sigma_{||}$ as function of $\tilde\mu$ at fixed $\tilde B$ for   $\gamma=2/\sqrt{3}$ (left) and $\gamma=0$ (right). Each color corresponds to a fixed value of $\tilde B$ given by $\tilde B=\{1/30,1/3,2/3,1\}$ (red, blue, purple, dark green).}
    \label{fig:ratiosigma}
\end{figure}

According to our numerical data, $\sigma_\perp/\sigma_{||}$ changes only slightly (at most by $10^{-2}$) in the range $\tilde B=0.0, ..., 1.0$ where our Kubo formula is valid; see derivation in section~\ref{sec:hydrodynamics}. Due to systematic and numerical errors for this particular ratio it is not possible to consider this change significant. 
In figure \ref{fig:ratiosigma} we depict the ratio $\sigma_\perp / \sigma_\parallel$ which is a measure of the anisotropy of the system. The left panel shows the case $\gamma=2/\sqrt{3}$. Note that in this case the ratio is always larger than one. In contrast, in the case of vanishing chiral anomaly which is shown in the right panel of figure \ref{fig:ratiosigma} the ratio may be also less than one, depending on the value of the chemical potential. In both cases, i.e. for $\gamma=0$ and for $\gamma=2/\sqrt{3}$ the ratio $\sigma_\perp / \sigma_\parallel$ deviates from one quadratically in $\tilde{B},$ i.e.
\begin{equation}
    \frac{\sigma_\perp}{\sigma_\parallel} = 1 + \tilde{c}(\tilde{\mu}) \, \tilde{B}^2 \, ,
\end{equation}
which holds up to magnetic fields of order $\tilde{B}=1.$

\subsubsection*{Novel transport coefficients $c_3$, $c_4$ and $c_5$ as well as $c_8$} 
\begin{figure}
    \centering
    \includegraphics[width=7.2cm]{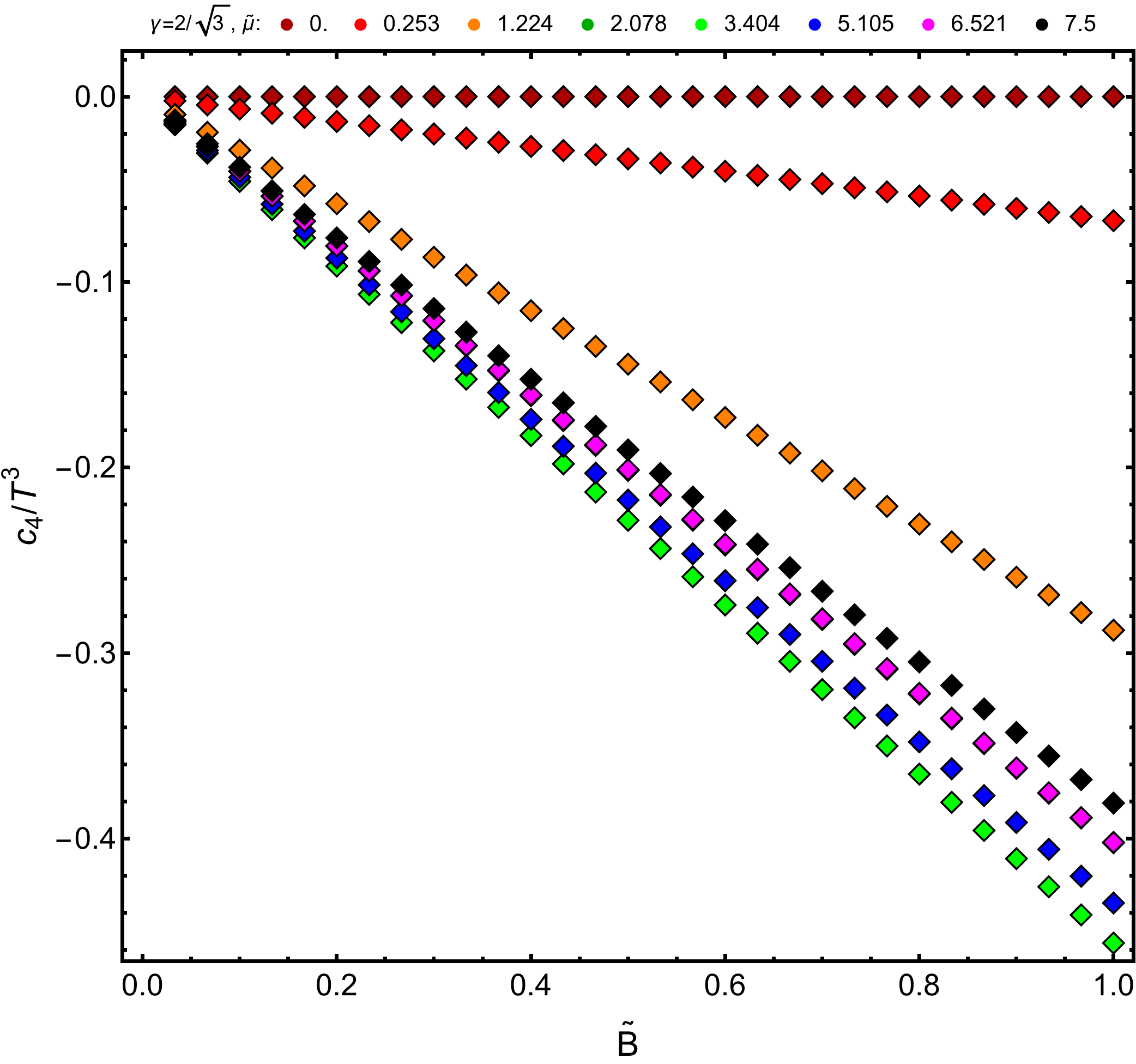}\hspace{0.5cm}
    \caption{The transport coefficient $c_{4}/T^3$ in presence of the chiral anomaly $\gamma=2/\sqrt{3}$. Without chiral anomaly or for vanishing chemical potential, the transport coefficients is zero.}
  \label{fig:c4versusB}
\end{figure}
\begin{figure}
    \centering
    \includegraphics[width=7.43cm]{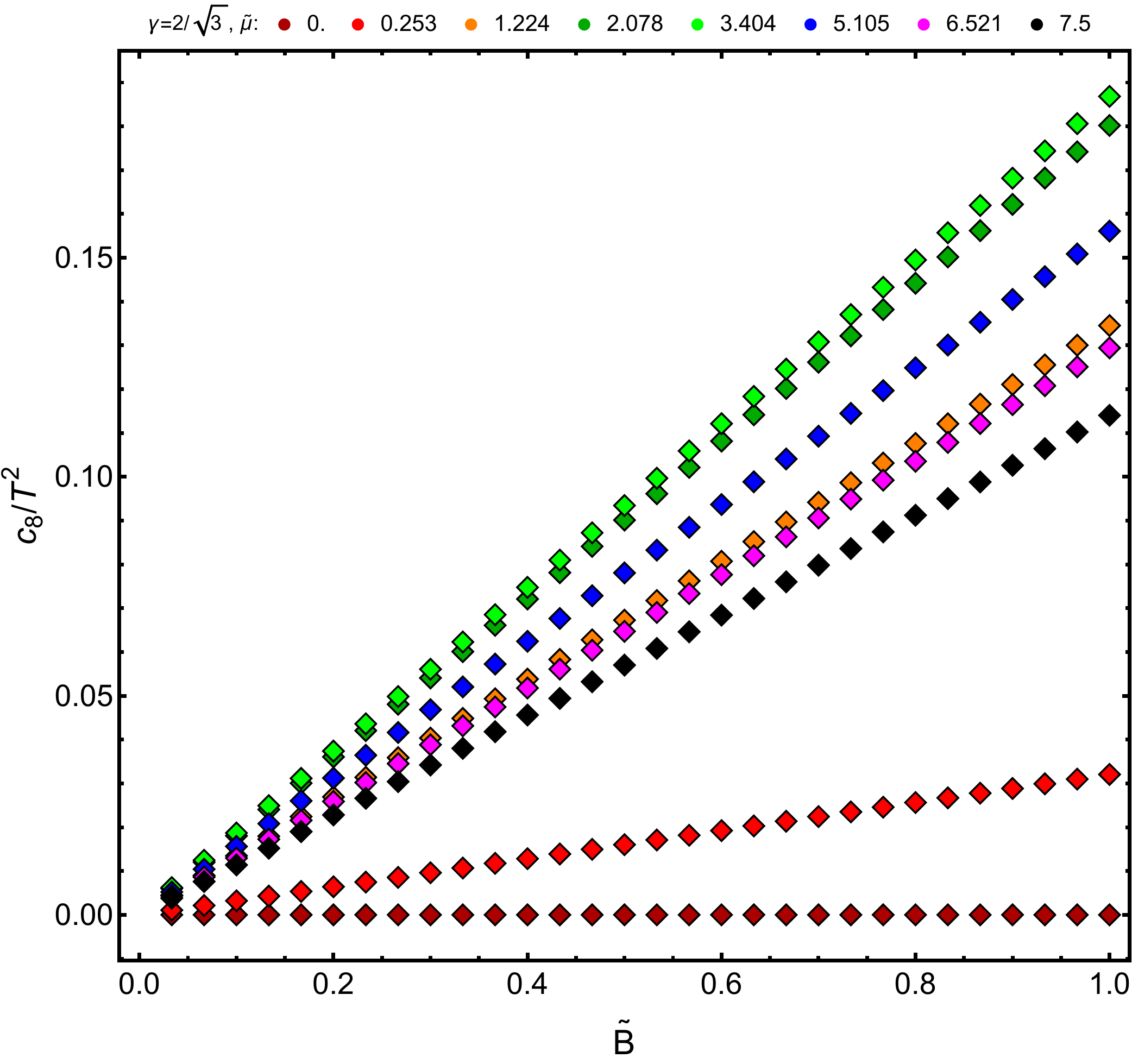}\hspace{0.12cm} \includegraphics[width=7.2cm]{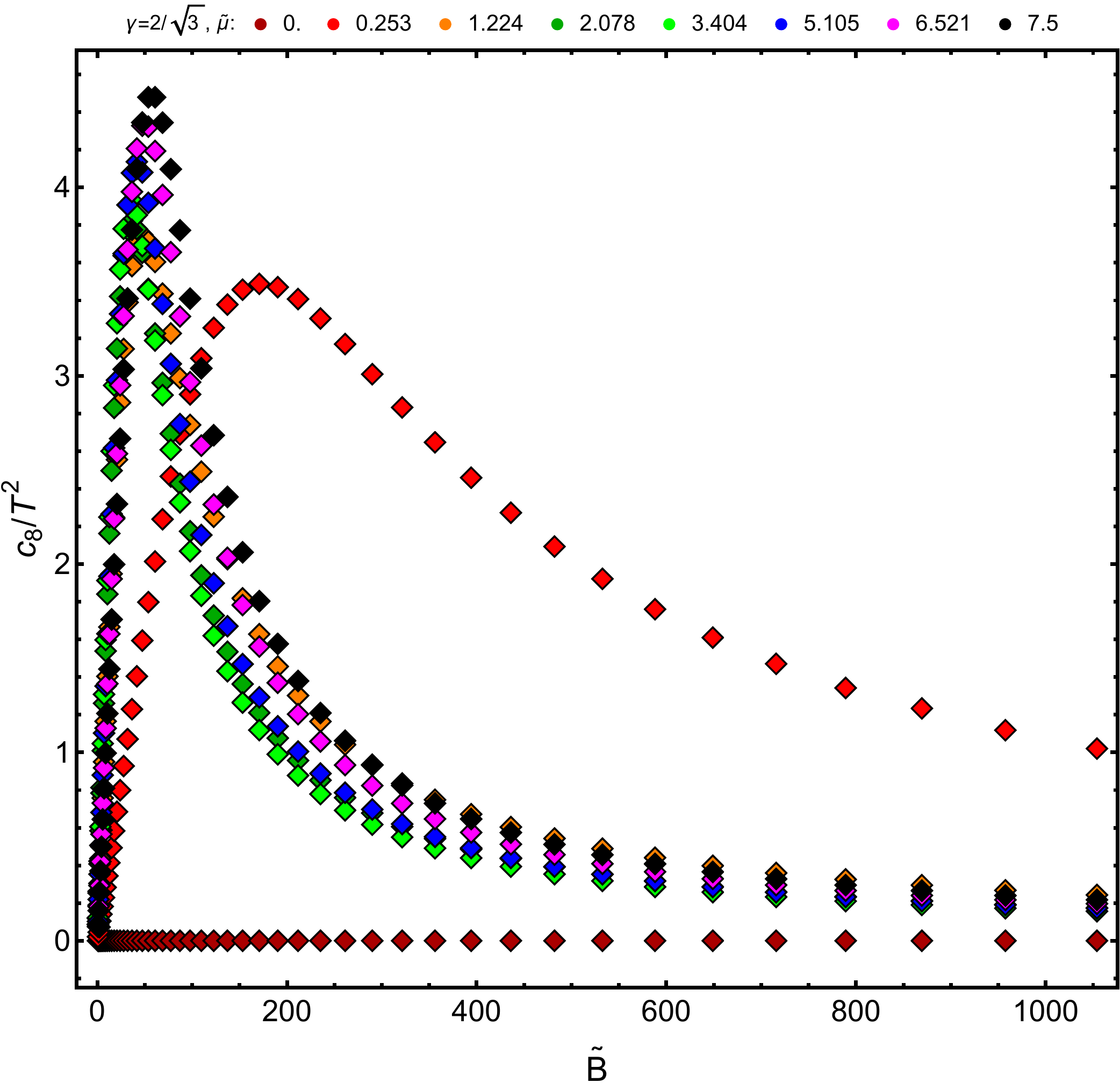}\vspace{-0.1cm}
    \caption{The transport coefficient $c_{8}/T^3$ for small and large magnetic fields in presence of the chiral anomaly $\gamma=2/\sqrt{3}$. Without chiral anomaly or for vanishing chemical potential, the transport coefficients is zero.}
  \label{fig:c8versusB}
\end{figure}

The novel transport coefficients $c_4$, $c_5$ as well as $c_8$ are nonzero in our holographic model. 
However, they vanish if either the anomaly is absent, i.e.~if $\gamma=0$, or if the magnetic field vanishes, or if the chemical potential vanishes. In other words, our holographic calculation is the first to reveal these transport coefficients, because they need $\mu\neq 0$, $B\neq 0$, and the chiral anomaly coefficient $C=-\gamma\neq 0$ simultaneously. 
Our numerical data shows that $c_3$ vanishes in our model. This is expected due to the conformal invariance of theory. 
The vanishing of $c_3$ implies that $c_4 = - c_5/3$ due to the Onsager relation \eqref{eq:ons-pv2}. 
The two relations $c_4 =- \frac{1}{3} c_5$ and $\eta_2 = - \frac{1}{3} \eta_1$ explain why the bound $\det M\ge 0$ with $M$ given by eq.~\eqref{eq:M} is saturated (within numerical accuracy $\det M\approx 10^{-6}$). 
The Onsager relation for $c_4$ and $c_5$ is satisfied up to errors of order $10^{-16}$ at worst ($\tilde B =1$, and $\tilde\mu=2.078$ corresponding to the dark green curve).  
We hence plot only the dimensionless quantity $c_4/T^3$, see figure \ref{fig:c4versusB}. 
It turns out that $c_4$ is negative which in turn implies that $c_5$ is positive. 
Furthermore, we find $c_4/T^3 \sim d(\tilde{\mu}) \tilde{B}$, where $d(\tilde{\mu})$ is negative. 
Note that $d(\tilde{\mu})$ first decreases as a function of the chemical potential $\tilde{\mu}$ and then increases again. 
The dimensionless quantity $c_8/T^2$ as a function of $\tilde{B}$ is depicted in figure~\ref{fig:c8versusB}. 
Note that $c_8/T^2$ is proportional to $\tilde{B}$ for small $\tilde{B}$, i.e.~the slope changes its sign. 
Finally, as observed in the other transport coefficients above, $c_8/T^2$ seems to go to zero for large magnetic fields. 

We make the observation that many of the dissipative transport coefficients vanish at large magnetic fields. 
This may be interpreted as dissipation being suppressed at large magnetic fields, as indicated for example by $\eta_{||}$, see figure~\ref{fig:eta2small}. 
A subset of the current authors has previously observed this behavior in the sound attenuation coefficient and the shear diffusion coefficient~\cite{Ammon:2016fru,Ammon:2017ded} computed in the holographic model we also consider here.

\section{Discussion}
\label{sec:discussion}
In this paper we have used effective field theory methods in order to construct a hydrodynamic description of chiral charged relativistic fluids subject to a strong external magnetic field. 
In particular, hydrodynamic constitutive relations are generated for the energy-momentum tensor, consistent, and covariant (axial) charge currents. 
All the independent transport coefficients appearing in the constitutive equations are summarized and classified in tables \ref{tab:transportCoeffsThermo} (thermodynamic), \ref{tab:transportCoeffsHydroNonDiss} (hydrodynamic, non-dissipative), and \ref{tab:transportCoeffsHydroDiss} (hydrodynamic, dissipative). A discussion and physical interpretation of these transport coefficients is provided in section~\ref{sec:allTransportCoeffs}. 
Instead of repeating the discussions already given above, in the following we highlight novel phenomena and interesting observations. 
Novel coefficients and transport effects arise due to the combination of magnetic field, axial charge, and chiral anomaly. These three quantities break time-reversal symmetry, rotational symmetry from $O(3)$ down to $O(2)$, parity symmetry, and chiral symmetry. This symmetry breaking has a drastic impact on the effective field theory description, leading to {\it odd} transport effects. 

Remarkably, one novel non-dissipative hydrodynamic transport coefficient is found: the shear-induced Hall conductivity $c_{10}$. This transport effect falls into the same category as the known Hall viscosity $\tilde\sigma_\perp$ and the transverse Hall viscosity $\tilde\eta_\perp$. This is because all three of them describe time-dependent but dissipationless transport effects. There exists no {\it longitudinal} Hall conductivity.   

In addition, three novel dissipative transport coefficients are found: the shear-induced conductivity $c_8$, as well as the expansion-induced longitudinal conductivities $c_{4}$ and $c_5$. 
Even the well-known charge conductivity and resistivity highlight the drastic changes to the fluid description. In standard hydrodynamic Kubo formulas the charge conductivity is the transport coefficient appearing in the lowest order of the current-current correlator. However, here the lowest transport coefficient is the charge resistivity, see e.g.~eq.~\eqref{eq:Kubo-r1b}.  The four novel coefficients, $c_4$, $c_5$, $c_8$ and $c_{10}$ can be nonzero in a strong magnetic field if either a chiral anomaly or an axial chemical potential is breaking the parity symmetry. 

Kubo relations are derived for 8 independent thermodynamic ($M_2,\, M_5$, $\xi$, $\xi_B$, $\xi_T$, $M_1$, $M_3$, $M_4$), for 4 independent non-dissipative hydrodynamic ($\tilde\eta_\perp$, $\tilde\eta_{||}$, $c_{10}$, $\tilde\sigma_\perp$), and for 10 independent dissipative hydrodynamic ($\eta_\perp$, $\eta_{||}$, $c_8$, $\sigma_\perp$, $\sigma_{||}$, $\eta_1$, $\eta_2$, $\zeta_1$, $c_4$, $c_5$) transport coefficients. For the explicit formulas refer to the equations listed in the third column in tables \ref{tab:transportCoeffsThermo} (thermodynamic), \ref{tab:transportCoeffsHydroNonDiss} (hydrodynamic, non-dissipative), and \ref{tab:transportCoeffsHydroDiss} (hydrodynamic, dissipative), respectively.    
In order to derive the thermodynamic transport effects, a field theory generating functional for time-independent $n$-point functions was constructed, see eq.~\eqref{eq:WA}.  
Five Onsager relations, \eqref{eq:ons} and \eqref{eq:ons-pv2}, are found and can be used for cross-checks in explicit computations of the 19 possible hydrodynamic transport coefficients. 
Kubo relations for the dependent hydrodynamic transport coefficients ($c_3$, $c_{14}$, $c_{15}$, $c_{17}$, $\zeta_2$) in a parity-violating microscopic theory (one with a chiral anomaly), are provided in eq.~\eqref{eq:KuboJJ}, \eqref{eq:KuboTT2an3}, and \eqref{eq:KuboTT2an4}. The coefficients $c_3,\, c_4, \, c_5$ have counterparts in a theory which does not feature an anomaly but which breaks parity through the axial chemical potential, those are given in~\eqref{eq:c-onsager} and~\eqref{eq:Kubo-pv}.  

Most prominently, our work discusses strong magnetic fields. The magnetic field breaks rotation symmetry leading to an anisotropic equilibrium state (in the rest frame of the fluid). 
Let us summarize here which transport effects newly arise at strong magnetic fields of order $\mathcal{O}(\partial^0)$, 
versus the effects appearing already in states with weak magnetic fields of order $\mathcal{O}(\partial)$. 
Consider the list of all transport coefficients provided in tables~\ref{tab:transportCoeffsThermo}, \ref{tab:transportCoeffsHydroNonDiss}, and \ref{tab:transportCoeffsHydroDiss}. 
First, some of the first order transport coefficients each split into two distinct coefficients, one for transport perpendicular and the other parallel to the magnetic field: $\eta_{||},\, \eta_\perp,\, 
\tilde\eta_{||},\, \tilde\eta_\perp,\, 
\sigma_{||},\, \sigma_\perp$ (or equivalently $\rho_{||},\, \rho_\perp$), 
$\tilde\sigma_{||}$, and $\tilde\sigma_\perp$. 
At weak magnetic field all $\alpha_{||} = \alpha_\perp$, where $\alpha$ stands for any transport coefficient within this group, and all tilded coefficients vanish. 
Second, at strong magnetic field new tensor structures in the constitutive relations give rise to $\eta_1,\, \eta_2,\, \zeta_{1}, \, \zeta_2$. 
At weak magnetic fields $2/3 \,\eta_2 = -2\eta_1 =\eta$, $\zeta_{1} =\zeta$, and $\zeta_2=0$. 
Third, two of the chiral conductivities, namely $\xi_B,\, \xi_{TB}$, are promoted from first order in derivative effects at weak magnetic fields to zeroth order effects at strong magnetic fields. 
Fourth, all of the $M_n$ vanish at weak magnetic field, as seen from the $s_n$ in table~\ref{tab:T1-2}. 
Finally, all $c_i$ with $i\ge 3$ vanish at weak magnetic field, as can be deduced from eq.~\eqref{eq:PV-non-eq}.

Interestingly, we find three incarnations of the Nernst effect. 
Strikingly, we confirm a previous claim~\cite{Chernodub:2017jcp}, that the standard Nernst coefficient, see~\eqref{eq:Nernst}, is proportional to the conformal anomaly, which in turn is then related to the magnetic susceptibility $\chi_B$.  
In addition, we find a magnetic version of the Nernst effect, see \eqref{eq:magneticNernst}, with the magnetic Nernst coefficient related to the perpendicular magnetic vorticity susceptibility $M_2$, and a  momentum Nernst effect, see~\eqref{eq:QNernst}, with momentum Nernst coefficient related to the magneto-vortical susceptibility $M_5$. 
        
As a proof of existence and consistency of this hydrodynamic description of chiral fluids we provide an explicit computation in a holographic model. This model is the well-known $\mathcal{N}=4$ Super-Yang-Mills~(SYM) theory minimally coupled to an external axial $U(1)_A$ gauge field. In order to probe a charged fluid in a magnetic field, we consider perturbations around charged magnetic black brane solution. A Chern-Simons term yields the desired anomalous (chiral) charge current. 
We find values for 25 of the transport coefficients, for which we verify the 5 Onsager relations~\eqref{eq:ons} and \eqref{eq:ons-pv2}. For the remaining two, $M_1$ and $M_3$ we find only their derivatives as the hydrodynamic theory gives only Kubo relations for those derivatives. However, as stated in section~\ref{sec:hydrodynamics}, $M_3=0$ in a conformal field theory. Of the 25 computed coefficients 20 are nonzero.
In our model the gravitational anomaly is absent, hence all effects associated with $c_1$, see eq.~\eqref{eq:gaugeGravitationalAnomalyCoeffs}, vanish in this model. 
The magneto-acceleration susceptibility $M_3$ vanishes due to conformal invariance. 
Also the transverse Hall viscosity $\tilde\eta_\perp$ vanishes, however, without any reason obvious to us.  

Due to conformal invariance the bulk viscosities $\zeta_1$, and $\zeta_2$ vanish. In addition, $c_3$ is zero.

As discussed in the introduction, we have restricted our attention to an axial $U(1)_A$ symmetry in this work. 
In Nature, however, an interplay between axial and vector symmetries is generic. In the standard model of particle physics, for example, currents of left- and right-handed fermions may be combined into axial and vector currents. Thus, it is natural to extend the hydrodynamic description to the combination, $U(1)_V \times U(1)_A$. As usual, the vector and axial currents can be related to the current of left- and right-handed particles by $J^\mu_{V/A} = (J^\mu_L\pm J^\mu_R)/2$.\footnote{It would also be interesting to have two U(1) gauge fields in the gravity theory~\cite{Jimenez-Alba:2014pea}, i.e.~introducing an axial and a conserved vector current in the dual field theory. This is relevant for testing some predictions for chiral magnetic waves and for Weyl semimetals and their surface states, see~\cite{Landsteiner:2015lsa,Landsteiner:2015pdh,Copetti:2016ewq,Ammon:2016mwa,Grignani:2016wyz}. } 
A different combination of currents may be interesting in the context of the $\Lambda$-hyperon polarization effect recently discovered in heavy ion collisions~\cite{STAR:2017ckg}: one may consider the helicity current instead of the axial current~\cite{Ambrus:2019khr,Ambrus:2020oiw}. This would require disentangling the axial current carried by particles from that carried by anti-particles, which is the definition of the helicity current. 

A next logical step is to restrict the vector gauge fields to satisfy Maxwell's equations and allow them to interact dynamically with the fluid. The resulting effective field theory may be labelled {\it magnetohydrodynamics}. However, previous constructions excluding anomalies~\cite{Grozdanov:2016ala,Grozdanov:2016tdf, Kovtun:2018dvd,Grozdanov:2017kyl} indicate that such a description goes well beyond what is known as textbook magnetohydrodynamics.\footnote{Important steps towards incorporating the effects of chiral anomalies into magnetohydrodynamics have been taken in~\cite{Boyarsky:2015faa}.} 
Although the setting with dynamical gauge fields~\cite{Grozdanov:2017kyl} is distinct from the one we consider here, some of the Kubo relations associated with energy momentum tensor correlators agree with ours. In order to understand this observation, recall that the constitutive equations for hydrodynamics coupled to external electromagnetic fields can be interpreted as those of hydrodynamics coupled to dynamical gauge fields~\cite{Hernandez:2017mch}. 

Over the past decade progress has been made towards a generating functional yielding non-equilibrium contributions to the constitutive relations and more generally to the $n$-point functions~\cite{Jensen:2012jh,Crossley:2015evo,Haehl:2015pja,Haehl:2015foa,Haehl:2016pec,Haehl:2018lcu,Glorioso:2018wxw}. This progress will eventually establish hydrodynamics as an effective field theory. 
Our work provides a step towards this goal by encoding the equilibrium response in our generating functional~\eqref{eq:WA}. Extending it to include non-equilibrium transport effects would be accomplishing that goal. 
In order to understand the fluid response to time-dependent or spatially modulated sources, frequency and wave length dependence of the transport coefficients should be studied, which can be computed with our methods. 
In the holographic context frequency- and wavelength-dependent transport coefficients have been computed before~\cite{Amado:2011zx,Hoyos2013}. 
One may also venture to extend hydrodynamics in strong electromagnetic fields to descriptions of fluids far-from-equilibrium. Recent developments applying insights from resurgence are promising in this regard~\cite{Romatschke:2017vte,Heller:2013oxa,Heller:2015dha,Heller:2020anv,Heller:2020uuy}. 
For an alternate expansion, linearizing in hydrodynamic fields, see~\cite{Bu:2016oba,Bu:2016vum,Bu:2018drd,Bu:2018psl,Bu:2019mow}.

An application of our fluid description is the hydrodynamic modelling of heavy ion collisions~\cite{Romatschke:2017ejr}, as well as the hydrodynamic description of astrophysical fluids in strong magnetic fields~\cite{RezzollaHydroTextbook,Joyce:1997uy, Boyarsky:2011uy,Tashiro:2012mf,Brandenburg:2017rcb} (see \cite{Boyarsky:2020ani, Boyarsky:2020cyk} for recent papers), 
and the description of condensed matter materials and cold atom gases subjected to magnetic fields, see~\cite{Fujita:2012fp,An:2020tkn,Hoyos:2020ywe,Baggioli:2020edn} for holographic models in this context. While these systems are subject to a strong \emph{dynamical} vector magnetic field, when the backreaction of the fluid on the electromagnetic field can be neglected, the description in terms of external magnetic fields as outlined in this paper may be useful. For example this is the case in (quantum) Hall effect experiments, ultracold atom gases in strong magnetic field traps, and possibly intermediate time phase of heavy-ion-collisions where the magnetic field changes from growing to decaying.

Moreover, external axial gauge fields are of importance in the context of Weyl semimetals.  Elastic  deformations therein couple  to  the fermionic low-energy excitations with different signs, giving rise to effective axial gauge fields \cite{Cortijo_2015, Cortijo:2016wnf}. Such deformations naturally arise at the surfaces of Weyl semimetals, inducing localised axial magnetic fields in their vicinity. In fact, the surface states (so-called Fermi arcs) may be understood  from  this  perspective  in terms of  lowest Landau levels generated by those axial magnetic fields (see e.g. \cite{Chernodub:2013kya,Ammon:2016mwa}).

In quantum chromodynamics~(QCD) the chiral magnetic effect and more generally QCD thermodynamics in the presence of an external magnetic field has been investigated using lattice simulations~\cite{Buividovich:2009wi,Buividovich:2010tn,Bali:2011qj,Bali:2012zg,Bali:2014kia,Astrakhantsev:2019zkr}. 
In particular, a result from lattice gauge theory gives motivation to the study of holographic models as a surrogate for QCD with strong magnetic fields:  the magnetoresponse of $\mathcal{N}=4$ SYM as computed in the same holographic dual we used in this work has been found to have universal similarities to the magnetoresponse of QCD~\cite{Endrodi:2018ikq}. 
The Kubo relations for thermodynamic transport coefficients and constitutive relations derived in the present paper facilitate a more detailed comparison of the equilibrium response in QCD and SYM subject to external magnetic fields. 

The time-evolution of (electro)magnetic fields in relativistic heavy-ion-collisions has also been considered~\cite{Voronyuk:2011jd,Inghirami:2019mkc}. It would be interesting to compare the hydrodynamic transport in our holographic model to the results from these other approaches. Even an analysis of the data from the other approaches using our Kubo relations may lead to interesting results in and out of equilibrium.

Especially with regard to condensed matter physics, we predict additional effects in the Hall response of materials. 
Hall effects of quantum or classical nature have been studied in condensed matter physics since Hall's discovery in 1879. 

For example, Hall  viscosity~\cite{Avron:1995fg,Avron:1997} is a dissipationless transport coefficient of relevance to topological  states  of  matter, e.g.~fractional  quantum Hall systems, see for example~\cite{Wen:2017,Hoyos:2014pba}. 
In (2+1) dimensions Hall viscosity in a hydrodynamic context has been found to be given by the angular momentum density in condensed matter~\cite{Read:2008rn,Read:2010epa}, and in holographic systems, e.g.~in application to $p$-wave superfluids~\cite{Son:2013xra}. In fractional quantum Hall systems the Hall viscosity is quantized and may be used to detect topological phase transitions~\cite{Hoyos:2014pba}. 

However, the Hall viscosities which we find in (3+1) dimensions prepared in an anisotropic state are novel. 
We predict two distinct Hall viscosities, one associated with the plane parallel to a strong magnetic field, the other with a plane perpendicular to it with generically distinct values. 
It would be very interesting to understand how the anomalous Hall effect~\cite{Nagosa2010} may fit into our hydrodynamic description. We assume that it would be measured in our Hall conductivity~$\tilde\sigma_\perp$. Potentially analogous effects may appear in the other Hall coefficients.   
Finally, the novel dissipationless coefficient $c_{10}$ is to be interpreted as a Hall response of a charge current to the shear of a fluid in a plane in the fluid creates a charge current perpendicular to that plane. 
This considerably extends the possibilities for Hall physics to be studied in the future and promises new technological applications. 

One promising testing ground are Weyl- or Dirac-semimetals~\cite{Nielsen:1983,Cortijo_2015,Cortijo:2016wnf,Pikulin:2016,Taguchi:2016,Breunig:2019kjd,Zhang:2018}. 
A quantum Hall effective action for the anisotropic Dirac semimetal is discussed in~\cite{Hoyos:2020ywe}. 
Experiments with Weyl semimetals report the observation of chiral transport effects in presence of magnetic fields~\cite{2014arXiv1412.6543L,2015arXiv150606577S,2015arXiv150407698Z,2015arXiv150600924W,2015arXiv150603190Y}. 
In these experiments, the relevant observable is the negative magnetoresistance. 
However, note that negative magnetoresistance cannot unambigously be related to the presence of a chiral anomaly (see e.g. the holographic computations in \cite{Baumgartner:2017kme,Ammon:2009jt}).\footnote{
It is not yet clear if Weyl- or Dirac-semimetals are governed by strong correlations. Thus an exploration from the strongly-interacting perspective is rather interesting. 
The interplay between external magnetic fields and a chiral anomaly within holographic models of Weyl-semimetals is discussed in~\cite{Landsteiner:2015lsa,Landsteiner:2015pdh,Copetti:2016ewq,Ammon:2016mwa,Grignani:2016wyz,Kharzeev:2016mvi}.}  
Hydrodynamic behavior has been measured most reliably or has been theoretically argued for in (2+1)-dimensional materials~\cite{Berdyugin162,Molenkamp:1994,Moll1061} where the parity anomaly leads to anomalous transport effects~\cite{DiSante:2019zrd,Tutschku:2020rjq,Tutschku:2020drw,parityAnomalyQHEWuerzburgPRL,Haldane:1988zza}. 
Our discussion of (2+1)-dimensional hydrodynamics in section~\ref{sec:relationTo2DHydro} will help relating our (3+1)-dimensional hydrodynamic transport effects to these lower dimensional experiments and theoretical descriptions.

All these examples, taken from fields as different as particle physics and condensed matter physics, illustrate the importance of a more detailed understanding of the response of (non)relativistic fluids in the presence of electromagnetic fields and anomalies, towards which this work is taking a step.

\bigskip

\begin{table}
\begin{center}
\begin{tabular}{|c|c|c|c|c|c|}
\hline
coefficient & name & Kubo formulas &  $\mathcal{C}$ & $\mathcal{P}$ & $\mathcal{T}$ \\
\hline
\multicolumn{6}{|c|}{Thermodynamic $\left (\lim\limits_{\mathbf{k}\to 0}\lim\limits_{\omega\to 0}\right )$,\, non-dissipative}\\
\hline
\multicolumn{6}{|l|}{momentum diffusion sector}\\
\hline
${M_2}$ & perp. magnetic vorticity susceptibility & 
$T^{xz}T^{yz}$ \eqref{eq:M2M5Kubo} & + & - & + \\
\hline
${M_5}$ & magneto-vortical susceptibility & $T^{tx}T^{yz}$ (\ref{eq:M2M5Kubo},\ref{eq:M5-Kubo})& +  & - & + \\
\hline
$\xi = \xi_{TB}$ & chiral vortical conductivity & $J^x T^{ty}$ (\ref{eq:chiral1},\ref{eq:chiral2}) & + & + & + \\
\hline
$\xi_B$ & chiral magnetic conductivity & $J^x J^y$ (\ref{eq:chiral1},\ref{eq:chiral2}) & +  & - & + \\
\hline
$\xi_{T}$ & chiral vortical heat conductivity & $T^{tx} T^{ty}$ 
(\ref{eq:chiral1},\ref{eq:chiral2})&  + & - & + \\
\hline
\multicolumn{6}{|l|}{scalar sector}\\
\hline
${M_1}$ & magneto-thermal susceptibility & $J^t T^{xx}$ \eqref{eq:M1M3M4Kubo}  &  + & + & - \\
\hline
${M_3}$  & magneto-acceleration susceptibility & $J^t T^{tt}$ \eqref{eq:M1M3M4Kubo} & +  & + & - \\
\hline
${M_4}$ & magneto-electric susceptibility & $J^t J^t$ \eqref{eq:M1M3M4Kubo}&  + & - & - \\
\hline
\end{tabular}
\end{center}
\caption{\label{tab:transportCoeffsThermo}
Independent thermodynamic transport coefficients in a charged chiral thermal plasma subjected to a strong $U(1)_A$ magnetic field. 
The column ``Kubo formulas'' points to the equations containing the relevant Kubo formulas. In that column, the operator combinations, e.g.~$J_x T_{ty}$ indicate the correlation function appearing in the Kubo formula for the spatial momentum to be aligned with the magnetic field. 
The susceptibilities $\chi_{11}$, $\chi_{13}$, $\chi_{33}$ and $\chi_B$ are not counted here as transport coefficients because they are thermodynamic derivatives of pressure. Here, ``perp.'' denotes ``perpendicular''. 
}
\end{table}
\begin{table}
\begin{center}
\begin{tabular}{|c|c|c|c|c|c|}
\hline
\multicolumn{6}{|c|}{Non-dissipative Hydrodynamic $\left (\lim\limits_{\omega\to 0}\lim\limits_{\mathbf{k}\to 0}\right )$}\\
\hline
coefficient & name & Kubo formulas &  $\mathcal{C}$ & $\mathcal{P}$ & $\mathcal{T}$ \\
\hline
\multicolumn{6}{|l|}{shear sector}\\
\hline
${\tilde\eta_\perp}$ & transverse~Hall viscosity& $T^{xy}(T^{xx}-T^{yy}) \eqref{eq:tildeEtaPerpKubo}$ 
& + & - & + \\
\hline

\multicolumn{6}{|l|}{momentum diffusion sector}\\
\hline

$\boxed{c_{10}}\propto c_{17}$ & shear-induced Hall cond. &  
$T^{tx}T^{xz},T^{tx}T^{yz}$ (\ref{eq:c10tildeIsolated},\ref{eq:KuboTT2an1},\ref{eq:KuboTT2an2}) & + & + & + \\
\hline
${\tilde\eta_{||}}$ & parallel Hall viscosity &  $T^{yz}T^{xz}$ \eqref{eq:KuboTT1an2Hall}
& + & - & + \\
\hline

${\tilde\sigma_\perp}$ & Hall conductivity & $J^xJ^x$,$J^x J^y$ (\ref{eq:Kubo-r2},\ref{eq:Kubo-r1b},\ref{eq:Kubo-r1c})  & + & - & + \\
\hline
\end{tabular}
\end{center}
\caption{\label{tab:transportCoeffsHydroNonDiss} 
Independent {\it non-dissipative hydrodynamic transport coefficients} in a charged chiral thermal plasma subjected to a strong $U(1)_A$ magnetic field. Similar to table~\ref{tab:transportCoeffsThermo}. Note that the Hall conductivity $\tilde\sigma_\perp$ is dissipationless, however, it is computed from the resistivity matrix including $\rho$ and $\tilde\rho$, which are both dissipative. Here, ``cond.'' denotes ``conductivity''. 
Boxed quantities are novel in this paper. 
}
\end{table}
\begin{table}
\begin{center}
\begin{tabular}{|c|c|c|c|c|c|c|c|}
\hline
\multicolumn{6}{|c|}{dissipative, hydrodynamic $\left (\lim\limits_{\omega\to 0}\lim\limits_{\mathbf{k}\to 0}\right )$}\\
\hline
coefficient & name & Kubo formulas & $\mathcal{C}$ & $\mathcal{P}$ & $\mathcal{T}$ \\
\hline
\multicolumn{6}{|l|}{shear sector}\\
\hline
$\eta_\perp$ & perp.~shear viscosity & $T^{xy}T^{xy}$ \eqref{eq:Kubo-Tmn}  & + & + & - \\

\hline

\multicolumn{6}{|l|}{momentum diffusion sector}\\
\hline
${\eta_{||}}$ & parallel shear viscosity &  $T^{xz} T^{xz}$ \eqref{eq:KuboTT1an2} 
& + & + & - \\
\hline
$\boxed{c_8} \propto c_{15}$    & shear-induced conductivity &  
$T^{tx}T^{xz}$, $T^{tx}T^{yz}$ \eqref{eq:Kubo-pv} & + & + & + \\
\hline
${\rho_\perp}$ & perp.~resistivity  & $J^x J^x$ \eqref{eq:Kubo-r2}& + & + & - \\
\hline
$\sigma_{||}$ & long.~conductivity & 
$J^z J^z$ \eqref{eq:Kubo-r1a}
& + & + & - \\
\hline
\rowcolor{Gray}
$\sigma_{\perp}$ & perp.~conductivity & 
$\rho_{ab} \equiv (\sigma^{-1})_{ab} = \rho_\perp \delta_{ab} + \tilde\rho_\perp\, \epsilon_{ab}$
& + & + & - \\
\hline
\multicolumn{6}{|l|}{scalar sector} \\
\hline
$\eta_1$  & bulk viscosity &  $\mathcal{O}_1\mathcal{O}_1$ \eqref{eq:Kubo-Tmng}  & + & + & - \\
\hline
$\eta_2$  & bulk viscosity & $\mathcal{O}_2\mathcal{O}_2$ \eqref{eq:Kubo-Tmnh}  & + & + & - \\
\hline
$\zeta_1$ & bulk viscosity & $T^{ij}(T^{xx}+T^{yy})$\eqref{eq:Kubo-Tmne}  & + & + & - \\
\hline
\rowcolor{Gray}
$\zeta_2$ & bulk viscosity &  $3\zeta_2-6\eta_1=2\eta_2$ &  + & + & - \\
\hline
$\boxed{c_4}$ & expan.-induced long.~cond. & 
$J^x T^{xx}$ \eqref{eq:Kubo-pv} & + & - & - \\
\hline
$\boxed{c_5}\propto c_{14}$ & expan.-induced long.~cond. & 
$J^z T^{zz}$ \eqref{eq:Kubo-pv} & + & - & - \\
\hline
\rowcolor{Gray}
$c_3$ &  & 
$c_5= -3(c_3+c_4)$  & + & - & - \\
\hline
\end{tabular}
\end{center}
\caption{\label{tab:transportCoeffsHydroDiss} 
Independent {\it dissipative hydrodynamic transport coefficients} in a charged chiral thermal plasma subjected to a strong $U(1)_A$ magnetic field. Similar to table~\ref{tab:transportCoeffsHydroNonDiss}. 
Abbreviations: perpendicular (perp.), longitudinal (long.), conductivity (cond.), expansion (expan.).  
Gray shaded rows display relation of dependent transport coefficients to the set of independent transport coefficients. 
Boxed quantities are novel in this paper.}
\end{table}

\acknowledgments
We would like to thank C.~Cartwright, A.~Jain, P.~Kovtun, K.~Landsteiner, A.~Shukla, L.~Yaffe for helpful discussions, as well K.~Landsteiner for detailed remarks on a draft of the paper. MA is funded by the Deutsche Forschungsgemeinschaft (DFG, German Research Foundation) under Grant No.406235073 within the Heisenberg program.
SG is in part supported by the `Atracci\'on de Talento' program (2017-T1/TIC-5258, Comunidad de Madrid) and through the grants SEV-2016-0597 and PGC2018-095976-B-C21.
SG furthermore gratefully acknowledges financial support by the Fulbright Visiting Scholar Program which is sponsored by the US Department of State and the German-American Fulbright Commission and the DAAD. JH is supported in part by the Natural Sciences and Engineering Research Council of Canada through a Postgraduate Doctoral Scholarship. This work was supported, in part, by the U. S. Department of Energy grant DE-SC-0012447. Research at Perimeter Institute is supported in part by the Government of Canada through the Department of Innovation, Science and Economic Development Canada and by the Province of Ontario through the Ministry of Colleges and Universities.
SG would like to thank the University of Victoria for their hospitality.

\appendix
\section{Details of the hydrodynamic calculation}
\label{hydroAppendix}
%
\begin{table}
\begin{center}
\begin{tabular}{|c|c|c|c|}
\hline
quantity & ${C}$ & ${P}$ & ${T}$ \\
\hline
$t$ & + & + & - \\
\hline
$x^i$ & + & - & + \\
\hline
$\varrho$ & + & + & + \\
\hline
$T,h_{tt},T^{tt}$ & + & + & + \\
\hline
$\mu_A,A_t,J^t$ & + & - & + \\
$\mu_V,V_t,J_V^t$  &-& + &+ \\
\hline
$A_i, J^i$ & + & + & - \\
$V_i, J_V^i$  & - & - & - \\
\hline
$A_\varrho$
& + & - & - \\
$V_\varrho$ 
& - & + & - \\
\hline
$u^i,h_{ti},T^{ti}$ & + & - & - \\
\hline
$h_{ij},T^{ij}$ & + & + & + \\
\hline
$B^i$ & + & - & - \\
$B_V^i$ & - & + & - \\
\hline
$E^i$ & + & + & + \\
$E_V^i$ & - &- & + \\
\hline
$dx^\mu\wedge dx^\nu\wedge dx^\rho\wedge dx^\sigma\wedge dx^\kappa$ & + & - & - \\
\hline
$\int\limits_i^f A\wedge F \wedge F$ & + & + & + \\
$\int\limits_i^f V\wedge F_V \wedge F_V$ & - & - & + \\
\hline
$u^t$ & + & + & + \\
\hline
generating functionals $W_s,\,W_{cons}, \, W_{cov}$ (axial $U(1)_A$) & + & + & + \\
\hline
\end{tabular}
\end{center}
\caption{
\label{tab:CPT} 
Transformation properties of the following quantities under charge-parity ${C}$, parity ${P}$, and time-reversal ${T}$: field theory coordinates $x^\mu = t,\, x,\, y,\,z$, field theory axial $U(1)_A$ (or vector $U(1)_V$) current $J^\mu$ (or $J^\mu_V$) and energy-momentum tensor $T^{\mu\nu}$, hydrodynamic variables $T, \, u^\mu,\, \mu$ (or $\mu_V$) and sources $h_{\mu\nu}$, $A_\mu$ (or $V_\mu$), as well as the extra spatial direction $r$ and Chern-Simons term, both appearing in the anomaly inflow formulation in section~\ref{sec:hydrodynamics}. In some entries we have split the time $t$ from the spatial components in the field theory directions, labeled by $i=x,\,y,\,z$, and the $\varrho$-component. Note that we consider here an axial gauge field $A_\mu$ which has transformation properties distinct from that of a vector gauge field. We here refer to the parity ${P}$ with respect to the field theory directions $x,\, y,\, z$. The integral boundaries are indicated as initial, $i$, and final, $f$. 
} 
\end{table}

\subsection{Consistent-consistent Kubo Formulas}
\label{app:jconsjcons}
The Kubo formulas in sections~\ref{sec:thermo} and \ref{sec:kubo} were written in terms of covariant-consistent correlation functions. The consistent-consistent connected correlation functions are the same as the covariant-covariant connected correlation functions since
\begin{eqnarray}\nonumber
    \langle J^\mu_{cov} J^\nu_{cov} \rangle &=& \langle (J^\mu_{cons} +J^\mu_{BZ})(J^\nu_{cons} +J^\nu_{BZ}) \rangle \\
    &=& \langle J^\mu_{cons} J^\nu_{cons} \rangle + J^\mu_{BZ} \langle J^\nu_{cons} \rangle+ \langle J^\mu_{cons} \rangle J^\nu_{BZ} + J^\mu_{BZ} J^\nu_{BZ} \, ,
\end{eqnarray}
and the last three terms are disconnected. 
The consistent-consistent connected correlation functions differ from the covariant-consistent connected correlation functions~\eqref{eq:cons-var} by the variation of the Bardeen-Zumino polynomial
\begin{equation}
\label{eq:JcovJ-JconsJ}
    G_{J_{cons}^\mu J_{cons}^\nu} = G_{J_{cov}^\mu J_{cons}^\nu} + \frac{\delta J^\mu_{BZ}}{\delta A_\nu} \, .
\end{equation}
Alternatively, the static consistent-consistent correlation functions can be found by varying the 3+1 dimensional generating functional~\eqref{eq:WA}
\begin{equation}
    G_{J_{cons}^\mu J_{cons}^\nu}(\omega=0,\mathbf{k}) = \frac{ \delta^2 W_{cons}}{\delta A_\mu(\mathbf{k}) \delta A_\nu(-\mathbf{k})}\,.
\end{equation}
The Kubo formula for $\xi_B$ in terms of static consistent-consistent correlation functions is then
\begin{subequations}
\label{eq:chiral3}
\begin{align}
    \langle J^x_{cons}(\mathbf{k}) J^z_{cons}(-\mathbf{k})\rangle  = -i k_y (\xi_B-\coeff13 C A_t) \,,
\end{align}
for fluctuations perpendicular to the magnetic field and
\begin{align}
\label{eq:kubo_xiB_BZterm}
    \langle J^x_{cons}(\mathbf{k}) J^y_{cons}(-\mathbf{k})\rangle  = -i k_z (\xi_B-\coeff13 C A_t) \,,
\end{align}
\end{subequations}
for fluctuations parallel to the magnetic field. These thermodynamic formulas are unchanged by the presence of strong magnetic fields, and agree with the expressions found in~\cite{Amado:2011zx,Jensen:2013vta}. We can also rewrite the current-current Kubo formulas in ~\eqref{eq:KuboJJ} in terms of the consistent currents 
\begin{subequations}
\begin{align}
\label{eq:KuboJJcons}
  &\coeff{1}{\omega} {\rm Im}\, G_{J^z_{cons} J^z_{cons}}(\omega,{\bf k}{=}0) =  \sigma_\parallel + \cdots \,,
\end{align}
\begin{align}
\label{eq:KuboJJconsb}
  &  \coeff{1}{\omega} {\rm Im}\, G_{J_{cons}^x J^x_{cons}}(\omega,{\bf k}{=}0) = \omega^2 \rho_\perp \frac{w_0(w_0 - \MO_{,\mu}B_0^2)}{B_0^4}  \,,
  \end{align}
\begin{align}
  & \label{eq:hall-conduct} \coeff{1}{\omega}{\rm Im}\, G_{J_{cons}^x J_{cons}^y}(\omega,{\bf k}{=}0) = 
     \frac{n_0}{B_0} - \omega^2 \tilde\rho_\perp \frac{w_0(w_0 - \MO_{,\mu}B_0^2)}{B_0^4}\, {\rm sign}(B_0)\,, 
     \end{align}
\begin{align}
\label{eq:KuboJJconsd}
     & \coeff{1}{\omega}{\rm Im}\, G_{J_{cons}^{z} O_1}(\omega,{\bf k}{=}0) = - c_4 \,{\rm sign}(B_0) + \cdots \,,\\[5pt]  \label{eq:KuboJJconse}
  & \coeff{1}{\omega}{\rm Im}\, G_{J_{cons}^{z} O_2}(\omega,{\bf k}{=}0) = - c_5\, {\rm sign}(B_0) + \cdots \,
  \end{align}
\begin{align}
\label{eq:KuboJJconsf}
    & \coeff{1}{\omega}\delta_{ij}{\rm Im}\, G_{T^{ij}J_{cons}^{z}}(\omega,{\bf k}{=}0) =  3 c_3 \,{\rm sign}(B_0) +\cdots \,,
    \end{align}
\begin{align}
\label{eq:KuboJJconsg}
  & \coeff{1}{\omega}{\rm Im}\, G_{O_2 J_{cons}^{z}}(\omega,{\bf k}{=}0) =  2 c_{14} \, {\rm sign}(B_0) + \cdots \,,
\end{align}
\end{subequations}
where $O_2$ is defined below equation~\eqref{eq:Kubo-Tmn}. The terms omitted vanish for $B_0 \ll T_0^2$ or when $M_1=M_3=M_4=0$.

\subsection{Eigenmodes}\label{app:eigenmodes}
In this appendix, we give more details about the gapless eigenmodes found in section~\ref{sec:eigenmodes-1}.

The velocities $v_0\,,\ v_+$ and $v_-$ appearing in eq.~\eqref{eq:omega-par} are the solutions to the cubic equation
\begin{equation}
    a_3 v^3 + a_2 v^2 + a_1 v + a_0 = 0\,,
\end{equation}
where 
\begin{align*}
    &a_0 = - B_0 C \frac{(s_0 T_0)^2}{\det(\chi)}\,,\\
    &a_1 = v_s^2 - \frac{(B_0 C \mu_0 T_0)^2}{\det(\chi)} \frac{ds}{dT}\,,\\
    &a_2 =  B_0 C \left( \frac{\mu_0^2}{w_0} -  2 n_0 \mu_0 \frac{\chi_{11} - \mu_0 \chi_{13}}{{\rm det}(\chi)} + w_0 \frac{\chi_{11} - \mu_0^2 \chi_{33}}{{\rm det}(\chi)} \right)\,, \\
    &a_3 = -1 + \frac{(B_0 C \mu_0 T_0)^2}{\det(\chi)} \frac{ds}{dT}\,,
\end{align*}
and the speed of sound expressed in terms of the elements of the susceptibility matrix is given by~\eqref{eq:speedOfSound}

These can be solved perturbatively in $B_0 C$. We first label
\begin{align*}
\alpha &= \frac{(s_0 T_0)^2}{{\rm det}(\chi)}\,, \quad \quad  \quad \quad \beta = \frac{(\mu_0 T_0)^2}{{\rm det}(\chi)} \frac{ds}{dT}\,,\\
\gamma & = \frac{\mu_0^2}{w_0}-2n_0 \mu_0 \frac{\chi_{11}-\mu_0\chi_{13}}{{\rm det}(\chi)}+w_0 \frac{\chi_{11}-\mu_0 \chi_{33}}{{\rm det}(\chi)}\,,
\end{align*}
from which we find
\begin{equation}
\begin{aligned}
    v_0 &= B_0 C \frac{\alpha}{v_s^2}+\cdots\,,\\
    v_\pm & = \pm v_s + B_0 C \frac{\gamma v_s^2 - \alpha}{2 v_s^2} \mp B_0^2 C^2 \frac{3\alpha^2 -2\alpha \gamma v_s^2 + v_s^4(4\beta(1-v_s^2)-\gamma^2)}{8 v_s^5}+\cdots\,,
\end{aligned}
\end{equation}
where we have omitted terms of order $B_0^3 C^3$ and higher.

The damping coefficient in eq.~\eqref{eq:Gamma-parallel} has the following corrections

\begin{equation}
\begin{aligned}
    \frac{w_0}{W_0} &= 1 + C B_0 \frac{\chi_{11} w_0 + 2 n_0 \mu_0(\mu_0 \chi_{13} - \chi_{11}) + \mu_0^2 \left( (3 \det(\chi) - \chi_{13}^2 w_0)/\chi_{11} - \det(\chi)/w_0 \right) - (s_0 T_0/v_s )^2}{v_s   \det(\chi)}\,,\\
    & \quad \quad \quad \Sigma_\perp  = \frac{8 \mu_0 w_0^2 (\mu_0 \chi_{13} - \chi_{11})(w_0 \chi_{13} - n_0 \chi_{11})}{v_s (n_0 \chi_{11} - 2 w_0 \chi_{13})^2 \det(\chi)}\,,
    \end{aligned}
\end{equation}
and $\Sigma_\parallel$ is a lengthy function of the susceptibilities, other thermodynamic derivatives of the pressure, the chemical potential and the temperature. We have omitted higher order terms in $C B_0$ in eq.~\eqref{eq:Gamma-parallel}.

The leading correction to the longitudinal diffusion constant in eq.~\eqref{eq:D-parallel} is 
$$
  D_\parallel = \frac{w_0^2 \sigma_\parallel}{v_s^2\det(\chi)}+B_0^2 C^2 \left(\left(\frac{ds}{d\mu}\right)_p \frac{s_0^4 T_0^6}{3 v_s^6 {\rm det}(\chi)^3} (3\zeta_1+10\eta_1+6\eta_2) + F \sigma_\parallel \right)+ \cdots\,,
$$
where $\left(\frac{ds}{d\mu}\right)_p$ is the derivative keeping pressure and magnetic field fixed, and 
\begin{align*}
F&= \frac{s_0^4\, T_0^6\, w_0}{v_s^8\, {\rm det}(\chi)^4}\Bigg[ 2 \left(\frac{dn}{d\mu} \right)_p^2 s_0^2 \mu_0^2 + 2 \left(\frac{ds}{d\mu} \right)_p^3n_0 T_0 \mu_0^2 \\& +  \left(\frac{ds}{d\mu} \right)_p^2 \left(3 s_0^2 T_0^2 - 2 \left(\frac{dn}{d\mu} \right)_p s_0 T_0 \mu_0^2 - \mu_0^2 n_0 \left(n_0 -2 \left(\frac{dn}{d\mu} \right)_p \mu_0\right)\right)\\
&+ 2\left(\frac{dn}{d\mu} \right)_p \left(\frac{ds}{d\mu} \right)_p s_0 \mu_0 \left(3 s_0 T_0 + \mu_0\left(n_0 -\left(\frac{dn}{d\mu} \right)_p \mu_0\right)\right)\Bigg]\,.
\end{align*}
The positivity of the diffusion constant implies $\sigma_\parallel >0$.

\subsection{A comment on magnetic susceptibilities}
\label{sec:noteOnChiB}
In this appendix, we comment on the choice of naming $\aBB=2p_{,B^2}$ as magnetic susceptibility. The thermodynamic function $\aBB$ appears in the constitutive relation for the magnetic polarization in front of $B^\mu$. To be more precise, we can define $\delta W = \frac12 \int \sqrt{-g} M^{\mu\nu} \delta F_{\mu\nu} = \int \sqrt{-g}\left( m^\mu \delta B_\mu + p^\mu \delta E_\mu\right)$ (see~\cite{Kovtun:2016lfw}). The polarization tensor $M_{\mu\nu}$ is related to the polarization and magnetization vectors $p_\mu$ and $m_\mu$ by

\begin{equation}
M_{\mu\nu} = p_\mu u_\nu - p_\nu u_\mu -\epsilon_{\mu\nu\rho\sigma} u^\rho m^\sigma\,.
\end{equation}

Varying the generating functional $ W =\! \int \sqrt{-g} {\cal F}(T,\mu,B^2,E^2,B{\cdot}E, B{\cdot}a, B{\cdot}\Omega, E{\cdot}a, E{\cdot}\Omega,\cdots)$ we find the constitutive relations for $p^\mu$ and $m^\mu$

\begin{equation}
\begin{aligned}
    m^\mu = 2{\cal F}_{,B^2} B^\mu + {\cal F}_{,B{\cdot}E} E^\mu + {\cal F}_{,B{\cdot}a} a^\mu+{\cal F}_{,B{\cdot}\Omega} \Omega^\mu + \cdots\,,\\
    p^\mu = 2{\cal F}_{,E^2} E^\mu + {\cal F}_{,B{\cdot}E} B^\mu+{\cal F}_{,E{\cdot}a} a^\mu+{\cal F}_{,E{\cdot}\Omega} \Omega^\mu+ \cdots\,.
\end{aligned}
\end{equation}
Now, if we focus on the zeroth order constitutive relations and with strong magnetic fields so that $B^\mu={\cal O}(1)$ while $E^\mu =a^\mu = \Omega^\mu = {\cal O}(\partial)$, we simply find
\begin{equation}
    m^\mu = 2p_{,B^2} B^\mu + \cdots\,, \quad p^\mu = \cdots\,,
\end{equation}
that is, the thermodynamic function relating $B^\mu$ to $m^\mu$ is simply $2p_{,B^2}$. When written in terms of magnetic field strength $H^\mu = B^\mu - m^\mu$, the relation reads
\begin{equation}
    m^\mu = \frac{2p_{,B^2}}{1-2p_{,B^2}} H^\mu\,.
\end{equation}

However, if we are interested in seeing how the magnetization changes as we change the magnetic field $\delta m^\mu = m^{\mu\nu} \delta B_\nu$, we find
\begin{equation}
    m^{\mu\nu} =4 p_{,B^2B^2} B^{\mu} B^\nu + 2 p_{,B^2} g^{\mu\nu}\,,
\end{equation}
where, in particular, the term aligned with $B^\mu$ is
\begin{equation}
    m^{\mu\nu} B_\mu B_\nu = 4 p_{,B^2 B^2} B^4 + 2p_{,B^2} B^2 = p_{,BB} B^2
\end{equation}
where $B = \sqrt{B^2}$. This is often quoted as the magnetic susceptibility at finite magnetic field. 

Note that the change in magnetization orthogonal to the existing magnetic field is given by $2p_{,B^2}$. That is, for
\begin{equation}
    \delta B^\mu = \delta B^\mu_\parallel  + \delta B^\mu_\perp\,,
\end{equation}
we have the following change in the magnetization
\begin{equation}
    \delta m^\mu = p_{,BB}\,\delta B^\mu_\parallel + 2p_{,B^2} \delta B^\mu_\perp\,.
\end{equation}
This suggests we could call $2p_{,B^2}$ the perpendicular magnetic susceptibility and $p_{,BB}$ the parallel magnetic susceptibility.
\section{Details of the holographic calculation}\label{sec:detailsapp}
In this section, we present more details of the holographic calculation. We also outline the numerical procedure used to construct the numerical solutions in this work and compute the transport coefficients. 
\subsection{Expansion close to horizon and conformal boundary}
Here we include the near-boundary and near-horizon expansions for the functions defined in the metric and gauge field ansatz, eq.~\eqref{eq:ansatzmetric} and eq.~\eqref{eq:ansatzF}, used in the holographic model. 
The solution close to the conformal boundary (i.e. for $\varrho=0$) reads
 \begin{eqnarray}\label{eq:boundaryExpansionBackground}
&& u(\varrho) = 1 + \varrho^4\left[  u_4 +{\cal O}(\varrho^2) \right]+ \varrho^4\ln(\varrho)\left[ \frac{B^2}{6}+ {\cal O}(\varrho^2) \right]  \nonumber \, ,\\
&& v(\varrho) = 1 + \varrho^4\left[- \frac{w_4}{2} +{\cal O}(\varrho^2)\right] + \varrho^4\ln(\varrho)\left[- \frac{B^2}{24}+ {\cal O}(\varrho^2) \right] \, , \nonumber\\
&& w(\varrho) = 1 + \varrho^4\left[  w_4 +{\cal O}(\varrho^2)\right] + \varrho^4\ln(\varrho)\left[ \frac{B^2}{12}+ {\cal O}(\varrho^2) \right] \, , \nonumber\\
&& c(\varrho) =\varrho^4 \left[ c_4 + {\cal O}(\varrho^2) \right] + \varrho^8\ln(\varrho) \left[ - \frac{B^2}{12}  {c_4} + {\cal O}(\varrho^2)\right]  \, ,\nonumber  \\	
&&A_t(\varrho)  =  \mu -  \frac{\rho}{2} \varrho^2  -  \frac{\gamma B  p_1}{8} \varrho^4  +  {\cal O}(\varrho^6)\, ,\nonumber\\
&&P(\varrho) = \varrho^2\left( \frac{p_1}{2}  + \frac{\gamma B  \rho}{8}  \varrho^2 + {\cal O}(\varrho^4 ) \right) \,,\label{eq:asymptexpbdy}
\end{eqnarray}
where $u_4,\, w_4,\, c_4, \, \rho, \, p_1$ are undetermined coefficients. 
Here we have chosen to set the non-normalizable mode for $P(\varrho)$ to zero at the conformal boundary. This choice is dual to switching off the source for the current in $z$-direction on the boundary. 
Near the horizon at $\varrho=1$, the expansion of the same functions reads
\begin{align}\label{eq:horizonExpansionBackground}
u(\varrho) &= (1-\varrho)\left[  \bar{u}_1 + {\cal O}(1-\varrho) \right], \qquad   & c(\varrho) &= (1-\varrho)\left[  \bar{c}_1  + {\cal O}(1-\varrho)\right], \nonumber \\
v(\varrho) &= \bar{v}_0 + {\cal O}(1-\varrho),  \nonumber  & A_t(\varrho) &= (1-\varrho)\left[  \bar{A}_{t \, 0} + {\cal O}(1-\varrho) \right] ,\nonumber\\
w(\varrho) &=  \bar{w}_0 + {\cal O}(1-\varrho), &  P(\varrho) &=  \bar{P}_0 + {\cal O}(1-\varrho) \,,
\end{align}
where $\bar{u}_1,\, \bar{c}_1,\, \bar{w}_0,\, \bar{v}_0,\, \bar{A}_{t \,0}$ and $\bar{P}_0$ are undetermined coefficients. At the horizon, we choose the coefficients of order $(1-\varrho)^0$ in $u(\varrho)$, $c(\varrho)$, and $A_t(\varrho)$ to vanish.

\subsection{Summary of thermodynamic details of the charged magnetic black brane}
In the following we display how to extract the expectation value of the energy-momentum tensor and of the covariant and consistent currents from the metric and gauge field solutions. 

In order to compute the expectation value of the energy momentum tensor in equilibrium for the charged magnetic brane we only have to evaluate the terms in the first line\footnote{The terms in the second line of \eqref{eq:Tmunu} will be important when computing correlation functions involving the energy-momentum tensor.} of \eqref{eq:Tmunu}. Given our ansatz for the metric \eqref{eq:ansatzmetric} and for the gauge field \eqref{eq:ansatzA} as well as the boundary expansion \eqref{eq:boundaryExpansionBackground}, the expectation value of the energy momentum tensor in equilibrium reads
\begin{equation}
\label{eq:EnergyMomTensor}
\langle T^{\mu\nu}\rangle =\left( \begin{array}{cccc}
-3 \, u_4 & 0 & 0 &  
-4 \, c_4 \\
0 & -\frac{B^{2}}{4}-u_4-4\,w_4  & 0 & 0\\
0 & 0 & -\frac{B^{2}}{4}-u_4-4\,w_4  & 0\\
-4 \, c_4  & 0 & 0 & 8 \, w_4 - u_4 \end{array} \right) \, .
\end{equation}
Again, let us stress that the components of the energy-momentum tensor are scheme-dependent. However, for any choice of the scheme, the trace of the energy momentum tensor is $\langle T_\mu{}^\mu \rangle = -B^2/2$.

The expectation value of the covariant current $J_{\text{cov}}^\mu$ for the charged magnetic black brane is given by
\begin{equation}\label{eq:CFTcurrents}
\langle J_{cov}^\mu \rangle = \left(\rho,\,0,\,0,\,p_1\right) \,,
\end{equation} 
where $\rho$ and $p_1$ are boundary coefficients defined in eq. \eqref{eq:boundaryExpansionBackground}. Note that the equations of motion imply $p_1 = - \gamma \mu B$ as shown in \cite{Ammon:2016szz} and hence $\langle J^z_{cov} \rangle = - \gamma \mu B$ assuming that $B$ is aligned along the (positive) $z$-axis.

In the following we will outline how to determine the other thermodynamic quantities, such as the entropy $s$, the temperature $T$ and the (density of the) grand canonical potential $\Omega.$ The entropy density $s$ is given by the function $v(\varrho)$ and $w(\varrho)$ evaluated at the horizon $\varrho=1$ using the expansion \eqref{eq:horizonExpansionBackground}
\begin{equation}
     s=4\pi \, v(1)^2 \, w(1)= 4 \pi \, \bar{v}_0^2 \, \bar{w}_0 .
 \end{equation}
while the temperature is given by
\begin{equation}\label{eq:temperature}
T = \frac{|u'(1)|}{4\pi} \, .
\end{equation}
Finally, the grand canonical potential or its density $\Omega$ is given by
\begin{equation}
    \Omega = \epsilon - s T - \mu \langle J_{cov}^t \rangle .
\end{equation}
This concludes the discussion concerning the thermodynamics of the charged magnetic black brane.

\subsection{Numerical details}\addtocontents{toc}{\protect\setcounter{tocdepth}{2}}

\subsubsection{Details for computing $\tilde{B}$ and $\tilde{\mu}$ derivatives of thermodynamic coefficients}
To extract the transport coefficients via Kubo formulas, we have to evaluate thermodynamic derivatives with respect to $\tilde B$ and $\tilde\mu$, respectively, while keeping the respective other quantity fixed. We compute the derivatives with spectral derivative matrices, where we discretize the parameter range in $\tilde B$ and $\tilde \mu$ in terms of a Chebychev grid. From a numerical point of view, we can only control $\mu$ in terms of a boundary condition on the temporal component of the gauge field and $B$ as external parameter. To compute the background for a given $\tilde B$, we vary $B$ in terms of an iterative solver until we find the desired value of $\tilde B$ (while keeping $\tilde\mu$ fixed).

\subsubsection{Spectral method for calculating the holographic Green's functions numerically}\label{sec:spectralMethod}

The numerical calculation of the transport coefficients in this paper are based on a pseudo-spectral method (see \cite{Boyd00,Ammon:2016fru,Grieninger:2017jxz,Grieninger:2020wsb,Baggioli:2019abx} for a more detailed introduction). In order to determine the thermo- and hydrodynamic (transport) coefficients we have to (numerically) compute Green's functions of the form $\lim\limits_{\omega \rightarrow 0} \coeff{1}{\omega}{\rm Im}\, G_{\mathcal O_a\mathcal O^c}(\omega,{\bf k}{=}0)$ and $\lim\limits_{k_z \rightarrow 0}  \coeff{1}{k_z}{\rm Im}\, G_{\mathcal O_a\mathcal O^c}(\omega{=}0,k \, {\bf e}_z)$ to very high accuracy. 

Green's functions of the form $G_{\mathcal O_a\mathcal O^c}(\omega,{\bf k})$ are determined by exploiting the relationship
\begin{equation}
    \delta \left\langle \mathcal{O}_a \right\rangle(\omega, \vec{k}) = G_{\mathcal{O}_a \mathcal{O}^c}(\omega, \vec{k}) \, \delta \phi_c(\omega, {\bf k}) \, ,
\end{equation}
where $\delta \phi_c$ is the source dual to the operator $\mathcal{O}_c$ and $\delta \left\langle \mathcal{O}_a\right\rangle$ is the response to the perturbation $\delta \phi_c.$

To directly compute two-point functions of the form $\lim\limits_{\omega \rightarrow 0} \coeff{1}{\omega}{\rm Im}\, G_{\mathcal O_a\mathcal O^b}(\omega,{\bf k}{=}0)$ and $\lim\limits_{k_z \rightarrow 0} \coeff{1}{k}{\rm Im}\, G_{\mathcal O_a\mathcal O^b}(\omega{=}0,k {\bf e}_z)$ we found it convenient to apply a three-step procedure which we explain in the following. As outlined in chapter \ref{sec:setup}, the background of our gravity model is the charged-magnetic black brane which we solve by means of a pseudo-spectral method as described in \cite{Ammon:2016fru,Grieninger:2017jxz,Grieninger:2020wsb,Ammon:2017ded}. 
On top of that background we calculate the fluctuations to linear order in $\omega$ and $k$ (were $k$ is the component along the $z$-axis, \emph{i.e.}, along the magnetic field), respectively, by doing an expansion in terms of them which read for the metric fluctuations
\begin{equation}
h_{mn}(\varrho,\omega)= h_{mn}^{(0)}(\varrho) + \omega\, h_{mn}^{(1)}(\varrho)\quad\text{and}\quad h_{mn}(\varrho,k)= h_{mn}^{(0)}(\varrho) + k\, h_{mn}^{(1)}(\varrho),
\end{equation} 
respectively. In order to compute a two-point functions of the form $\coeff{1}{\omega}{\rm Im}\, G_{T^{ab}T^{cd}}(\omega,{\bf k}{=}0)$, for example, we have to source the fluctuation $h_{cd}^{(0)}$ and read of the vacuum expectation value of the fluctuation $h_{ab}^{(1)}$. This may be done by plugging the background solution into the fluctuation equations of zeroth order in $\omega$ and then plugging the background solution and zeroth order solution into the first order equations and solve for the fluctuations of first order in $\omega$. For the metric fluctuations the near boundary expansion is schematically of the form
\begin{eqnarray}
&h_{mn}^{(0)}(\varrho)&=h_{mn,\bm{s}}^{(0)}+\varrho^4\,h_{mn,\bm{v}}^{(0)}+B^2 h_{mn,\bm{s}}^{(0)}\varrho^4 \log(\varrho)\nonumber\\
&h_{mn}^{(1)}(\varrho)&=h_{mn,\bm{s}}^{(1)}+\ldots+\varrho^4\,h_{mn,\bm{v}}^{(1)}+(B\,h_{mn,\bm{s}}^{(0)}+B^2\, h_{mn,\bm{s}}^{(1)})\,\varrho^4 \log(\varrho),\label{eq:fluctap1}
\end{eqnarray} 
where the prefactors $h_{mn,\bm{s}}^{(0)}$ and $h_{mn,\bm{s}}^{(1)}$ are related to the source of the energy momentum tensor. Depending on the operator $\mathcal O^a$ under considerations, we switch on the corresponding zeroth order sources $h_{ab,\bm{s}}^{(0)}$. The first order sources $h_{ab,\bm{s}}^{(1)}$ are always zero whereas the vacuum expectation value of the operator $\mathcal O_b$ is encoded in the corresponding $h_{ab,\bm{v}}^{(1)}$.

\subsubsection{Convergence and numerical accuracy}

In this subsection, we discuss the convergence and numerical precision of the numerical procedure we used to compute the transport coefficients. The asymptotic boundary expansions of the background fields \eqref{eq:asymptexpbdy} and the fluctuations \eqref{eq:fluctap1} contain logarithmic contributions proportional to the magnetic field $B$. To improve the numerical accuracy close to the boundary and simplify the process of reading of the expectation values, we introduce auxiliary functions, schematically given by
\begin{align}
   & h^\text{old}_{mn}=  h_{mn,\,\bm{s}}+\varrho^4\,h^\text{new}_{mn}+c_1\,\varrho^4\,\log(\varrho)\\
  & a^\text{old}_{m}=  a_{m,\,\bm{s}}+\varrho^2\,a^\text{new}_{m}+c_2\,\varrho^2\,\log(\varrho)
\end{align}
where $c_1,\,c_2$ is the coefficient of the logarithmic term proportional to the expectation value. We found it convenient to apply these kind of redefinition of the functions for all background fields and fluctuations.

\begin{figure}[h]
    \centering
     \includegraphics[width=7cm]{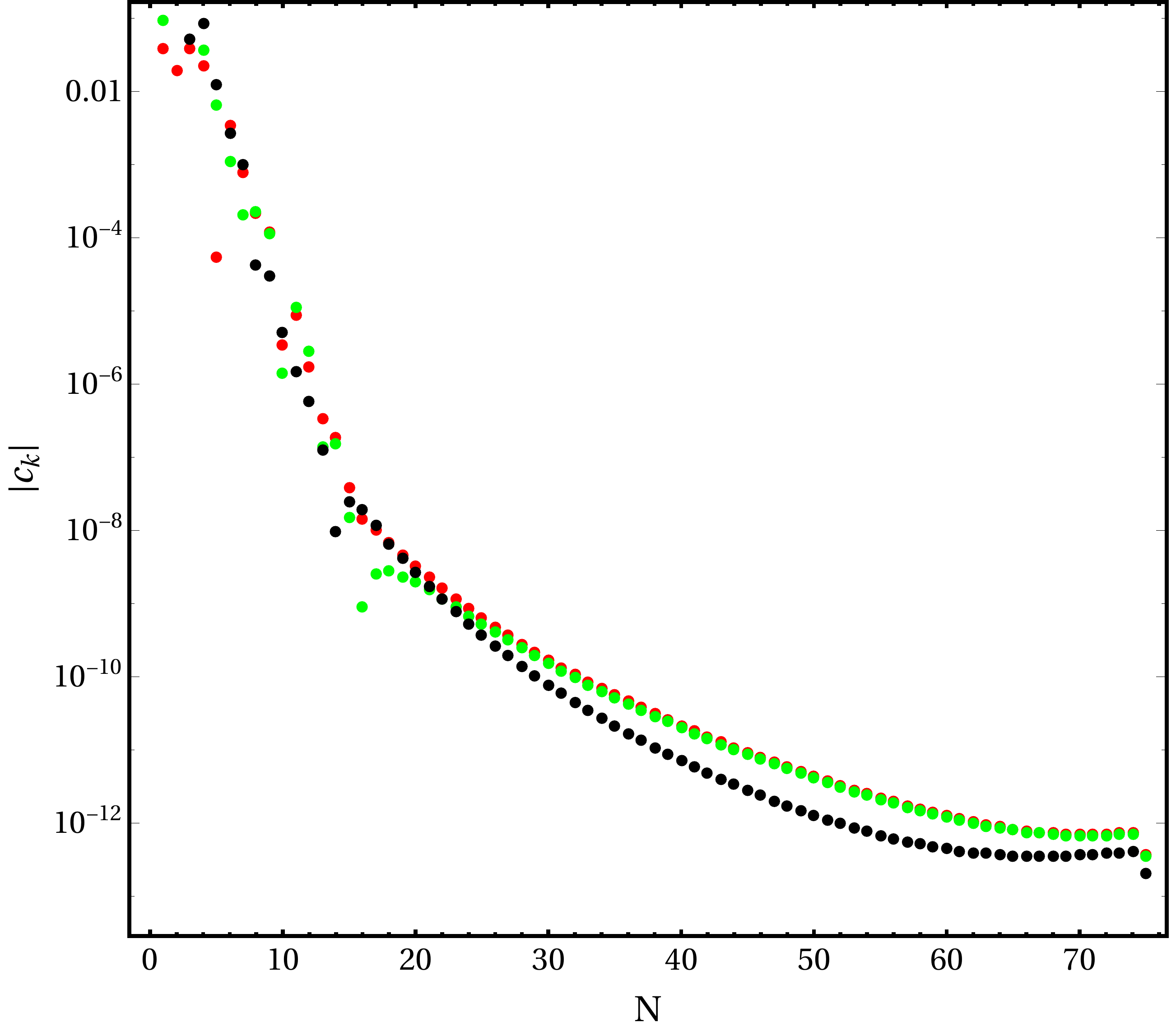}\quad
 \includegraphics[width=7cm]{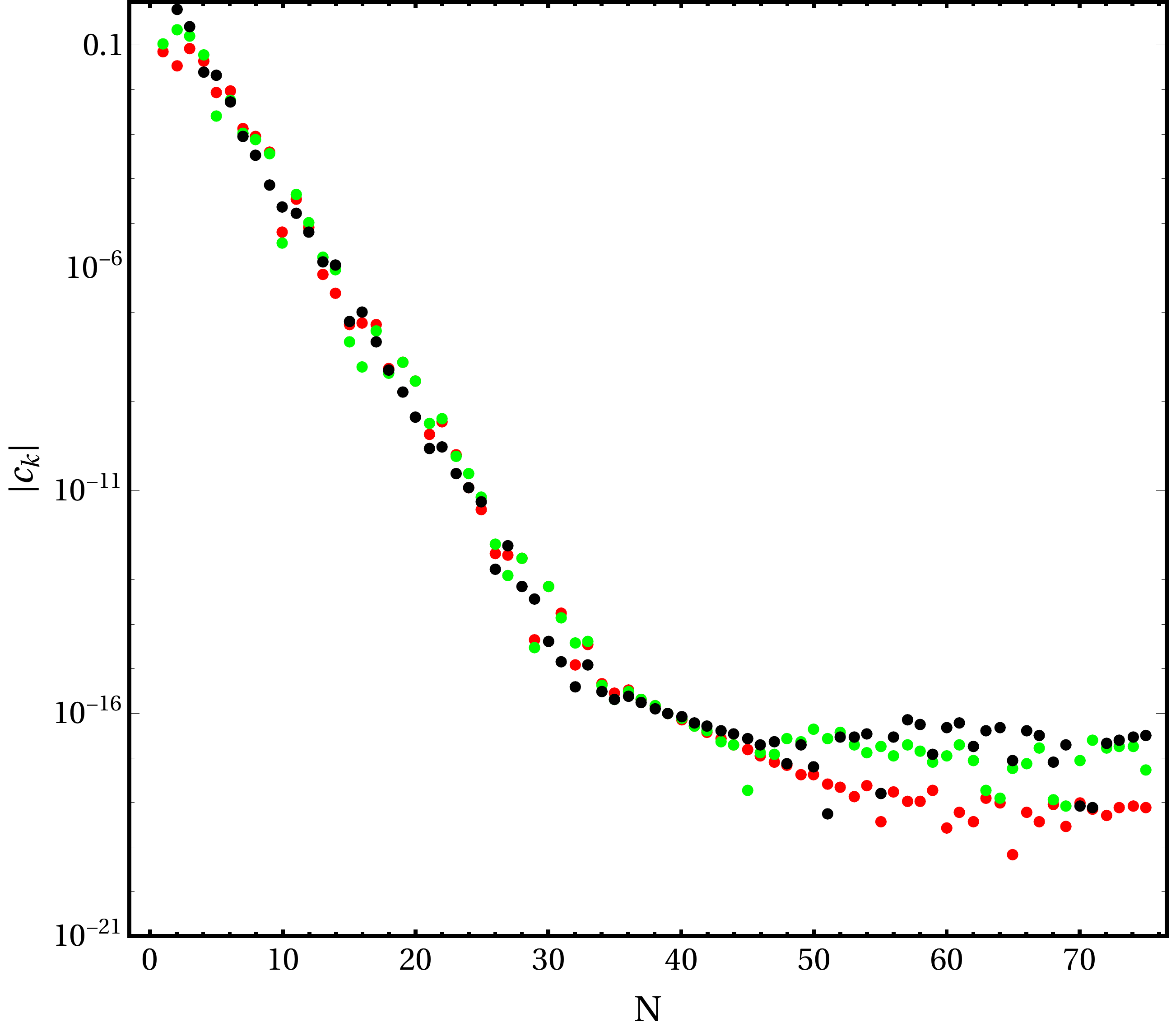}
        \caption{Convergence in the momentum diffusion sector without (left) and with (right) the coordinate mapping $\varrho\mapsto\varrho^2$. The $c_k$ are the Chebychev coefficients of the numerical solution and $N$ is the number of gridpoints of the Chebychev grid used to discretize the radial direction. The parameters are $\gamma=2/\sqrt{3},\, \tilde B=109.658,\,\tilde\mu=7.5$ and the fluctuations are $h_{tx},\,h_{zx},\,a_x$ (red, green, black).}\label{voncergencecm}
\end{figure}

Furthermore, to improve the numerical precision we shift the divergent logarithmic coefficients at the boundary to higher powers in terms of the radial coordinate by introducing the coordinate mapping $\varrho\mapsto\varrho^2$
as introduced in~\cite{Ammon:2016fru,Grieninger:2017jxz} and  more extensively discussed in~\cite{Grieninger:2020wsb}. The improvement of the numerical solution after applying this coordinate mapping may be seen in figure \ref{voncergencecm}, where we depict the convergence for the fluctuations $h_{tx},\,h_{zx},\,a_x$ in the helicity-one sector without coordinate mapping (left) and with coordinate mapping (right). We note that with coordinate mapping, the coefficients fall off geometrically to machine precision before they reach a plateau.

\begin{figure}[ht]
    \centering
   \includegraphics[width=7cm]{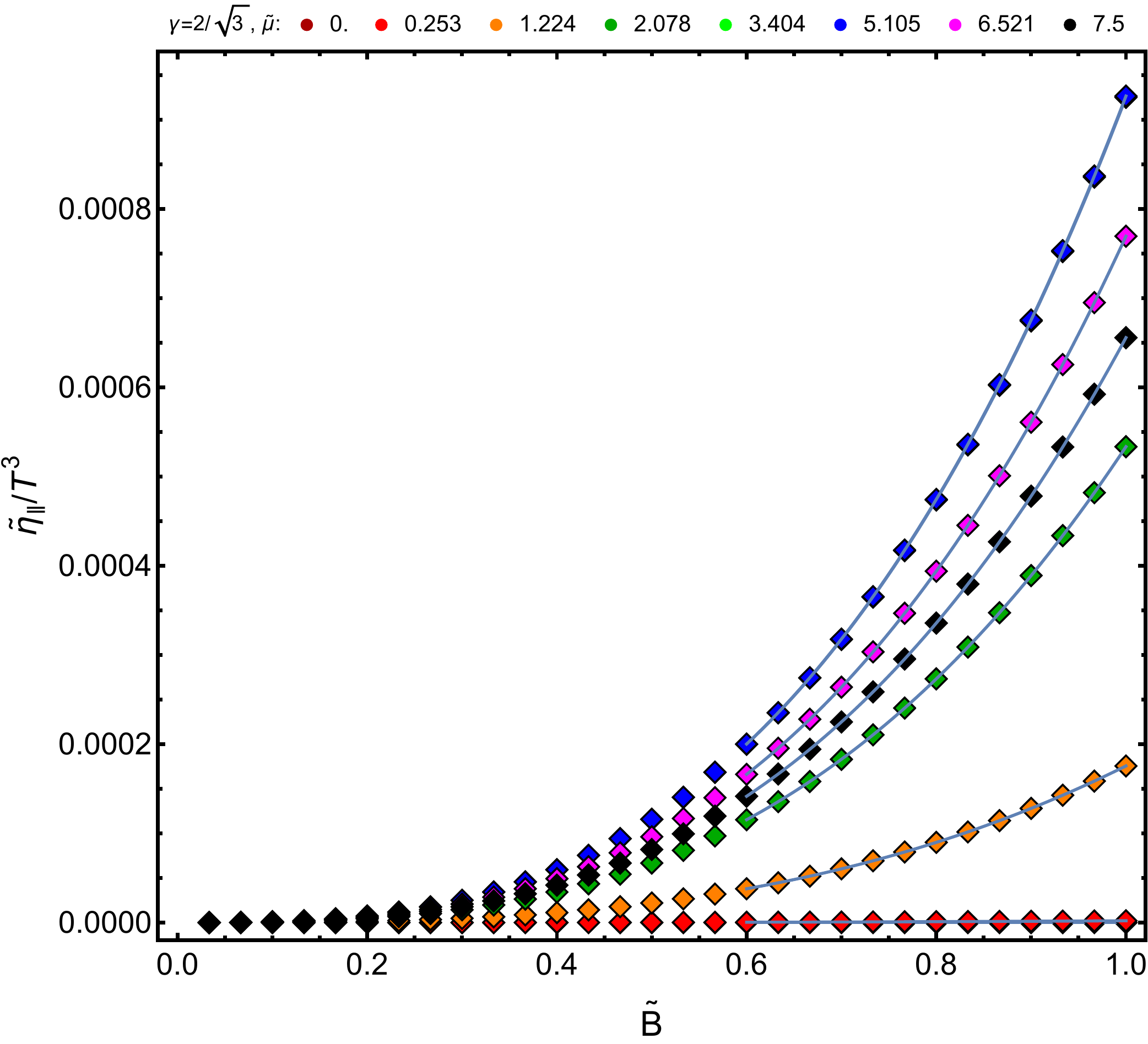}\hspace{0.5cm} \includegraphics[width=7cm]{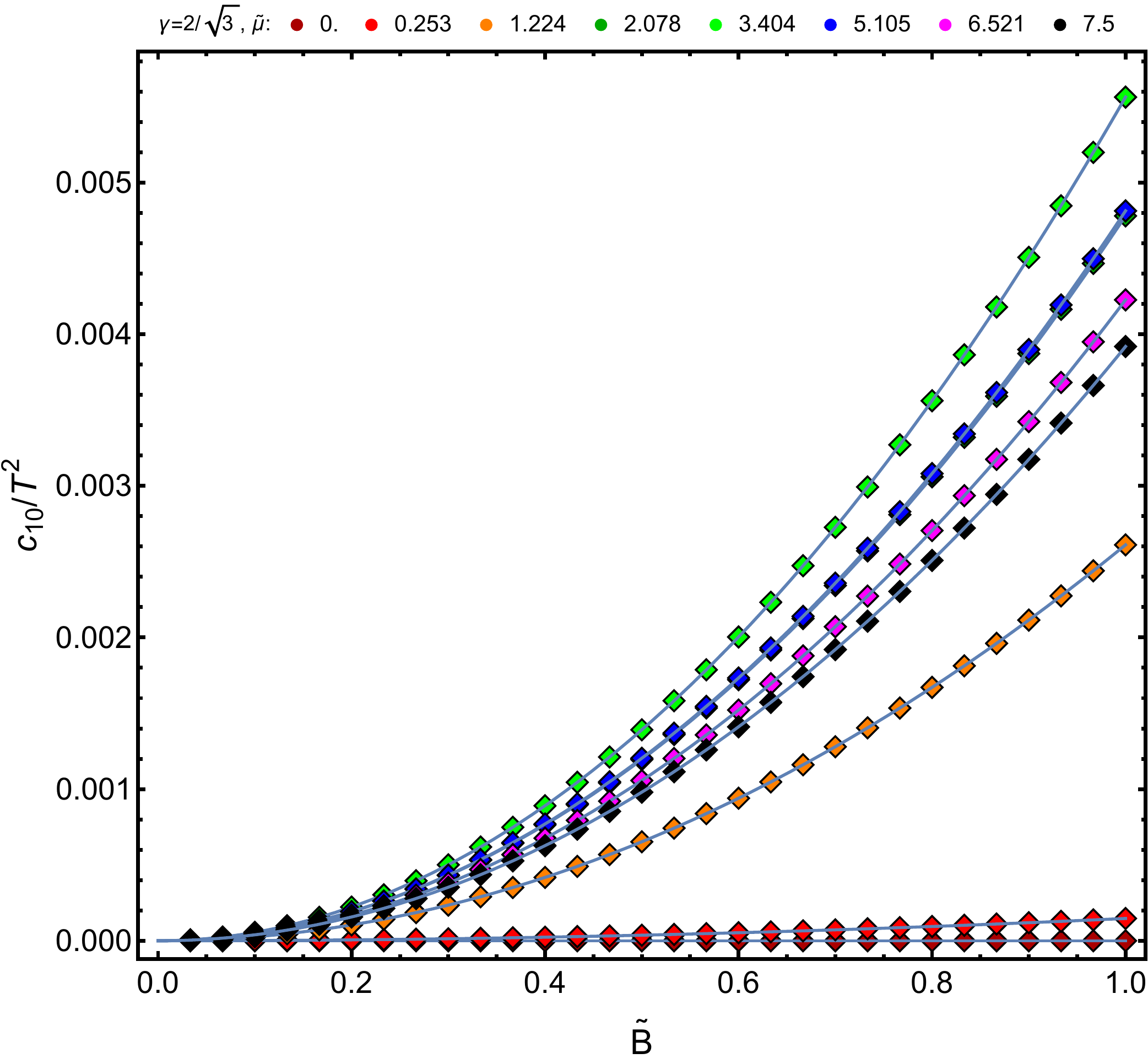}\vspace{-0.1cm}
        \caption{\label{fig:fitPlotAppendix}
         Fit of our numerical data (dots) for $c_{10}$ and $\tilde\eta_{||}$ to the form given in eq.~\eqref{eq:c10tildeEtaFits} (solid lines). 
        }
\end{figure}

In section~\ref{sec:dissipationlessTransportResults} we have mentioned a fit of our numerical data for $c_{10}$ and $\tilde\eta_{||}$ to the form given in eq.~\eqref{eq:c10tildeEtaFits}. The fit is visualized here in figure~\ref{fig:fitPlotAppendix}. 
\subsection{Computing thermodynamic derivatives of dimensionless quantities}\label{app:dimderiv}
We compute dimensionless quantities of the (conformal) field theory using holography. In particular, we use appropriate powers of the temperature $T$ to introduce the dimensionless chemical potential $\tilde{\mu}=\mu/T$ and the dimensionless magnetic field $\tilde{B}=B/T^2.$ In general, for a (thermodynamic) quantity $\mathcal{O}$ with energy dimension $\alpha$ we may introduce the dimensionless quantity $\tilde{\mathcal{O}}= \mathcal{O}/T^\alpha.$ Assuming $\mathcal{O}$ is only a function of $(T,\mu,B)$, the dimensionless quantity $\tilde{\mathcal{O}}$ can only depend on the dimensionless quantities, i.e. $(\tilde{\mu}, \tilde{B})$. In particular this implies 
\begin{equation}
    d\mathcal{O}(T,\mu,B) = \left( \frac{\partial\mathcal{O}}{\partial T} \right)_{\mu,B} dT + \left( \frac{\partial\mathcal{O}}{\partial \mu} \right)_{T,B} d\mu +\left( \frac{\partial\mathcal{O}}{\partial B} \right)_{T,\mu} dB \, , 
\end{equation}
as well as
\begin{equation}
d \mathcal{O}(T,\mu,B) = d\left(T^\alpha \, \tilde{\mathcal{O}}(\tilde\mu,\tilde B)\right) = \alpha T^{\alpha-1} \tilde{\mathcal{O}}(\tilde \mu, \tilde B) \, dT + T^\alpha d \tilde{\mathcal{O}}(\tilde \mu, \tilde B) \, ,
\end{equation}
with 
\begin{equation}
d\tilde{\mathcal{O}}(\tilde\mu,\tilde B) = \left( \frac{\partial\tilde{\mathcal{O}}}{\partial \tilde\mu} \right)_{\tilde B} d\tilde\mu +\left( \frac{\partial\tilde{\mathcal{O}}}{\partial \tilde B} \right)_{\tilde\mu} d\tilde B \, .
\end{equation}
Equating both expressions for $d\mathcal{O}$ and using $d\tilde{\mu} = d\mu/T - \mu \, dT/T^2$ as well as $d\tilde{B}= dB/T^2 - 2 B \, dT/T^3$, we may relate derivatives of $\mathcal{O}$ with respect to $T$, $\mu$ and $B$ to derivatives of $\tilde{\mathcal{O}}$ with respect to $\tilde{\mu}$ and $\tilde{B}$ as follows
\begin{eqnarray}
\left( \frac{\partial \mathcal{O}}{\partial \mu}\right)_{T,B} &=& T^{\alpha - 1}  \left( \frac{\partial \tilde{\mathcal{O}}}{\partial \tilde\mu}\right)_{\tilde{B}} \, , \\
\left( \frac{\partial \mathcal{O}}{\partial B}\right)_{T,\mu} &=& T^{\alpha - 2}  \left( \frac{\partial \tilde{\mathcal{O}}}{\partial \tilde B}\right)_{\tilde{\mu}} \, , \\
\left( \frac{\partial \mathcal{O}}{\partial T}\right)_{\mu,B} &=& T^{\alpha - 1} \left( - \tilde{\mu} \, \left( \frac{\partial \tilde{\mathcal{O}}}{\partial \tilde \mu}\right)_{\tilde{B}}   -2 \tilde{B} \, \left( \frac{\partial \tilde{\mathcal{O}}}{\partial \tilde B}\right)_{\tilde{\mu}} + \alpha \, \tilde{\mathcal{O}}  \right) \, .
\end{eqnarray}
We may apply those expressions to the case of a grand canonical potential, i.e. $\mathcal{O}(T,\mu,B) = \Omega(T,\mu,B)$ which for a four-dimensional relativistic field theory has energy dimension four, i.e. $\alpha=4$. In this case we obtain the usual relations for the charge density $\rho = - \left( \frac{\partial \Omega}{\partial \mu}\right)_{T,B}$, the magnetization $M = - \left( \frac{\partial \Omega}{\partial B}\right)_{T,\mu}$ and the entropy density $s = - \left( \frac{\partial \Omega}{\partial T}\right)_{\mu,B}$ in terms of their dimensionless counterparts.

%
\bibliographystyle{JHEP-2}
\bibliography{EMBranes}

\end{document}